\documentclass[12pt,aasms4]{article}
\usepackage{aasms4}
\usepackage[dvips]{graphicx}

\def\lapprox{\hbox{\lower .8ex\hbox{$\,\buildrel < \over\sim\,$}}}
\def\gapprox{\hbox{\lower .8ex\hbox{$\,\buildrel > \over\sim\,$}}}

\def\OI{O\,{\sc i}}

\def\NaI{Na\,{\sc i}}

\def\SII{S\,{\sc ii}}

\def\SiII{Si\,{\sc ii}}

\def\CaII{Ca\,{\sc ii}}

\begin{document}

\title{Type Ia SNe along redshift: the $\mathcal{R}$(\SiII) 
ratio and the  expansion velocities in intermediate z supernovae}

\author{G. Altavilla \altaffilmark{1}, 
P. Ruiz--Lapuente \altaffilmark{1,2}, 
A. Balastegui \altaffilmark{1},
J. M\'endez \altaffilmark{1,3},
M. Irwin \altaffilmark{4}, 
C. Espa\~na--Bonet \altaffilmark{1}, 
K. Schahmaneche \altaffilmark{5}, 
C. Balland \altaffilmark{5}, 
R. S. Ellis \altaffilmark{4,6}, 
S. Fabbro \altaffilmark{7},  
G. Folatelli \altaffilmark{8}, 
A. Goobar \altaffilmark{8}, 
W. Hillebrandt \altaffilmark{2},
R. M. McMahon \altaffilmark{4},
M. Mouchet \altaffilmark{5},
A. Mourao \altaffilmark{7},
S. Nobili \altaffilmark{8},  
R. Pain \altaffilmark{5}, 
V. Stanishev \altaffilmark{8}, 
N. A.Walton \altaffilmark{4}}

\altaffiltext
{1}{Departament of Astronomy, University of Barcelona, Diagonal
647, E--08028 Barcelona, Spain}
\altaffiltext
{2}{Max--Planck Institut f\"ur Astrophysik, Karl 
Schwarzschildstrasse 1, D-85741 Garching, Germany}
\altaffiltext
{3}{Isaac Newton Group of Telescopes, 38700 Santa Cruz de La Palma, Islas
Canarias, Spain}
\altaffiltext
{4}{Institute of Astronomy, University of Cambridge, Madingley
Road,Cambridge. CB3 0HA, United Kingdom}
\altaffiltext
{5}{LPNHE, CNRS--IN2P3 and Universities of Paris  \& 7, F--75252
Paris Cedex 05, France}
\altaffiltext 
{6}{California Institute of Technology, Pasadena. CA 91125, USA}
\altaffiltext 
{7}{CENTRA--Centro M. de Astrofisica and Department of Physics,
IST, Lisbon, Portugal}
\altaffiltext 
{8}{Department of Physics, Stockholm University, SE--10691
Stockholm, Sweden}

\slugcomment{{\it Running title:} Type Ia SNe along redshift}

\begin{abstract}
 
 We  study intermediate--z  SNe Ia using the empirical 
 physical diagrams which enable to learn about those
 SNe explosions. 
 This information can be very useful to reduce systematic
 uncertainties of the Hubble diagram of SNe Ia up to high z.
 The study 
 of the expansion velocities and  the measurement of the ratio
 $\mathcal{R}$(\SiII) allow to subtype SNe Ia as done in nearby samples. The
 evolution of this ratio as seen in the diagram $\mathcal{R}$(\SiII)--(t)
 together with $\mathcal{R}$(\SiII)$_{max}$ versus (B-V)$_{0}$ indicate 
 consistency of the properties at intermediate z compared with the nearby
 SNeIa.  At intermediate--z, expansion velocities of Ca II and
 Si II are found similar to those of the nearby sample.
 This is found in a sample of 6 SNe Ia  in the range 0.033$\leq z \leq$0.329 
 discovered within the  {\it International Time Programme} (ITP) of SNe Ia 
 for Cosmology and Physics in the spring run of 2002 
 \footnote{The programme run under {\it Omega and Lambda
 from Supernovae and the Physics of Supernova Explosions} within the
 {\it International Time Programme} at the telescopes of the {\it European
 Northern Observatory} (ENO) at La Palma (Canary Islands, Spain)}.
 Those supernovae  were identified using the 4.2m William Herschel Telescope. 
 Two SNe Ia at intermediate z were of the cool FAINT
 type, one being a SN1986G--like object highly reddened. The 
 $\mathcal{R}$(\SiII) ratio as well as subclassification
 of the SNe Ia beyond templates help to place SNe Ia in their sequence
 of brightness and to distinguish between reddened and
 intrinsically red supernovae. This test can be done with
 very high z SNe Ia and it will help to reduce systematic uncertainties due 
 to extinction by dust. It should allow to map the high--z sample 
 into the nearby one.

\end{abstract}

               

\section{Introduction}

The measurements using Type Ia Supernovae of the expansion rate of 
the Universe led to the discovery of its acceleration (Riess et al. 1998; 
Perlmutter et al. 1999) and has opened a new field in the identification 
of the driving force of the accelerated expansion, the so called dark energy. 
A large local supernova sample was first studied in the Calan--Tololo survey 
(Hamuy et al. 1996) and nowadays in a series of campaigns at low redshift by 
various collaborations. At high-$z$, the first supernova samples were gathered 
by the Supernova Cosmology Project (Perlmutter et al. 1999; Knop et al. 2003; 
Hook et al. 2005) and the High--Z SN team/ESSENCE (Riess et al. 1998; Tonry et 
al. 2003; Barris et al. 2004; Krisciunas et al. 2005; Clocchiatti et al. 2005).
In the last years, the high--z redshift range has been targeted as well by the 
Supernova Legacy Survey (Astier et al. 2005). The combination of the 
discoveries made by all these collaborations will provide hundreds of SNe Ia 
at z$>$ 0.2.  At very high--redshift, the Higher--Z Team using the {\it Hubble 
Space Telescope} concentrates in the discovery of supernovae at z$>$ 1 to 
better constraint dark energy (Riess et al. 2005). This is also the target of 
the latest runs of the Supernova Cosmology Project which is presently studying 
SNe Ia in galaxy clusters at very high z. 
 
 While the low and high redshift intervals are often targeted,
 the intermediate redshift (0.1$\la$z$\la$0.4) region is still an 
 almost unexplored zone.
 We started a programme to have a well covered sample of SNe Ia between 
 z$\sim$ 0.1 and z$\sim$ 0.4. In this paper, we present  the
 spectroscopic results of the observations  done in spring
 2002, of the ITP project on supernovae for their physics and
 cosmology (P.I. Ruiz--Lapuente)\footnote{The programme run in 2002
 under {\it Omega and Lambda
 from Supernovae and the Physics of Supernova Explosions} within the
 {\it International Time
 Programme} at the telescopes of the {\it European
 Northern Observatory} (ENO) at La Palma (Canary Islands, Spain)}.
 We discuss where these supernovae stand 
 in the empirical physical diagrams used to describe the
 supernova density profile and temperature. 
 The results of the photometric follow--up and their cosmological
 implications will be presented in a forthcoming paper.
 There are prospects that SDSS--II (Sako et al. 2005a,b) will provide 
 a large sample of SNe Ia at those z while the SNFactory 
 (Aldering et al. 2004) will concentrate in  supernovae at z $\sim$ 0.1.
 At ENO, our own collaboration plans to move to
 very high--z to carry campaigns that will explore the physics of
 SNe Ia in detail at those high--z in a similar way as done for the 
 nearby sample. Physical properties of SNe Ia can be better studied
 within intensive supernova campaigns by collecting a large database
 of spectra and photometry for each individual supernova. This task
 has been the aim of the RTN on Physics of Type Ia supernovae
 which compares each single SNIa with model spectra to better
 understand  SNe Ia explosions 
 (Hillebrandt et al. 2005). Detailed spectral evolution provides 
 a complete probe of the SNe Ia ejecta: chemical composition, velocity
 and other physical characteristics of the layers that successively
 become transparent.

\noindent
 At high--z, the 
  observing time per supernova
 has to be optimised. Long exposure times are needed to 
 obtain good S/N spectra for the large amount of candidates in the
 supernova searches. This prevents from having a complete sequence of
 spectra. However, as we will show here, one can go a step beyond what
 has been done up to now with high--z SN spectra and do a finer 
 classification.  
 In this intermediate--z campaigns (see Ruiz--Lapuente 2006 for a
 review), we have seen how 
 information similar to the one gathered in nearby SNe Ia can be 
 gathered at all z.

\noindent
  Spectral studies of intermediate z supernovae have started to
  incorporate the study of expansion velocities of the material within
  the ejecta (Altavilla et al. 2005; Balastegui et al. 2005; 
  Mendez et al. 2005; Balland et al. 2006).
The intermediate redshift spectra gathered offer here the possibility of 
investigating where the SNe Ia stand in distribution of chemical 
elements in the velocity space and the temperature within the ejecta
(Branch et al. 1993a; Hatano et al. 2000;
Benetti et al. 2004; Benetti et al. 2005; Branch et al. 2006). 
This opens a new window
inside the supernova ejecta and allows to test the existence of
continuity in the temperature and spatial gradient characteristics of 
SNe Ia. 
Ultimately, these intermediate redshift Type Ia SNe will help to 
fill the gap between the local and high--z SN samples, reducing the
statistical uncertainties by means of an evenly sampled Hubble diagram. These
supernovae are to be used in cosmology in conjunction with those gathered by
other surveys.

The outline of the paper is as follows. In section 2 we present the
observations  and the data reduction procedure,  and in section 3 the
SN candidates classification. In section 4 we comment the results  on
individual objects and bring them into comparison with the nearby
sample. Matches of the spectra
to nearby SNe Ia are examined. In section 5
the physical diagrams for intermediate z SNe Ia are first
buildt up in a way similar to what is done in nearby SNeIa. The
prospects to use these diagrams to reduce systematic
uncertainties are shown. Further discussion 
is presented in section 6. A brief summary of the run and
 conclusions are reported in section 7.


\section{Observations and data reductions}

Spectra of the SN candidates were obtained using the 4.2-m William
Herschel Telescope (WHT)\footnote{The WHT is
 operated by the Isaac Newton Group of
Telescopes (ING), located at the Roque de Los Muchachos Observatory,
La Palma, Spain}. Observations were 
done using the spectrograph ISIS
 on  June 10$^{th}$ and 11$^{th}$, 2002 (Table 1). A
dichroic allowed to carry
 simultaneous observations  in  the blue and red channels, which are 
optimised for their respective wavelength ranges ($\sim$3000--6000 \AA, 
$\sim$5000--10000 \AA). In the blue, 
the R158B grating was used in conjunction with the EEV12 detector. 
In the red, we used the  R158R grating  + GG495 filter\footnote{An
order-separation filter designed to filter out second order blue from
the first order red in usual applications, in the case of ISIS, where
a dichroic splits the beam into red and blue channels, it helps define
the 'red' channel short wavelength} and the MARCONI2 detector. In the red
channel fringing begins at about 8000 \AA\ and increases to $\sim$10
per cent at 9000 \AA. A longslit of 1.2 arcseconds was used in the
first night, under good weather conditions, and a longslit of 1.03
arcseconds was adopted  in the second night, under excellent weather
conditions (except for SN 2002lk and one spectrum of SN 2002lj, which
were observed at the beginning of the night with a 1.2'' slit). 
A journal of the spectral observations is given in Table 1.

Spectra were reduced following standard {\sc IRAF}
\footnote{http://iraf.noao.edu/ \\ 
{\sc IRAF} is distributed by the National Optical Astronomy
Observatories (NOAO), which are operated by the Association of
Universities for Research in Astronomy (AURA), Inc., under cooperative
agreement with the National Science Foundation.} procedures. All
images were bias subtracted and then flat fielded using dome
flats. The one dimensional spectrum extractions  were weighted by
variance based on the data values and a Poisson/CCD model using the
gain and read noise parameters. The background was interpolated by
fitting two regions beside the spectra  and then subtracted. Reference
spectra  of  Cu-Ne-Ar lamps were used for the wavelength
calibration. The results were checked measuring the position of bright
[\OI] sky lines at 5577 \AA\ and 6300 \AA\ and, when  necessary, a
rigid shift was applied to the spectrum to be consistent with these
values. The spectra were flux calibrated using  spectrophotometric
standard stars observed at the start and at the end of each
night. Correction for atmospheric absorption was applied to the red
arm spectra. The blue and the red sections were
joined in a single spectrum and multiple spectra of the same object were 
then combined in order to improve the signal to noise ratio. Spectra with 
different exposure times were weighted accordingly.


\section{Classification}
\subsection{Object, redshift and phase 
determination}\label{objectidentification}
Spectra  can be inspected visually in order to give a rough
classification of the SN candidate, where the
main feature used to discriminate between Type Ia, Type II SNe or
`other' sources (typically QSOs or AGN) is the presence/absence of the
strong \SiII\ absorption at  $\sim$6150 \AA\ (rest-frame)  and  the
typical  \SII\  $\sim$5400 \AA\  (rest--frame) `W' feature for Type Ia
SNe. Type II SNe show 
the characteristic Balmer series P--Cygni profiles, most noticeably
the H$_\alpha$ spectral feature. For a correct
 classification, one has to be aware of
the possible confusion between SNe Ia and SNe Ic. This possibility 
increases its chances for cool SNIa events when only one spectrum is
available. As it will be shown in this paper, such possibility can be 
reduced if one uses the physical diagrams (expansion velocity and  
$\mathcal{R}$(\SiII)) in intermediate and 
high--z SNe Ia.  
For all those reasons and to find the best match in a library of
SNe Ia, it is required to follow
an automatic procedure that will enable to size the difference of the
spectra of the SNe Ia with those of an archive from a large sample of nearby
SNe Ia at all phases. 

The redshift of the supernova is obtained from  
the redshift of the galaxy lines. In case of no emission lines, it
can be obtained by the algorithm. 
In our sample, the 
spectra were inspected  looking for typical narrow  galaxy lines:
Balmer lines, [O\,{\sc ii}] $\lambda3727$, [O\,{\sc iii}]
$\lambda5007$, [N\,{\sc ii}] $\lambda6583$, [S\,{\sc ii}]
$\lambda6716, 6731$). Images of the host galaxy of the supernovae 
are shown in Figures 1, 3, 5, 7, 9, and 11 and figures of the 
supernova spectra with the emission lines of the underlying host galaxy
 are shown in Figures 2, 4, 6, 8, 10, and 12 (top
panels). The lines used for the redshift determination (when present) are 
shown. Uncertainties on the redshifts are of the order of 0.001 and they 
have been estimated measuring the dispersion of the redshift determinations 
obtained from  each identified galaxy line.

The following step was taken to refine the Type Ia SN classification by means
of two different classification algorithms developed to this aim. The
first classification program transforms the spectra into rest--frame
and compares it to a set of Type Ia supernovae spectral templates
originally prepared by Nugent, Kim, \& Perlmutter (2002), 
and later adapted by Nobili et al. (2003). These spectral templates range 
from 19 days before maximum to 70 days after maximum, and the wavelength 
coverage is from 2500 \AA\  to 25000 \AA. Both the spectra and the templates 
are normalized in the wavelength range selected for the comparison, and
then the spectra are again rescaled and shifted in the flux axis
until the best match is found, see Figs. 2, 4, 6, 8, 10, and 12 (middle
panels) This procedure uses the whole spectrum and the result is not based 
on a few key features only.

In an interactive mode, the algorithm asks
 for a smoothing length and a $\sigma$
level to clean spikes in the observed spectrum. In this second mode, 
 the procedure 
allows the user to specify the wavelength interval  or to reject
wavelength intervals of the comparison spectrum. This can be useful to
reject a region of the spectrum highly contaminated with atmospheric
 absorption lines. Finally, the user can
 select the interval of redshifts and epochs used in the comparison.

Mathematically the algorithm works by finding the minimum $\chi^{2}$
 of the observed spectrum compared with spectra of
 all the possible values of the epoch, $j$, and the redshift, $z$:

\begin{equation}
\chi^{2}_j(z)=\sum_{i=1}^n \frac{[f(\lambda_i)-F_j(\lambda_i,z)]^2}
{{\sigma_i}^2}
\end{equation}
where $n$ is the total number of data points of the observed SN spectrum, 
$f(\lambda_i)$ is the SN normalized flux at wavelength $\lambda_i$, and 
$F_j(\lambda_i,z)$ is the template normalized flux at wavelength $\lambda_i$, 
epoch $j$, and redshift $z$. The algorithm delivers rest--frame epoch, 
flux scale and redshift as parameters. If the redshift is known from the 
narrow galaxy lines, the algorithm takes it as given and the redshift is not 
used in the minimization.

\noindent
Once the first classification is obtained
 based on templates, a second analysis
is done using an algorithm that
 compares the SNe Ia with those of the Padova
SN Catalogue and other spectra available in the literature 
(we also made use of the SUSPECT 
database\footnote{http://bruford.nhn.ou.edu/$\sim$suspect/index1.html}).
This allows to find a real template which best fits the SN Ia
(Figs. 2, 4, 6, 8, and 10, lower panel). 
The two algorithms {\it Genspecphase} and {\it Genspecsubtype}
were developed for this programme and can be used in campaigns at all
redshifts.  Results with the redshift z and phase determination 
can be seen in Table 2. 

\noindent
We estimated the spectral epoch and its uncertainty  by comparing the
 phases obtained matching the spectra  with the synthetic and real
 templates. We assumed a minimum error of $\pm$2 days (see also
 Riess et al. 1997). The comparison of the spectral phase
 $\tau_{spec}$ with the phases  $\tau_{pho}$ determined from the
 photometric data (see Table 2) shows that the spectral 
 epochs are correct within a few days, with a scatter of $\sigma=3.5$ days
 (Fig. 13). This value is  consistent with the
 adopted spectroscopic phase error bars.

\noindent
 SN 2002lk presented an interesting test for classification. 
 It was an intrinsically red (cool) SNIa and in addition to
 it, it is highly reddened. A rough approach might have mistaken it
 for a SNIc at another phase. We included a number of intrinsically
 red (cool) SNeIa in our database as well as Type Ic supernovae to increase 
 the quality of the classification. Consistency is found with the color 
 light curves and the spectral classification. 

\noindent
Observed and template spectra have been smoothed and 
tilted with an absorption amount derived empirically within the
classifying procedure. This `reddening factor' may not be totally
 due to absorption, and it is applied  to correct for light losses due
to observations with the slit out of the parallactic angle and/or to
take into account different colours between targets and templates,
and/or to take into account reddened templates. In some cases the
spectral templates have been slightly red- or blueshifted to account
for the small differences in the expansion velocities between the template 
and the observed SN.

\noindent
The peculiar Type Ia SN 2002lk and Type II supernovae  have been compared  
with real templates only (Figs. 12, 14, 15).

Table 3 summarizes the results of the classification of the host galaxies 
of all SN candidates.

\subsection{Host galaxy morphology}

Host galaxy morphology identification is usually done
 exploiting good imaging and spectra (Sullivan et al. 2003). In our sample,
 a certain visual identification of the host galaxy morphology is
 feasible for two objects only: SN 2002lq (Fig. 7), and
 SN 2002lk (Fig. 11). The latter is a Type Ia SN similar to the underluminous 
 SN 1986G, which exploded in a spiral galaxy (for a discussion on the host 
 galaxies of subluminous SNe, see Howell 2001). This galaxy has been 
 classified as spiral since its ellipticity, larger than 7, is not consistent 
 with an elliptical galaxy and, although it is observed  almost edge on, the
 existence of some structure, as a dust lane, can be detected. An
 image of the galaxy about one year after explosion reveals also the 
 bulge of the galaxy, hidden in Fig.11 by the SN light.

The rest of the galaxy images  do not give hints on the galaxy
morphology,  being compatible with both elliptical and spiral galaxies
(in the latter case the possible spiral structure is unresolved) 
(Figs. 1, 3, 5, and 9). Since we have no spectra of the host galaxies
alone, the identifications given are made exclusively
on the basis of the galaxy lines contaminating the SNe spectra
(Kennicutt 1992). Thus, we will only distinguish between spheroidal 
galaxies (ellipticals and lenticulars) and spiral galaxies (without giving 
any further subclassification).

\noindent
SN 2002li and SN 2002lq host galaxies are  classified as  spiral
galaxies, since they have a good number of lines from the Balmer series. 
The host galaxy morphology of the SN 2002lq host can be also inferred from 
visual inspection, while SN 2002li host galaxy structure is not as clear as 
the previous case.

\noindent
SN 2002lj lacks galaxy lines. Without any clear Balmer emission line,
 we have classified it as a spheroidal galaxy.

\noindent
SN 2002lp has H$\alpha$, H$\gamma$ and H$\theta$ in emission, but very
 faint. Its host galaxy is possibly spiral.
 The spectrum of SN 2002lr presents also narrow emission lines.

\noindent
The Type II SNe, SN 2002ln and SN 2002lo, show narrow emission 
lines of the Balmer series. Their host galaxies have been classified 
as  spirals (Table 3).

\section{Spectroscopic classification}

\subsection{Type Ia Supernovae}

\noindent
{\bf SN 2002li}

\noindent
This is the farthest supernova discovered in this search ($z=0.329$).
Fig. 2 (bottom panel) shows the comparison with SN
2000E  (Valentini et al. 2003), a SN spectroscopically  almost identical
to  SN 1990N, i.e. a typical Type Ia SN. Both the comparison with SN
2000E and Nobili's spectral templates (Fig. 2,
middle panel) suggest that the SN 2002li phase corresponds to a few days 
before maximum. 

The blue-shifted  H and K lines at 3950 \AA\ indicate that the  expansion 
velocity of \CaII\ in SN 2002li is $\sim$17200 km~s$^{-1}$, 
lower than for SN 2000E ($\sim$21000 km s$^{-1}$). The signal to noise ratio 
does not allow a reliable determination of  the expansion 
velocity of \SiII\ at $\sim$6150 \AA\ . Photometric data suggest that  
SN 2002li is a slow decliner object ($\Delta m_{15}(B)\sim0.79\pm 0.22$), but 
being spectroscopically normal. In the local sample, slow decliners are
often normal spectroscopically (Hamuy et al. 2002). 
\\

\noindent
{\bf SN 2002lj}

\noindent
The redshift of SN 2002lj has been derived from SNe features due to
 the absence of measurable galaxy emission lines. Both the match
with a real SN (SN 1994D, Patat et al. 1996)
(Fig. 4, bottom panel) and a template spectrum (Fig. 4, middle panel)
 suggest that the observed epoch for SN 2002lj is about a week past
maximum. Since SNeIa still show \SII\ features between 5500 \AA\ and
5700 \AA\ 7 days after maximum, the absence of these features in the SN 2002lj 
spectrum  suggests a phase between 7 and 10 days past maximum.
The redshift was found by comparison with the template SN features,
thus most of them have the same expansion velocity, except \SiII\ at 
6355 \AA, whose expansion velocity for SN 2002lj is lower ($\sim$8900 km 
s$^{-1}$) than the one measured for SN 1994D ($\sim$9800 km s$^{-1}$).
\\

\noindent
{\bf SN 2002lp}

\noindent
The finding chart for this SN (Fig. 5) shows the large offset from the 
host galaxy. The comparison of the spectra of SN 2002lp with that of 1989B at
maximum (Barbon et al. 1990; Wells et al. 1994) shows that they are
quite similar  (Fig. 6, bottom panel). The match with a template at day 3 
after maximum (Fig. 6, middle panel), confirms that the spectrum of SN 2002lp 
was taken close to maximum. The expansion velocity measured from the \SiII\ 
unresolved doublet 6355 \AA\  ($\sim$10400 km~s$^{-1}$) is almost coincident 
with the one measured for SN 1989B ($\sim$10000 km~s$^{-1}$). The
absorption dip at $\sim$5800 \AA, identified as \SiII\ 5972 \AA,  is
stronger than in SN 1989B. This is quantified by the 
$\mathcal{R}$(\SiII) parameter, or ratio of the two Si II absorptions
dips, i.e. the one at $\sim$5800 \AA\ and the one at $\sim$ 6150 \AA\ 
(Nugent et al. 1995). In SN 2002lp, $\mathcal{R}$(\SiII) is 0.50 $\pm$
0.05 and in SN 1989B is 0.29 $\pm$ 0.05 for the same period. 

The match with a template at day 3 after maximum, confirms that the
spectrum of SN 2002lp was taken near maximum. The expansion velocities
measured from the \SiII\ unresolved doublet (6355 \AA) and \CaII\
lines (3950 \AA) ($\sim$10400 km~s$^{-1}$, $\sim$13800 km~s$^{-1}$
respectively) are almost coincident with the ones measured for SN
1981B ($\sim$10400 km~s$^{-1}$, $\sim$14000 km s$^{-1}$). While the
velocities for SN 2002lp are in the range of those found in SN1981B, 
the absorption dip at $\sim$5800 \AA, identified as
\SiII\ 5972 \AA,  is stronger than in SN 1981B. In SN 1981B, 
 $\mathcal{R}$(\SiII) is 0.16 $\pm$ 0.05 near maximum, a value which
is typical for normal SNe Ia. 
The spectral comparison with the complete SNe Ia library finds that
SN 1989B is the best match for this supernova (Fig. 6, bottom panel).
\\

\noindent
{\bf SN 2002lq}

\noindent
The \SiII\ (6355 \AA) feature if existent, is very faint but the
comparison with SN 1990N, a  typical SN Ia with good spectral coverage
(Leibundgut et al. 1991), suggests that SN 2002lq is a Type Ia SN observed
about one week before maximum (Fig. 8, bottom panel). Matching synthetic 
templates gives a phase of 7 days before maximum for this supernova 
(Fig. 8, middle panel). The only well visible feature in 
this low S/N spectrum is \CaII\ line at 3950 \AA, and its expansion
 velocity ($\sim$18900 km s$^{-1}$),  measured with large uncertainty,
 seems somewhat  lower than  that of SN 1990N ($\sim$21000 km s$^{-1}$). 
\\

\noindent
{\bf SN 2002lr}

\noindent
This Type Ia supernova was observed about 10 days after maximum, 
as derived from fitting the templates and fitting to real
SNe Ia (Fig. 10, middle and bottom panels). SN 2002lr spectrum is similar 
to one of SN 1994D  (Patat et al. 1996) 10 days past maximum. The expansion 
velocity derived from \SiII\ (6355 \AA) ($\sim$8600 km s$^{-1}$) is similar 
to the one derived for SN 1994D ($\sim$9500 km s$^{-1}$). The expansion 
velocity measured from the \CaII\ line at 3950 \AA ~ is $\sim$12800 km 
s$^{-1}$ but it is not possible to compare this value with the corresponding 
one in SN 1994D because the limited spectral range of the template does not 
allow the measurement of the \CaII\ line.
\\

\noindent
{\bf SN 2002lk}

\noindent
For SN 2002lk  two different spectra are available since it was
observed in two consecutive nights (Fig. 12). 
The strong similarities between the spectra of SN 2002lk and SN 1986G 
(Phillips et al. 1987); Padova SN Catalogue), suggest that SN 2002lk is a 
SN 1986G--like event observed a few days before maximum. Due to the spectrum  
peculiarities, a comparison with Nobili's spectral templates is not 
appropriate. In Fig. 12, the spectra of SN 1986G have been
shifted in wavelength in order to match better that of SN 2002lk. This
small shift takes into account the different expansion velocities of
the two objects. For SN 2002lk the velocity determined from the
\SiII\ line at 6355 \AA~ is $\sim$14500 km s$^{-1}$, while for SN
1986G is  $\sim$10800 km s$^{-1}$. The significant  difference between
the two values  may be attributed to different explosion energy.
SN 2002lk is characterized by  higher expansion velocity, slower decline rate 
in the $B$ light curve ($\Delta m_{15}(B)\sim 1.23\pm 0.02$), and  similar 
$\mathcal{R}$(\SiII) with respect to SN 1986G. Thus, SN 2002lk, in its turn, 
can be considered a less extreme object  with respect to SN 1986G, suggesting 
a continuous transition from object to object. We notice that  similar high  
expansion velocity was found for SN 1998de, a  1991bg-like SN (Modjaz et al. 
2001), whose expansion velocity (6 days before maximum) derived from the 
\SiII\ (6355 \AA) minimum was 13300 km s$^{-1}$ (Garnavich, Jha, \& Kirshner 
1998).

As far as reddening is concerned, the detection of the \NaI D
($\lambda\lambda$5890, 5896 \AA) allows us to estimate the host galaxy
component of the color excess $E(B-V)$. The host galaxy is a spiral
galaxy and the supernova is highly reddened by the galaxy dust lane. 
By means of an empirical relation between the equivalent width (EW) of the 
\NaI D lines and $E(B-V)$ (Barbon et al. 1990; Richmond et al. 1994; 
Turatto, Benetti, \& Cappellaro 2003) we estimated $E(B-V)\simeq 0.15-0.5$ 
($A_V\simeq 0.46-1.55$), where the two values correspond to  two different 
slopes of the  relation found by Turatto, Benetti, \& Cappellaro (2003). 
The colour excess $E(B-V)=0.56\pm 0.04$ derived from the  photometric data is 
consistent with the value obtained from the spectroscopic analysis. The 
Galactic component of the absorption is quite negligible: $A_V=0.025$ 
(Schlegel, Finkbeiner, \&  Davis 1998). The spectrum of SN2002lk has 
a very flat bottom  Ca II and Si II absorption troughs. The shape of those 
absorptions suggest that this is a highly asymmetric supernova, like SN 2004dt.

\subsection{Type II Supernovae}

\noindent
{\bf SN 2002ln and SN 2002lo}

\noindent
Both Type II SNe are matched with SN 1999em as template, a normal Type II 
Plateau (Elmhamdi et al. 2003). The phase is not so well defined for 
SN 2002ln since the blue part of the spectrum has a low S/N. 
A comparison of SN 2002ln  with SN 1999em at 9, 14 (Hamuy et al. 2001), 
and 24 (Leonard et al. 2002) days since $B$ maximum, shown in Fig.14, 
suggests a phase of about 2 weeks. The H$_{\alpha}$ minimum is not 
measurable, the peak of the P-Cygni profile, in principle expected to be 
at null velocity, is slower in SN 2002ln ($\sim$220km s$^{-1}$ for SN 2002ln, 
$\sim$2500km s$^{-1}$ for SN 1999em).

A comparison with SN 1999em $\sim$35 days since $B$ maximum 
(Leonard et al. 2002) shows that SN 2002lo is 
about one month old (Fig. 15). The expansion velocity 
measured  from the H$_{\alpha}$ minimum in SN 2002lo ($\sim$8980km s$^{-1}$) 
is slightly higher than the one measured for SN 1999em ($\sim$6100km s$^{-1}$) 
(Fig. 16), while the peak of the P-Cygni profile is faster 
for SN 2002lo ($\sim$1500km s$^{-1}$) with respect to SN 1999em 
($\sim$950km s$^{-1}$). Uncertainties for our measurements are of the order 
of 500 km s$^{-1}$. However Type II Plateau SNe, and Type II SNe in general, 
are a very heterogeneous class, showing a wide range in the photometric and 
spectroscopic properties (Patat et al. 1994; Filippenko 1997).

\section{$\mathcal{R}$(\SiII) parameter \& Expansion velocities}

Several diagrams can help us to learn about these intermediate--z SNe Ia as
explosions and allow a comparison with the nearby sample.  
Expansion velocities for \CaII\ (3950 \AA) and \SiII\ (6355 \AA) lines
have been measured from the blueshift of the lines, as in the local
sample. The ratio of the two Si II lines, i.e. \SiII\ (6355 \AA) and
\SiII\ (5890 \AA) defines the  $\mathcal{R}$(\SiII) parameter as
introduced by Nugent et al. (1995). We measure in this exploration the
$\mathcal{R}$(\SiII) parameter following 
these authors: drawing segments between adjacent
continuum points of the absorption lines and measuring the difference
in flux between the higher excitation transition of the two
(those transitions are the {\rm 4P--5S} and the {\rm 4S--4P}). 
If the photosphere has a higher effective temperature {\it Teff}, one would
expect that the 5800 \AA \ trough would increase towards higher {\it
Teff}. The behavior in SNe Ia is, however, twofold at very early
phases (Fig. 17). In one group, $\mathcal{R}$(\SiII) is seen to increase
significantly towards
the premaximum, hotter phase. This is found in SNe Ia like
 SN2002bo and SN2004dt which present significant 5800
\AA\ troughs compared to the one at 6150 \AA\ in this premaximum phase, while
in another group it is the opposite: $\mathcal{R}$(\SiII) decreases
towards the premaximum hotter phase in SNe Ia like SN 1990N or stays
constant like in SN2003du. Such diversity 
of behaviors could be linked to the presence of Fe and Co in the outer
layers. According to Nugent et al (1995), the Si II lines interact
with line blanketing from Fe III and Co III at premaximum when the 
temperature is high and Fe and Co are substantially 
present in the outer layers. This effect washes out the 
 5800 \AA\ trough. The decrease of $\mathcal{R}$(\SiII) at premaximum
 in SN 1990N  could be due to Fe and Co in the outer
layers, as this supernova showed these elements in the premaximum
spectra. The supernovae which behave like SN 2004dt and SN2002bo
do not have substantial Fe in the outer layers. 
An independent analysis points in that direction for SN2004dt.
In SN 2002bo, intermediate--mass elements were most abundant in the
outer layers (Stehle et al. 2004). The high $\mathcal{R}$(\SiII)
is consistent with the expectation of minor line blanketing by FeIII
and CoIII.

While we have discussed this ratio at premaximum, which could be an
indicator of the physics of the outer layers, we have  that at maximum 
the time evolution in $\mathcal{R}$(\SiII) is quite flat, and one can
define $\mathcal{R}$(\SiII)$_{max}$. The blanketing by Fe III
and Co III at this epoch is likely negligible as such ionization
stages are not found with the lower temperatures at maximum. 
In that epoch $\mathcal{R}$(\SiII)$_{max}$ correlates well with 
absolute magnitude: the more luminous the SNe Ia the lower 
the $\mathcal{R}$(\SiII)$_{max}$ and the fainter the SNe Ia the higher 
the $\mathcal{R}$(\SiII)$_{max}$. We have investigated in Figure 18 the 
$\mathcal{R}$(\SiII)$_{max}$ ratio versus intrinsic color of the supernova. 
We find that the bluer SNe Ia have smaller
$\mathcal{R}$(\SiII)$_{max}$. We have placed our intermediate z 
SNe Ia in that Figure and obtained that those intermediate--z SNe Ia
are within the trend defined by nearby SNe Ia. For consistency of our 
comparison between the intermediate z sample and the nearby sample, 
$\mathcal{R}$(\SiII) ratio has been measured in a similar way. In both cases,
we follow Nugent et al. (1995) in defining the continuum and the minimum 
of the absorption throughs. The E(B-V) in the nearby sample is measured using
the Lira--Phillips relation (Lira 1995; Phillips et al. 1999)
 for the tail of the B-V color, which allows 
to determine the excess E(B--V) by comparison of the tail from 30 to 60 days
after maximum with the tail of unreddened SNe Ia (the tail in that epoch 
shows very low dispersion). In the intermediate 
z SNe Ia, the same has been done when observations 
after maximum were available
and, in the absence of those, we use the color curve for the stretch of the 
SNeIa (Nobili et al. 2003).

We have been able to measure $\mathcal{R}$(\SiII)$_{max}$ for all the
SNe Ia found at maximum. In our sample, spectra of two of the SNe Ia
 were taken one week or more after maximum. Then the Si II 5800 \AA\ trough is
replaced by iron lines and  $\mathcal{R}$(\SiII)$_{max}$ can not be
measured. In the premaximum
 SN 2002lq the S/N level does not allow to measure this ratio. 
It has been possible to obtain a good measurement of
 this ratio for SN 2002lk and  SN 200lp and  measured with larger
error bars for 
SN2002li (see Table 4). Fig. 17 shows the ratio
  $\mathcal{R}$(\SiII) of the depth 
of the \SiII\ 5972 and \SiII\  6355 \AA\ absorption troughs
(see Nugent et al. 1995; Benetti et al. 2005).
SN2002lp and SN2002lk show remarkable high $\mathcal{R}$(\SiII) values
a few days before maximum in consistency with what is found for red,
faint SNIa. SN 2002li has values for $\mathcal{R}$(\SiII)$_{max}$
consistent with what is found from SNe Ia of the same brightness in
the nearby sample.

\section{Discussion}

The spring run of the ITP 2002 on SNe Ia for Cosmology and Physics  gave 
 6 Type Ia SNe,  2 Type II SNe, 7 quasars and 2 Seyfert galaxies. The
 redshift range of all the sample is $0.033\leq z \leq 2.89$ while the
 SN redshift range is $0.033\leq z \leq 0.329$. As expected in a
 magnitude--limited survey, 
the peak in the redshift distribution of Type II SNe
 is closer than the peak in redshift of Type Ia SNe. On the other
 hand, QSO
 are observed up to very high--redshift (Fig. 19).

\noindent
Our sample allowed a careful inspection of SNeIa. We have made 
a comparison of the kinematic and temperature observables in these
intermediate--z SNe Ia with those of the nearby sample. Some SNe Ia
were intrinsically red and, as in the nearby sample, showed a large
($\sim$ 0.5 near maximum)
 $\mathcal{R}$(\SiII)$_{max}$ ratio. Other SNe Ia were bluer with
$(B-V)_{0}$ $<$ 0 near maximum light. Those showed a small
 $\mathcal{R}$(\SiII)$_{max}$ ratio ($\sim$ 0--0.2 near maximum), as 
in the nearby sample. Amongst the intrinsically red SNe Ia, also
called cool SNeIa, we find SN 2002lk and SN2002lp.
 Those cool SNe Ia are similar to SN 1986G 
but with less extreme characteristcs (higher expansion velocity,
slower decline rate of the $B$ light curve). SN 2002lp is a cool
SNIa similiar to SN 1989B. It is a SNIa in between SN1989B and SN1986G. 
It has a deeper \SiII\ $\sim$5800 \AA\ line and larger 
$\mathcal{R}$(\SiII)$_{max}$ than SN1989B. 
Both SN 2002lk and SN 2002lp show a similar $\mathcal{R}$(\SiII) value
($\sim$0.5). 
SN 2002lk and SN 2002lp have some other common characteristics: they have
the largest $\Delta m_{15}(B)$, 
they are the closest and the faintest Type Ia SNe. Such trend is
in accordance with what could happen under cosmological selection effects. 

\noindent
The interesting aspect shown here is that the intrinsic 
faintness of the two SNe Ia 
is not only revealed by a low stretch factor, i.e. high $\Delta m_{15}(B)$,
but also from the large
$\mathcal{R}$(\SiII) values  (Nugent et al. 1995;
Benetti et al. 2005) which correlate with intrinsic color (see Fig. 18),  
This fact opens new avenues for a better control over reddening in
SNe Ia samples gathered in cosmological surveys. 

\noindent
The expansion velocities measured
for all the SNe Ia in this intermediate z sample
are consistent with the values measured for nearby SNe (Fig. 20).

\noindent
 From Table 3 we observe that most Type Ia SNe
 are in spiral galaxies. The 2 Type II observed exploded in spiral
 galaxies as well. The
fraction of  Type Ia  exploded in spirals amounts to
  83 per cent if we take into account the uncertain  classification
of the host galaxy of SN 2002lp and SN 2002lr.
Even if limited by the small statistics and by  
 rough discrimination between spheroidal and spiral galaxies,
 our result is consistent with the ones  derived in a similar
redshift range by Valenti et al. (2005) and Balland et al. (2006)  (94 and
83 per cent of Type Ia host galaxies are respectively classified as
spirals). These values can be compared to the results derived by
Sullivan et al. (2003 and references therein) at low-redshift ($z<0.01$, 
88 per cent) and high-redshift ($0.19\leq z \leq 0.83$, 71 per cent).

\noindent
The comparison between the spectral phase determined from the spectral
analysis and from the photometric analysis shows a good agreement
between the two estimates ($\sigma=3.5$ days). The agreement gets significantly
 worse if the time dilation correction (1+z) is not taken into account in the
light curve fitting ($\sigma=5.6$ days). 

The time dilation effect is non--neglible at this z range 
and measurable in those spectra as in high--z SNe Ia.

\section{Conclusions}

In this paper we present SNe Ia in the desert area of intermediate z, 
in the redshift range 0.033$\leq$ z $\leq$0.329.
The comparison of intermediate $z$ SN spectra with high
signal-to-noise ratio spectra of nearby SNe has not revealed
significant differences in the spectral features and kinematics. In
particular,
 for all the Type Ia SNe of our sample, included a peculiar one, it
has always been possible to find a  nearby Type Ia SN counterpart, and
the expansion velocities derived from the \SiII\ and \CaII\ lines are
within the range observed for nearby Type Ia SNe. The spectral
concordance seems to correspond as well in color and  decline of the 
light curve.
 The comparison between the epochs determined for each spectrum from the
spectral analysis and the photometric epochs derived from fits of the 
light curves show values in agreement within a few days.

\noindent
The main conclusion of this research is that 
the $\mathcal{R}$(\SiII) parameter is useful to investigate the
physics of SNe Ia explosions at high redshift
 and serve as meaningful comparison 
with the nearby SNe Ia sample. It can help to distinguish in the
sequence of bright---faint SNe Ia (sequence correlated with slow--fast
and with bluer--redder at maximum). It 
provides a control over reddening that can be very useful to reduce 
systematic uncertainties. While in the side of large 
$\mathcal{R}$(\SiII)$_{max}$ one finds faint and
intrinsically redder SNe Ia, in the side of low
$\mathcal{R}$(\SiII)$_{max}$ one finds the intrinsically bluer SNe Ia.
We have found that intermediate and nearby SNe Ia follow a
similar $\mathcal{R}$(\SiII)$_{max}$--(B---V)$_{0}$ and the
intermediate z SNe Ia occupy the same space in the evolution
of  $\mathcal{R}$(\SiII) along time than nearby ones. Similar 
results are found for the expansion velocity, 
though a larger sample of SNe Ia at all epochs  
needs to be investigated.

While the behavior of $\mathcal{R}$(\SiII) along time can be linked
to luminosity and color properties, there is no trend correlating
$\mathcal{R}$(\SiII)$_{max}$ with expansion velocities of Si II or Ca
II. In the side of large $\mathcal{R}$(\SiII)$_{max}$ one can have 
SNe Ia with a large Si II velocity or with a lower one. Therefore,
velocity of the intermediate mass elements is not a tracer of the 
luminosity according to the present sample. We also have some doubts
on the correlation of asymmetries in the ejecta with overall
luminosity of the SNIa. SN2002lk is a faint and red SNIa which shows
high asymmetries in the Si and Ca lines, while SN 2004dt, a normally
luminous SNIa, has those asymmetries as well.

\noindent
Moreover, 
the $\mathcal{R}$(\SiII) at epochs well before maximum is an observable
quantity of great value and accesible to high--z SNeIa
searches. While $\mathcal{R}$(\SiII)$_{max}$ correlates well with
intrinsic color, $\mathcal{R}$(\SiII)$_{premax}$ can 
identify the composition of the outer layers of the 
SNeIa observed in cosmological searches. Thus, with the same
spectra obtained in the discovery runs one can investigate 
the nature of those explosions.

\section*{Acknowledgments}

This research was carried through the International Time Programme
{\it Omega and Lambda from Supernovae and the Physics of Supernova
Explosions} at 
the ENO Observatory and it is based on observations made with the 4.2-m 
William Herschel Telescope, operated on the island of La Palma by the Isaac 
Newton Group in the Spanish Observatorio del Roque de los Muchachos of the 
Instituto de Astrofisica de Canarias.
Thanks are given to the scientific staff of the Observatory in Padova,
and specially to Stefano Benetti. This work is supported in part by the
European Community's Human Potential Programme under contract 
HPRN--CT--2002--00303, {\it The Physics of Type Ia Supernovae}.

\clearpage

\appendix 
\onecolumn
\section{Spectral templates}\label{spectra}
The spectral analysis has been performed using both synthetic and real spectra.
In particular we used 90 synthetic SN Ia spectra by Nobili et al. (2003),
based on Nugent's spectral templates (Nugent, Kim, \& Perlmutter 2002, 
{\tt http://supernova.lbl.gov/$\sim$nugent/nugent$_{-}$templates.html}). 
Phases range from -19  to +70 days since
  $B$ maximum, with 1 day step. Wavelength coverage
 is from 2500 \AA\  to 25000 
\AA, but in our analysis the useful range is limited up to $\sim$10000
\AA. Fig. 21 shows the spectral sequence.

We also made use of real spectral templates, mainly collected from  the 
Padova-Asiago SN Catalogue (Pd-As Cat.) 
({\tt http://web.pd.astro.it/supern/snean.txt}, 
Barbon et al. 1990) and SUSPECT 
database ({\tt http://bruford.nhn.ou.edu/$\sim
$suspect/}),  
spanning a wide range in the  decline
rates ($\Delta m_{15}(B)$) of the corresponding $B$ band light curves. In 
particular we used the  spectra listed in Table 5.  Table 5 
shows the spectra available of the comparison SNe Ia that best
match our intermediate z ones, while Fig. 22 shows their  phase 
distribution.

\clearpage

   \begin{table*}
     \centering
      \caption[]{Journal of spectroscopic observations.}
       \label{tab1}
     $$ 
         \setlength\tabcolsep{5pt}
         \begin{tabular}{l c c c c c c c c  }
            \hline
	    \hline
            \noalign{\smallskip}
  SN   &Date      & JD & Telescope & \multicolumn{2}{c}{Exposure. 
time (s)} & Slit \\
            &   (UT)       & -2400000    & Instrument            &    Blue Arm   &   Red arm  & arcsec.  \\
		\hline
            \noalign{\smallskip}
2002li      & 2002 Jun. 10 & 52436.6  & WHT+ISIS  &    900        &    900    &  1.20   \\
	    & 2002 Jun. 10 & 52436.7  & WHT+ISIS  &    900        &    900    &  1.20   \\
	    & 2002 Jun. 10 & 52436.7  & WHT+ISIS  &    900        &    900    &  1.20   \\
2002lj      & 2002 Jun. 11 & 52437.4  & WHT+ISIS  &    900        &    900    &  1.20   \\
	    & 2002 Jun. 11 & 52437.4  & WHT+ISIS  &    900        &    900    &  1.03   \\
2002lp      & 2002 Jun. 10 & 52436.5  & WHT+ISIS  &    900        &    900    &  1.20   \\
	    & 2002 Jun. 10 & 52436.5  & WHT+ISIS  &    900        &    900    &  1.20   \\
2002lq      & 2002 Jun. 10 & 52436.4  & WHT+ISIS  &    900        &    900    &  1.20   \\
	    & 2002 Jun. 10 & 52436.5  & WHT+ISIS  &    900        &    900    &  1.20   \\
2002lr      & 2002 Jun. 10 & 52436.7  & WHT+ISIS  &    900        &    900    &  1.20   \\
	    & 2002 Jun. 10 & 52436.7  & WHT+ISIS  &    900        &    900    &  1.20   \\
2002lk      & 2002 Jun. 10 & 52436.4  & WHT+ISIS  &    300        &    300    &  1.20   \\
	    & 2002 Jun. 10 & 52436.4  & WHT+ISIS  &    600        &    600    &  1.20   \\
            & 2002 Jun. 11 & 52437.4  & WHT+ISIS  &    600        &    600    &  1.20   \\
2002ln      & 2002 Jun. 10 & 52436.6  & WHT+ISIS  &    900        &    900    &  1.20   \\
	    & 2002 Jun. 10 & 52436.6  & WHT+ISIS  &    900        &    900    &  1.20   \\
2002lo      & 2002 Jun. 11 & 52437.6  & WHT+ISIS  &    900        &    900    &  1.03   \\
	    & 2002 Jun. 11 & 52437.6  & WHT+ISIS  &    900        &    900    &  1.03   \\
	    
            \noalign{\smallskip}
            \hline
            \hline
         \end{tabular}
     $$ 
\end{table*}

   \begin{table*}
     \centering
      \caption[]{Summary of observations.}
         \label{tab2}
     $$ 
         \setlength\tabcolsep{5pt}
         \begin{tabular}{l l l l l  l  r  r}
            \hline
	    \hline
            \noalign{\smallskip}

 Object   & Date & R.A.(2000.0)                    & ~~$\delta$(2000.0)     & Redshift~ & Object    & \multicolumn{2}{c}{Notes}\\
 name     & (UT)   &   ~  $^{hh}$ $^{mm}$ $^{ss}$    & ~~~~~$^{\circ}$ ' ''    &           & Type     & $\tau_{spec}^{\star}$  & $\tau_{pho}^{\diamond}$     \\
            \hline
            \noalign{\smallskip}
  
SN 2002li  & 2002 Jun. 10             & 15:59:03.08 & +54:18:16.0  ~  & 0.329   & SN Ia     &  -3$\pm$2  &  -3.31$\pm$1.0\\
   
SN 2002lj  & 2002 Jun. 11             & 16:19:19.65 & +53:09:54.2    & 0.180    & SN Ia     &  +7$\pm$2  &  11.95$\pm$0.9 \\
  
SN 2002lp  & 2002 Jun. 10             & 16:40:11.45 & +42:28:30.2    & 0.144    & SN Ia     &  +1$\pm$3  &  -2.45$\pm$0.1  \\
   
SN 2002lq  & 2002 Jun. 10             & 16:40:28.83 & +41:14:09.1    & 0.269    & SN Ia     &  -7$\pm$2  &  -10.56$\pm$1.3 \\
  
SN 2002lr  & 2002 Jun. 10             & 22:33:12.59 & +01:05:56.7    & 0.255    & SN Ia     &  +10$\pm$2 &   13.47$\pm$3.5\\
 
SN 2002lk  & 2002 Jun. 11             & 16:06:55.92 & +55:28:18.2    & 0.033    & SN Ia .&  -2$\pm$2  &   -2.90$\pm$0.1\\
     
SN 2002ln  & 2002 Jun. 10             & 16:39:24.93 & +41:47:29.0    & 0.138    & SN II     &  $\sim$14        &  \\

SN 2002lo  & 2002 Jun. 11             & 16:39:56.42 & +42:19:20.5    & 0.136    & SN II     &  $\sim$35        &  \\

            \noalign{\smallskip}
            \hline
            \hline
         \end{tabular}
$$ 
$^{\star}$ spectral epoch from the spectroscopic analysis\\
$^{\diamond}$ spectral epoch (rest-frame) from the photometric analysis\\

\end{table*}

   \begin{table*}
     \centering
      \caption[]{Host galaxy morphology.}
         \label{tab3}
     $$ 
         \setlength\tabcolsep{5pt}
         \begin{tabular}{l c l l l l  c}
            \hline
	    \hline
            \noalign{\smallskip}

SN Ia    &   $z$   & Galaxy  Type    &  \multicolumn{2}{c}{Offset}& Identified lines & $^{\star}$Imaging  \\

             \hline
             \noalign{\smallskip}
  
2002li        & 0.329 &spiral      &     0".1 W & 0".2 S    & H$_\alpha$, H$_\beta$, H$_\delta$, [S\,{\sc ii}], [N\,{\sc ii}], [O\,{\sc iii}], [O\,{\sc ii}] & y  \\
   	   	 		     		        		     
2002lj        & 0.180 &spheroidal  &     0".2 W &           & - &  n \\
  	   	 		     		        		     
2002lp        & 0.144 &spiral?     &     0".2 E & 2".0 N    & H$_\alpha$, H$_\gamma$, H$_\theta$, [O\,{\sc ii}] & n \\
   	   	 		     		        		     
2002lq        & 0.269 &spiral      &     4".5 E & 0".7 S    & H$_\alpha$, H$_\beta$, H$_\gamma$, H$_\delta$, H$_\zeta$,  H$_\eta$, H$_\theta$, [S\,{\sc ii}], [O\,{\sc iii}] &  y \\
  	   	 		     		        		     
2002lr        & 0.255 & spiral?   &     3".0 S &           & H$_\beta$, [S\,{\sc ii}], [N\,{\sc ii}], [Ne\,{\sc v}] &  n\\
 	   	 		     		        		     
2002lk        & 0.033 &spiral (Sb) &     0".7 W&0".2 N      & H$_\alpha$, [S\,{\sc ii}], [N\,{\sc ii}], Na\,{\sc D} &  y\\
                    			         	   	 		     		        		     
2002ln        & 0.138 &spiral      &     1".1 W & 8".3 S    & H$_\alpha$, H$_\beta$, H$_\zeta$,  H$_\theta$, [S\,{\sc ii}] &  n  \\
     		        		     
2002lo        & 0.136 &spiral      &     0".6 E &  1".3 S   & H$_\alpha$, H$_\beta$, H$_\theta$, [N\,{\sc ii}], [O\,{\sc iii}], [O\,{\sc ii}]  &  y \\

            \noalign{\smallskip}
            \hline
            \hline
\noalign{\smallskip}
 \multicolumn{7}{l}{$^{\star}$  Host galaxy morphology can be inferred (y)  or not (n) from visual inspection} \\
         \end{tabular}
     $$ 
\end{table*}

   \begin{table}
     \centering
      \caption[]{Expansion velocities for \CaII\ (3950 \AA) and \SiII\ (6355 \AA) lines and  $\mathcal{R}$(\SiII) ratio. }
         \label{tab4}
     $$ 
        \setlength\tabcolsep{1pt}
	\begin{tabular}{l c c c c }
	\hline
	    \hline
            \noalign{\smallskip}
Object & phase$^{\star}$ & Exp. velocity & ~Exp. velocity &  $\mathcal{R}$(\SiII)\\
name &since $B$ max.   &\CaII\ (km s$^{-1}$) & ~\SiII\ (km s$^{-1}$)& \\

	\noalign{\smallskip}
            \hline
            \noalign{\smallskip}

SN 2002li & -3.31$\pm$1.0   & 17200$\pm$500   &	                &  0.2$^{+0.05}_{-0.10}$ \\ 
SN 2002lj &  11.95$\pm$0.9  & 12300$\pm$300   & 8900$\pm$300    &  \\
SN 2002lp & -2.45$\pm$0.1   & 13800$\pm$900   & 10400$\pm$400   & 0.50$\pm$0.05  \\
SN 2002lq & -10.56$\pm$1.3  & 18900$\pm$1400  &	                &  \\
SN 2002lr &  13.47$\pm$3.5  & 12800$\pm$300   & 8600$\pm$400    &  \\
SN 2002lk & -2.90$\pm$0.1   & 		      & 14500$\pm$500   &0.50$\pm$0.05\\
SN 2002lk & -1.94$\pm$0.1   & 		      & 13850$\pm$500   &0.46$\pm$0.05\\

            \hline
            \hline
         \end{tabular}
     $$ 
$^{\star}$ from the light curve fitting
\end{table}


   \begin{table*}
     \centering
      \caption[]{SN Ia templates}
       \label{tab5}
     $$ 
         \setlength\tabcolsep{1.5pt}
          \begin{tabular}{l c c c c l c}
            \hline
            \hline
            \noalign{\smallskip}
 SN &  Type  & N.               & Time             &   $\Delta m_{15}(B)$    &\multicolumn{2}{c}{Reference(s)} \\
    &        & spectra          & coverage$^\star$ &       &                 &  
                   \\
                \hline
            \noalign{\smallskip}
1986G   &   Ia pec.   &  39         &  -4..+324 & 1.69(0.05)
          & Phillips et al. (1987)   
 & Pd-As Cat. - SUSPECT\\
1989B   &   Ia        &   8         &   0..+52  & 1.28(0.05) 
& Barbon et al. (1990) & Pd-As Cat. - SUSPECT  \\
1990N   &   Ia        &   6         &  +2..+333 & 1.05(0.05) 
& Leibundgut et al. (1991)  & Pd-As Cat. - SUSPECT\\
 & & & & & Mazzali et al. (1993) &  \\
1994D   &   Ia        &  31         & -11..+106 & 1.31(0.05) &
          Patat et al. (1996) & 
 Pd-As Cat. - SUSPECT \\
1999em  &   II P      &  40         &  -1..+515 &  -         & 
Hamuy et al. (2001)   &      Pd-As Cat. - SUSPECT \\
 & & & & & Leonard et al. (2002) &  \\
2000E   &   Ia        &   7         &  -6..+124 & 0.94(0.05) &
          Valentini et al. (2003)
 &  Pd-As Cat. - SUSPECT \\
            \noalign{\smallskip}

            \hline
            \hline
         \end{tabular}
     $$ 
$^\star$ since $B$ maximum
\end{table*}

\clearpage

\onecolumn

   \begin{figure}
   \centering
   \includegraphics[width=0.6\columnwidth]{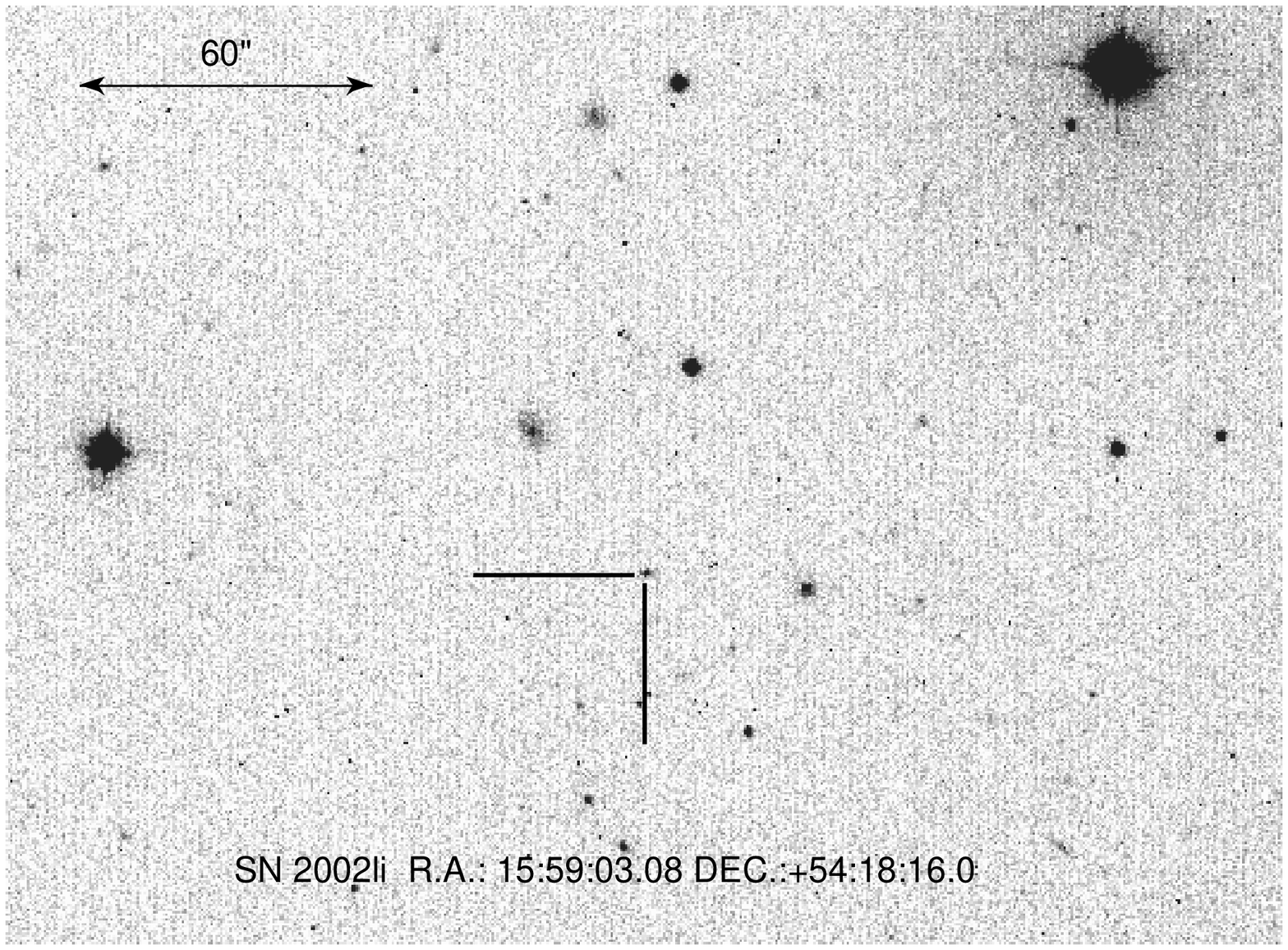}
  
   \centering

   \centering 
   \includegraphics[width=0.6\columnwidth]{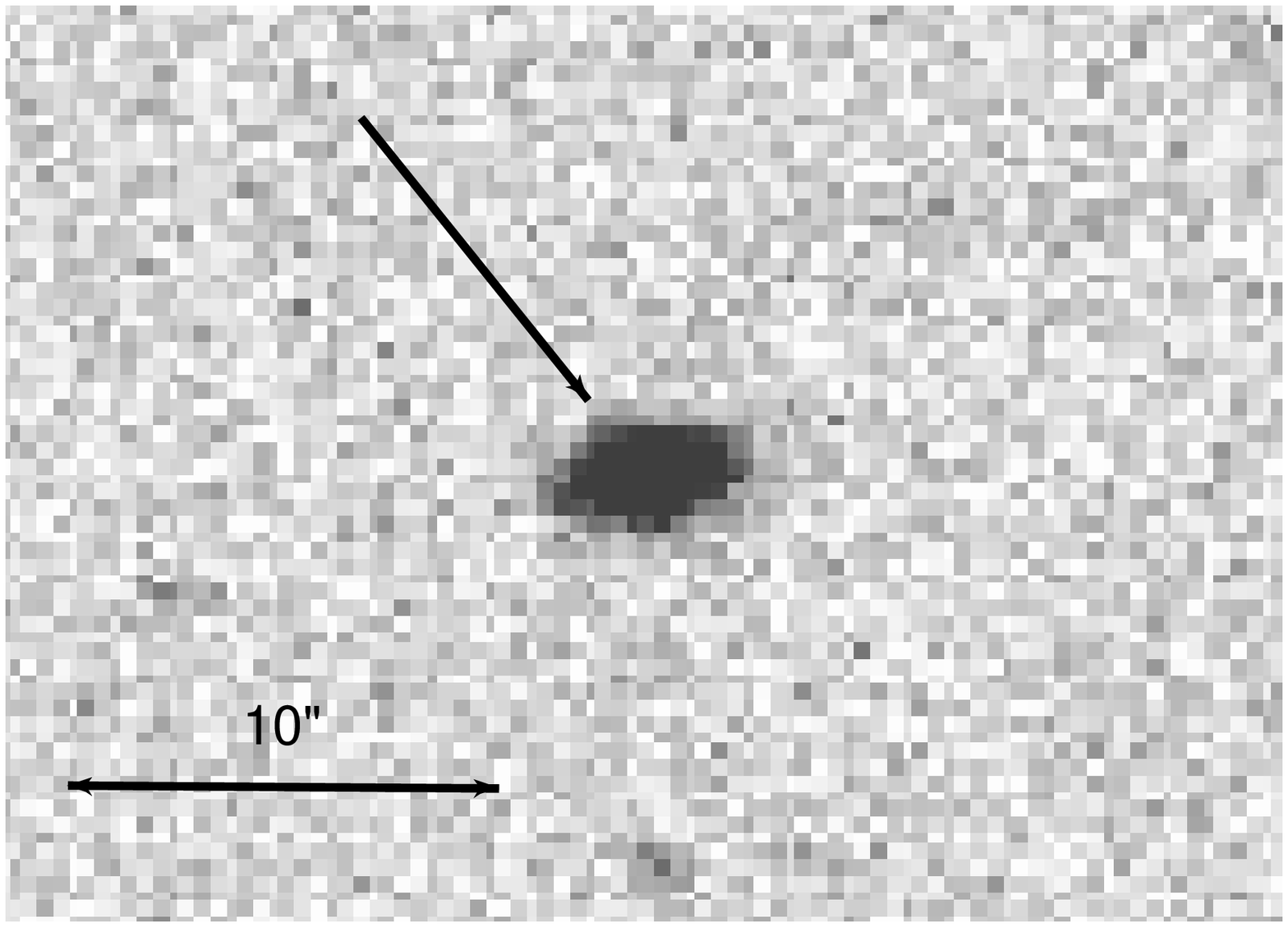}

    \caption{
Top: Finding chart of SN 2002li.  g' band image (exp. time : 600 sec)
obtained on  June 11, 2002 with the JKT 1.0-m telescope + JAG-CCD).
Bottom: SN 2002li. INT+WFC, Jun 7, 2002, exp. time: 240 sec, g' band.}
         \label{Figure 1}
\end{figure}

   \begin{figure}
   \centering
   \includegraphics[angle=-90,width=0.45\columnwidth]{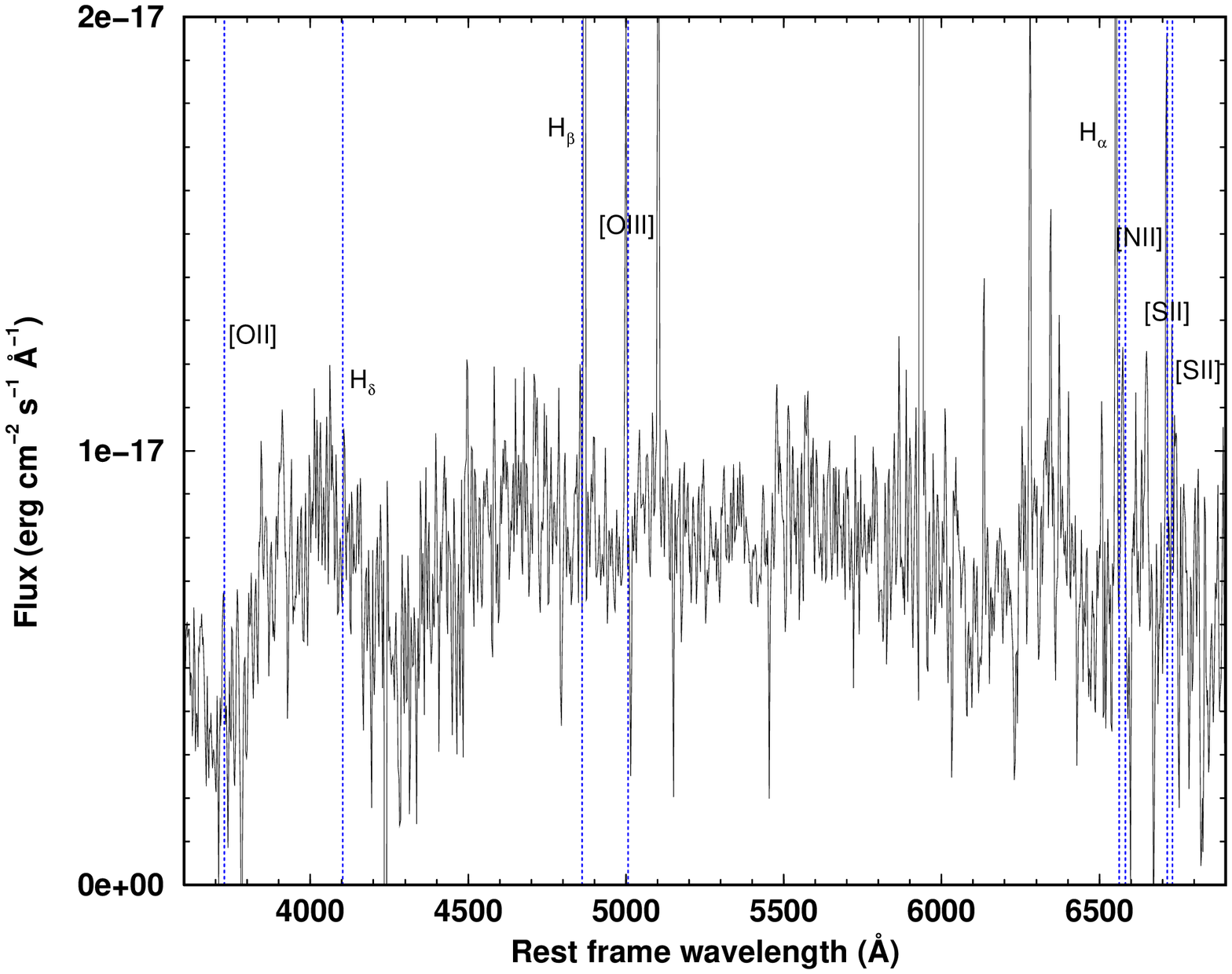}
     
   \centering
   \includegraphics[angle=-90,width=0.45\columnwidth]{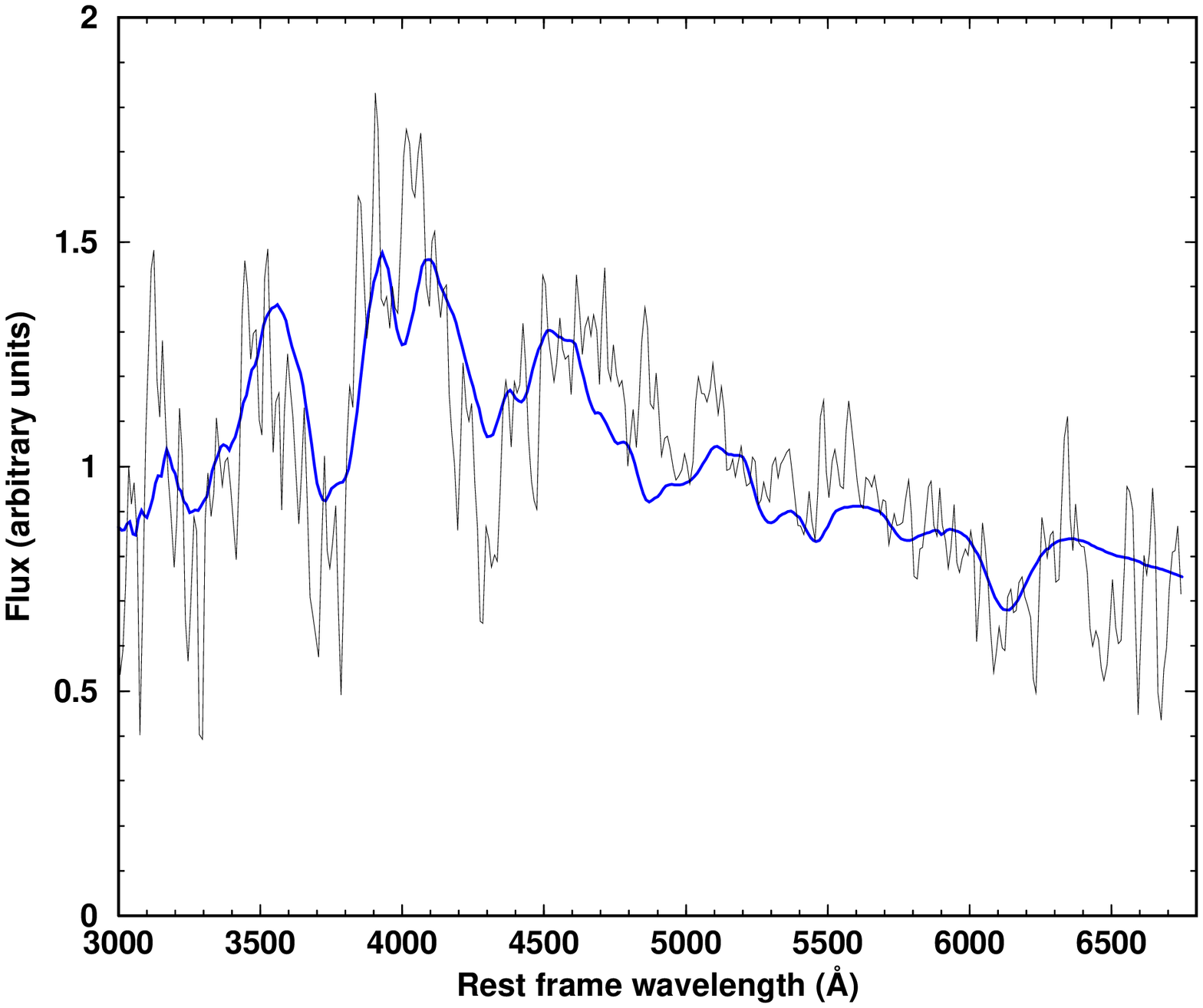}

   \centering
   \includegraphics[angle=-90,width=0.45\columnwidth]{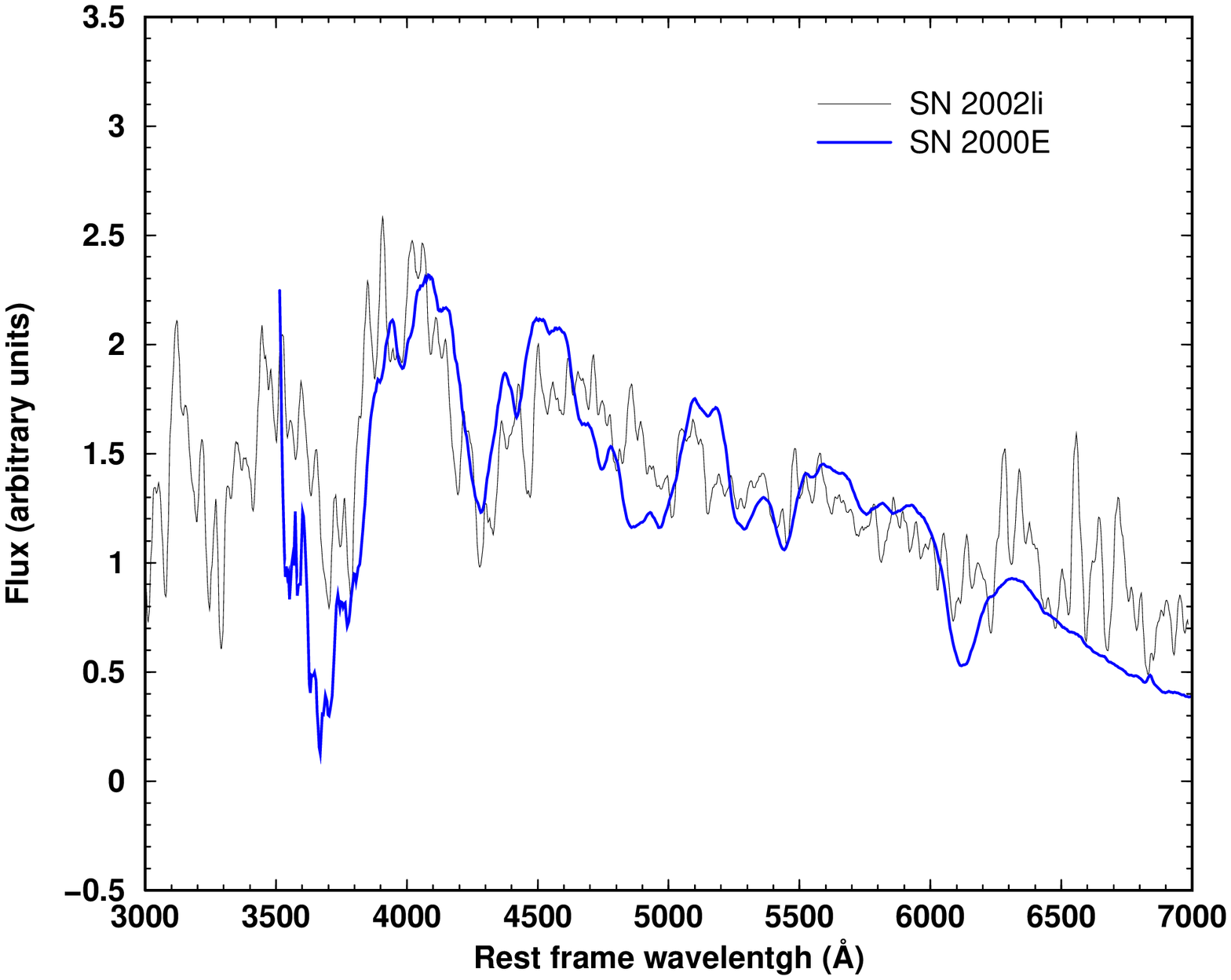}

      \caption{
Top: Spectrum of SN 2002li, including galaxy lines for the redshift
determination.
Middle: Template fitting of SN 2002li spectrum. Template epoch
is 2 days before maximum. SN 2002li spectrum has been dereddened and smoothed.
Bottom: 
Comparison of SN 2002li smoothed spectrum with that of SN 2000E 4 days 
before maximum. Both SN 2002li and SN 2000E spectra have been
dereddened and smoothed.}
         \label{Figure 2}

   \end{figure}

\clearpage

   \begin{figure}
   \centering

   \centering
   \includegraphics[width=0.8\columnwidth]{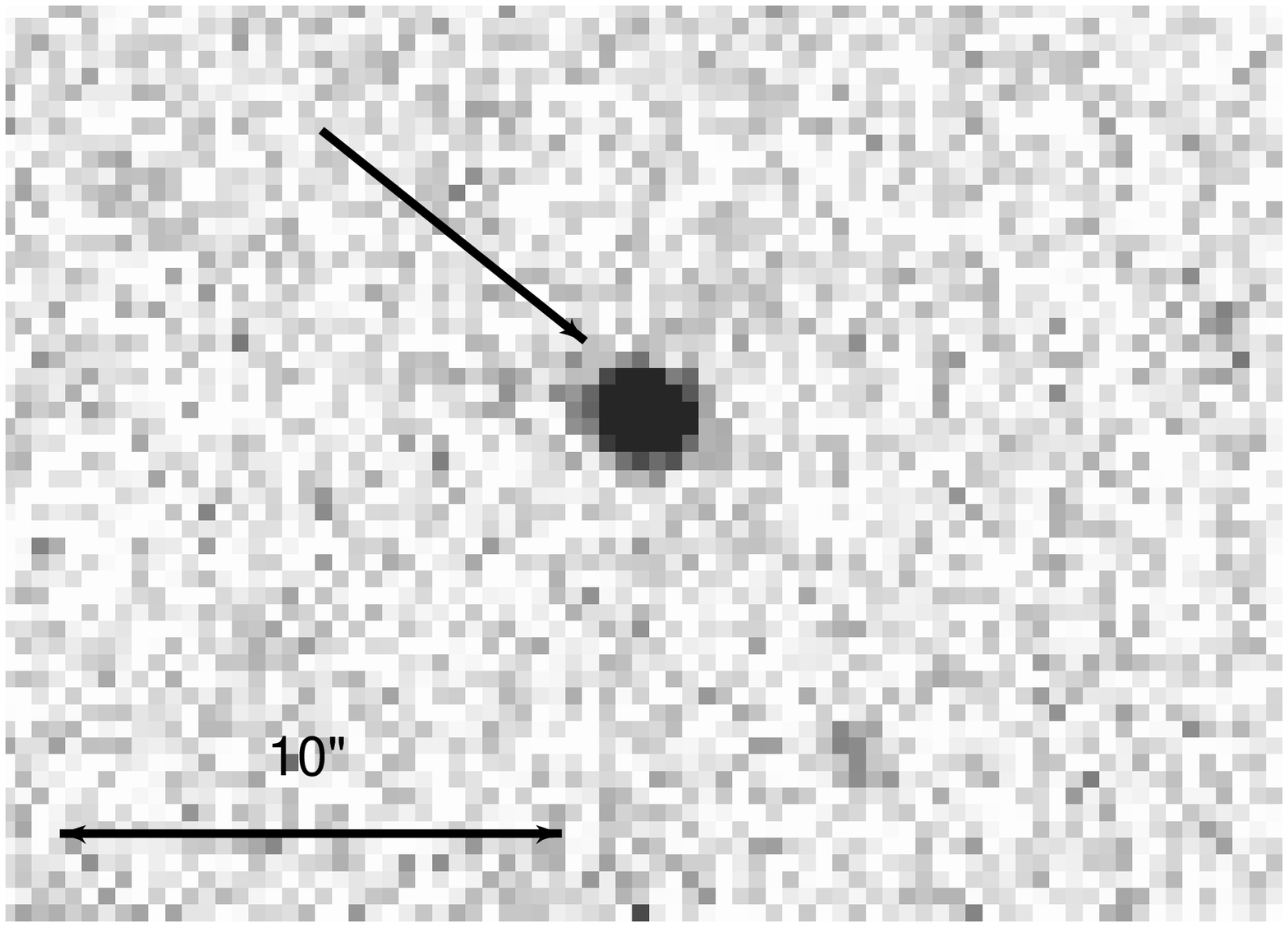}
      \caption{
Top: 
Finding chart of SN 2002lj. g' band image (exp. time : 240 sec)
      obtained on 
June 5, 2002 with the INT telescope + WFC.
Bottom: SN 2002lj, INT+WFC, Jun. 7, 2002, exp. time:  240 sec, g' band.}
         \label{Figure 3}
   \end{figure}

\clearpage

   \begin{figure}
   \centering
   \includegraphics[angle=-90,width=0.45\columnwidth]{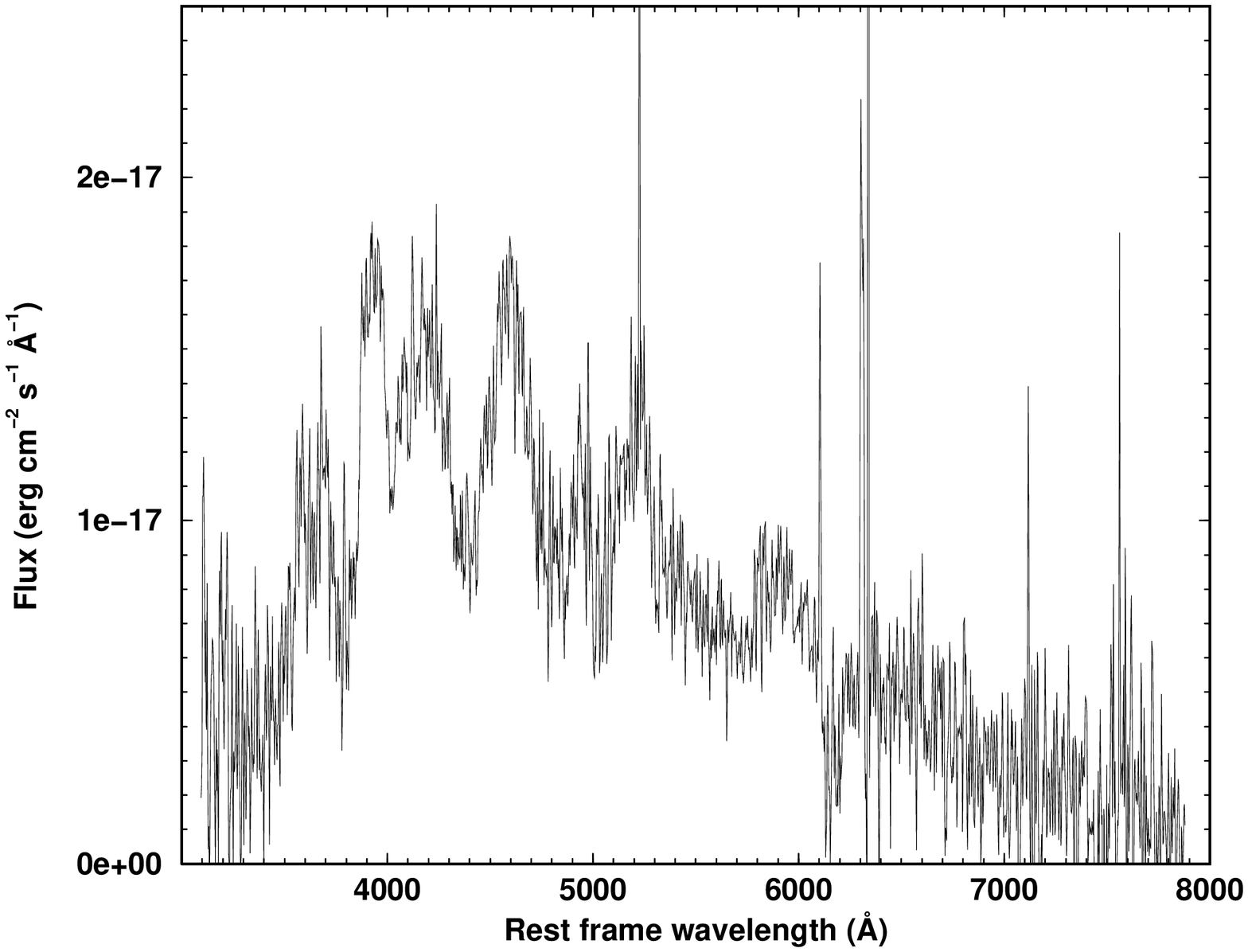}

   \centering
   \includegraphics[angle=-90,width=0.45\columnwidth]{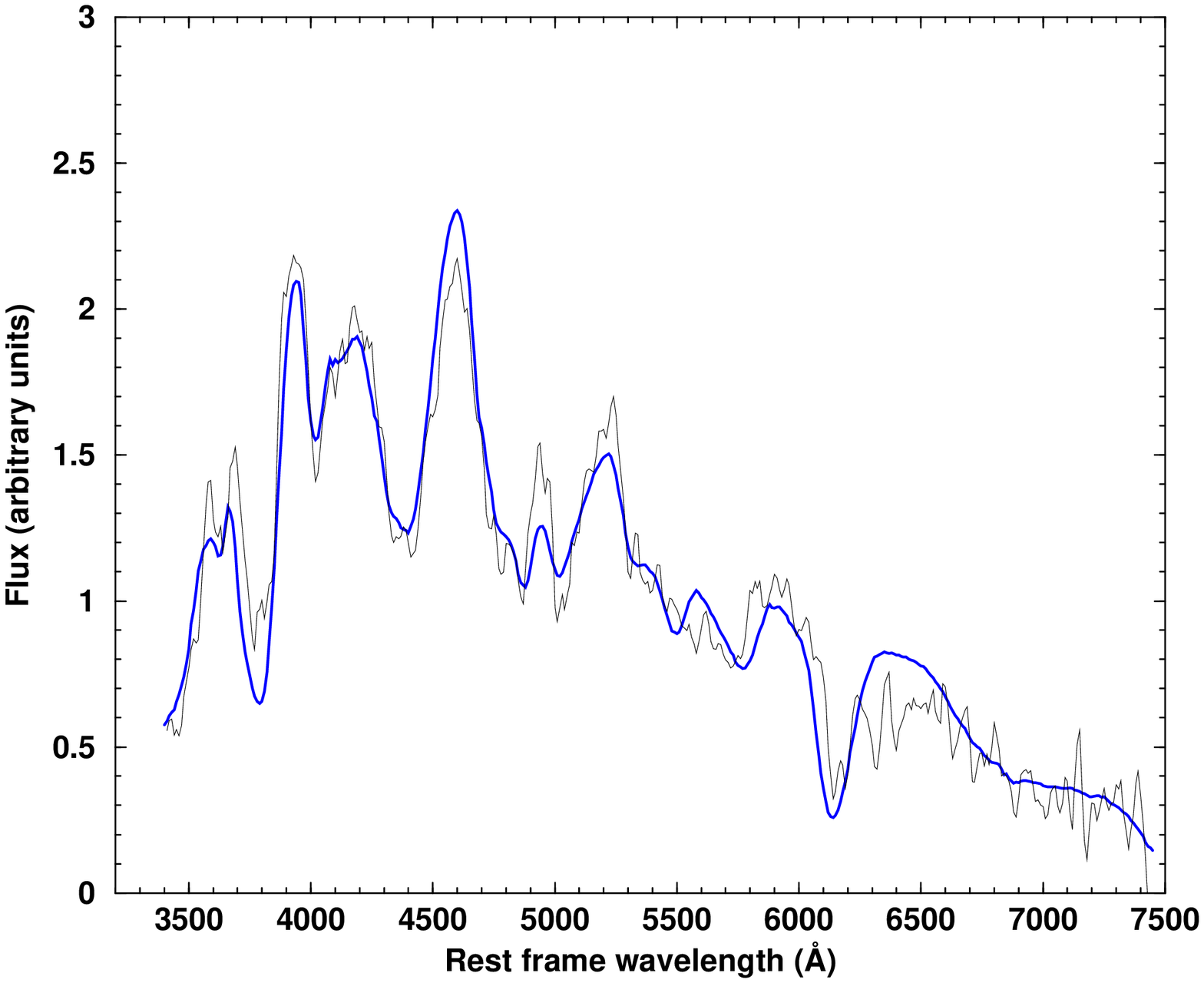}

   \centering
   \includegraphics[angle=-90,width=0.45\columnwidth]{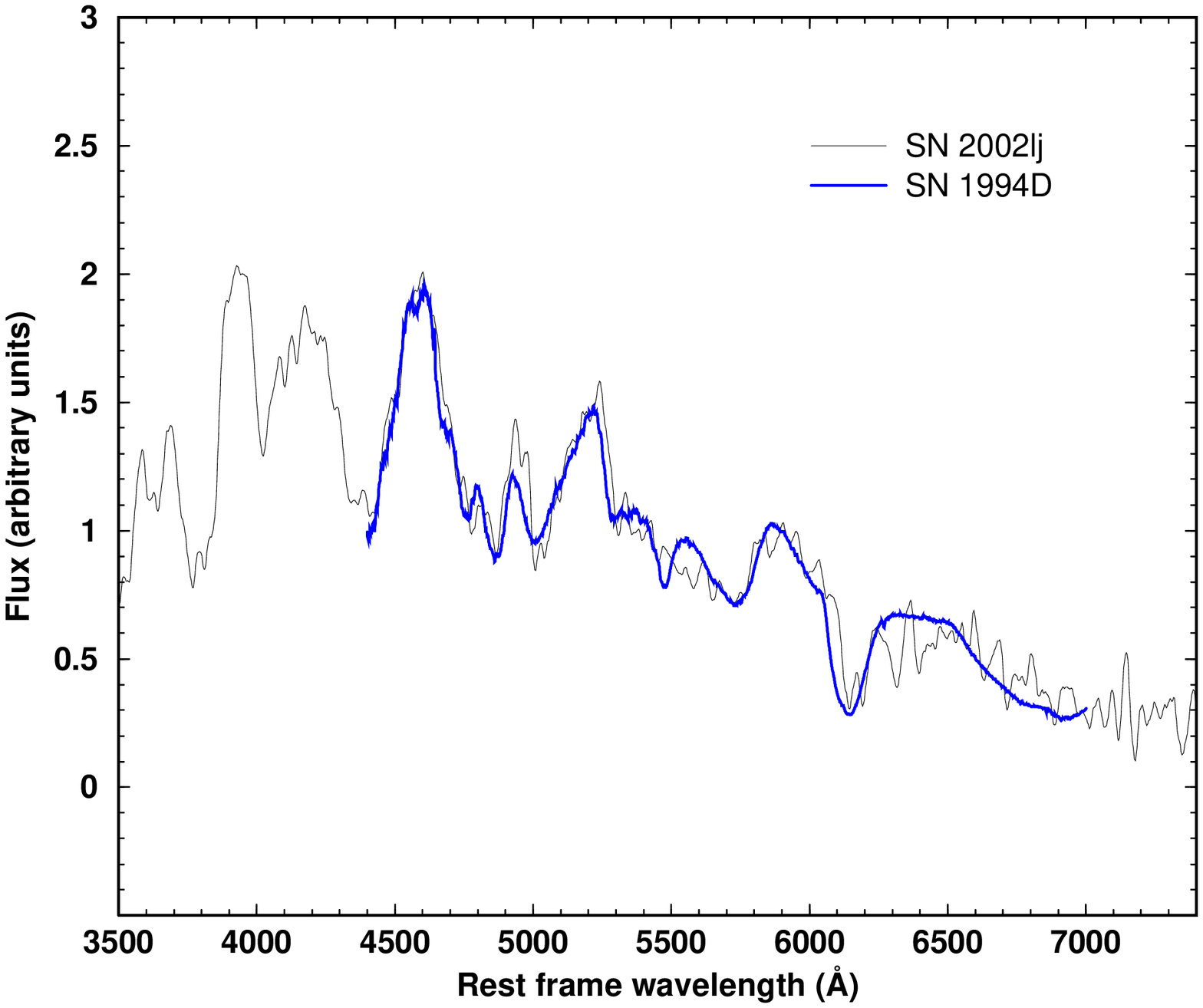}
   
\caption{Top: Spectrum of SN 2002lj. No galaxy lines have been
   found. Redshift determined from supernovae features.
Middle: Template fitting of SN 2002lj smoothed spectrum. Template epoch is 
7 days past maximum. Bottom: Comparison of SN 2002lj spectrum with that of 
SN 1994D 7 days past  maximum. Both spectra have been smoothed.}
         \label{Figure 4}
   \end{figure}

   \begin{figure}
   \centering

\includegraphics[width=0.8\columnwidth]{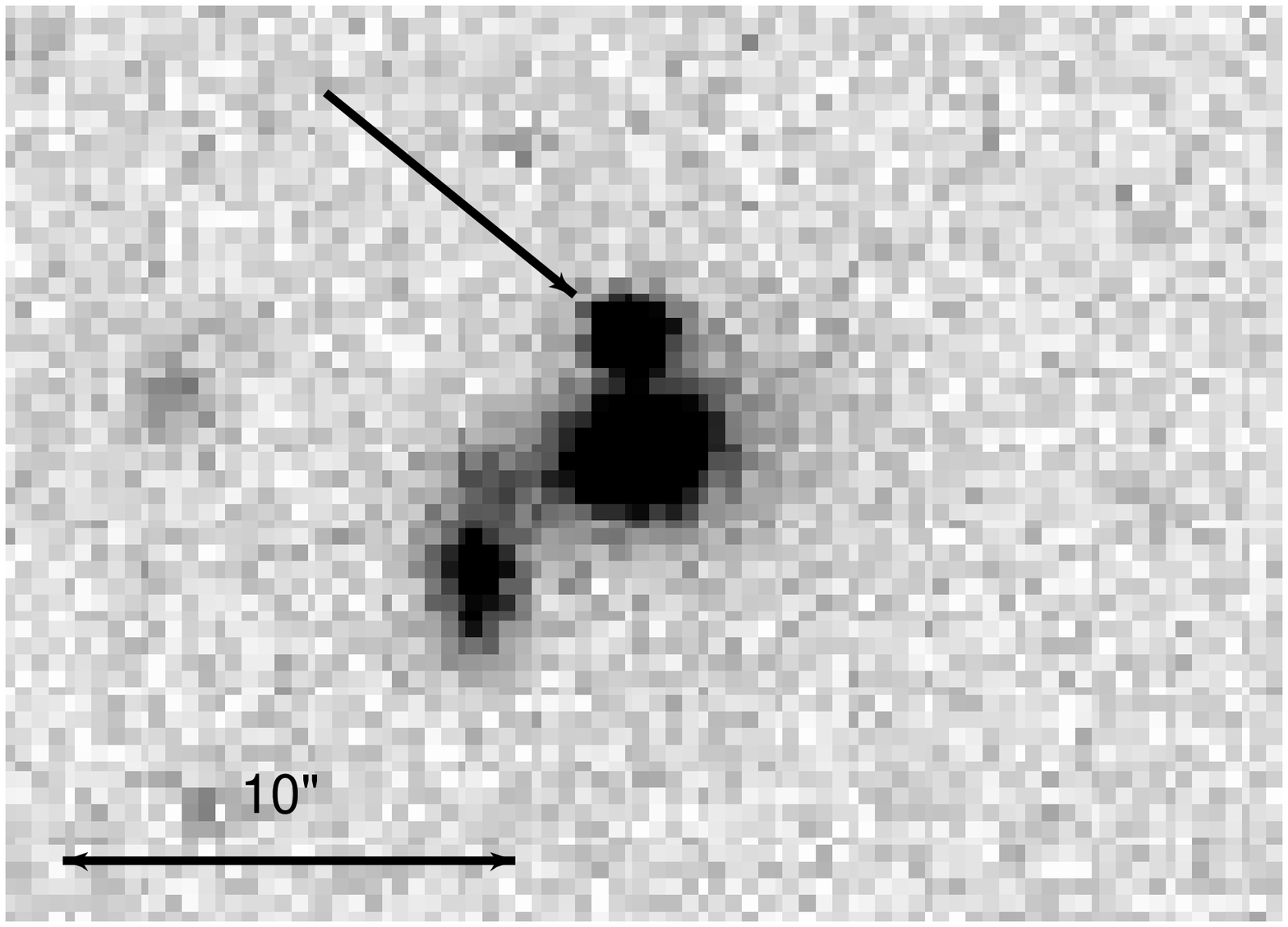}

\caption{
Top: Finding chart of SN 2002lp. g' band image (exp. time : 240 sec)
obtained on  June 5, 2002 with the INT telescope + WFC).
Bottom: SN 2002lp, INT+WFC, Jun. 6, 2002,  exp. time: 240 sec, g' band .}

\label{Figure 5}

   \end{figure}

\clearpage

   \begin{figure}
   \centering
   \includegraphics[angle=-90,width=0.45\columnwidth]{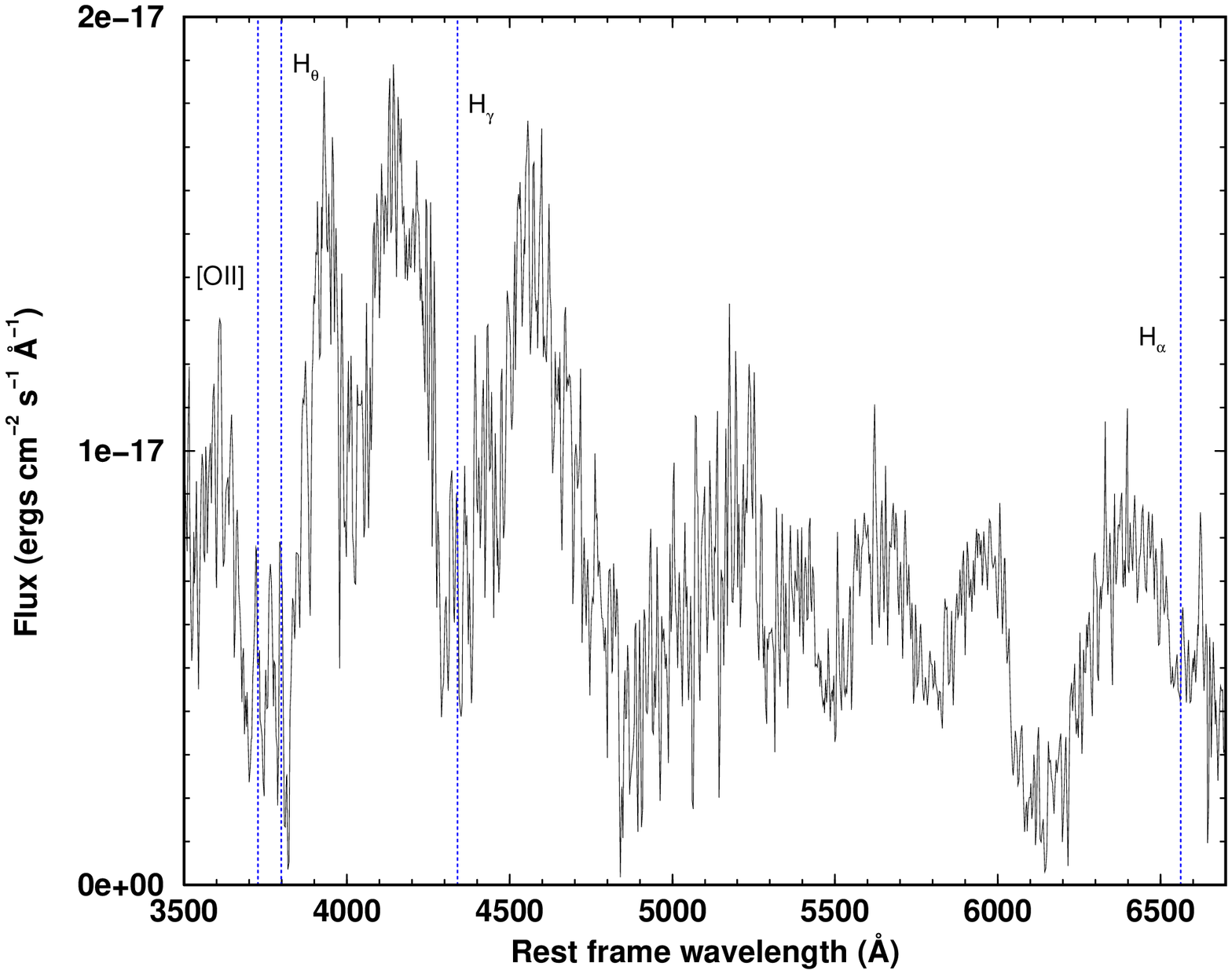}

   \centering
   \includegraphics[angle=-90,width=0.45\columnwidth]{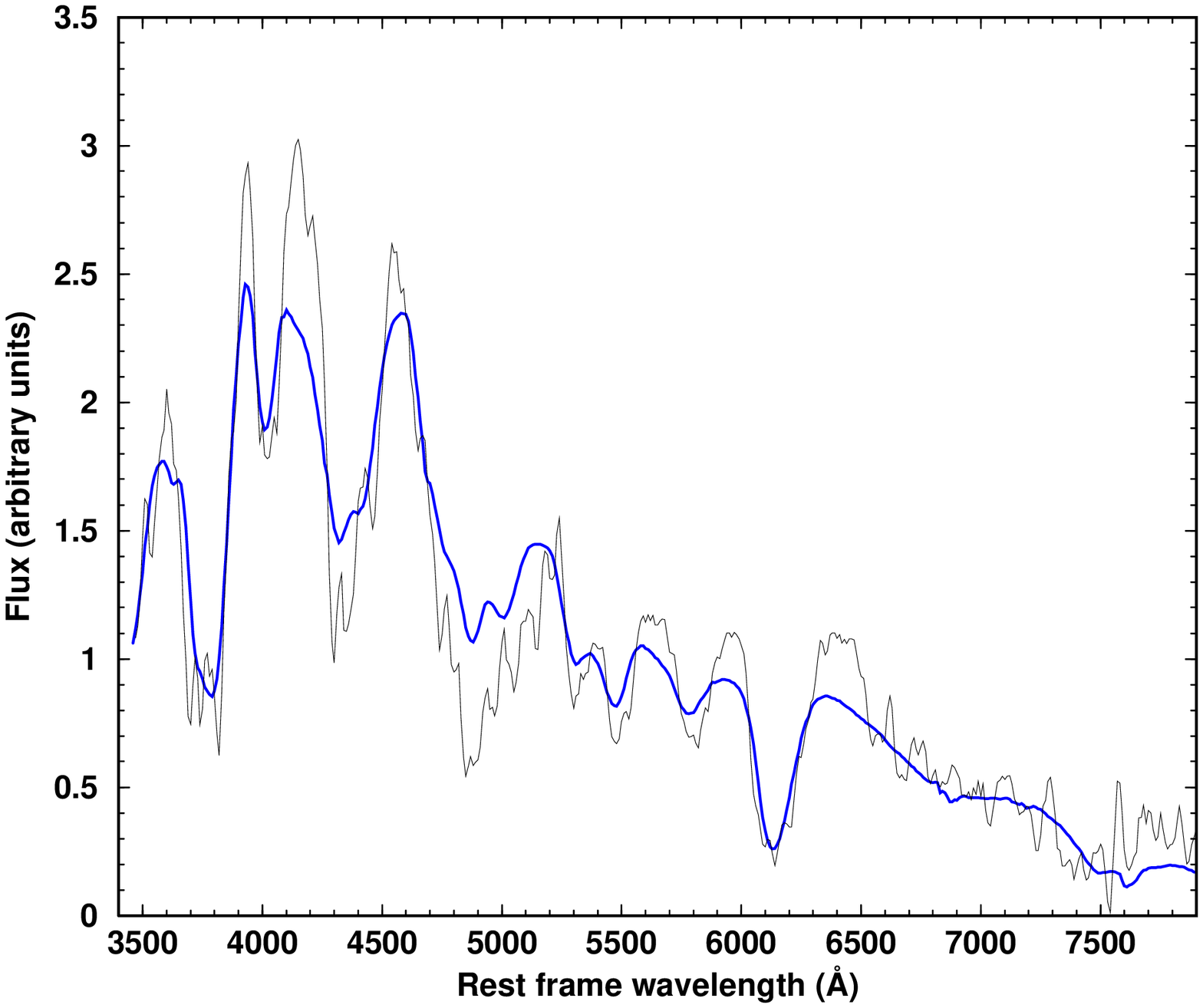}

 \includegraphics[angle=-90,width=0.45\columnwidth]{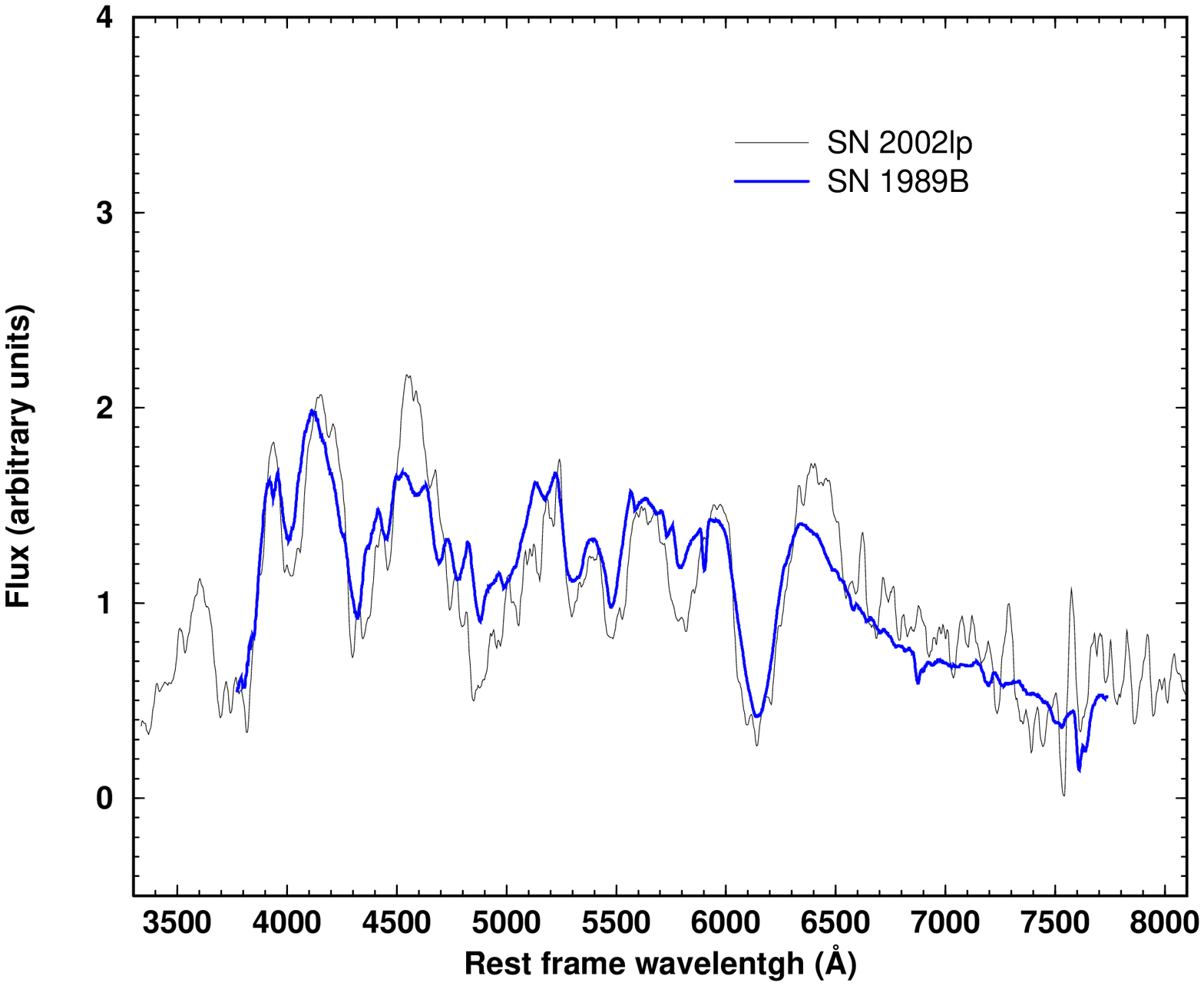}
 \caption{
 Top:
 Spectrum of SN 2002lp, including galaxy lines for the redshift determination.
Middle: 
Template fitting of SN 2002lp spectrum. Template epoch is 3 days past
 maximum. SN 2002lp spectrum has been smoothed and dereddened.
Bottom: Comparison of SN 2002lp spectrum with that of SN 1989B at
 maximum.
 SN 2002lp spectrum has been dereddened. Both spectra have been smoothed.}
         \label{Figure 6}
   \end{figure}

\clearpage

   \begin{figure}
   \centering

   \centering
   \includegraphics[width=0.8\columnwidth]{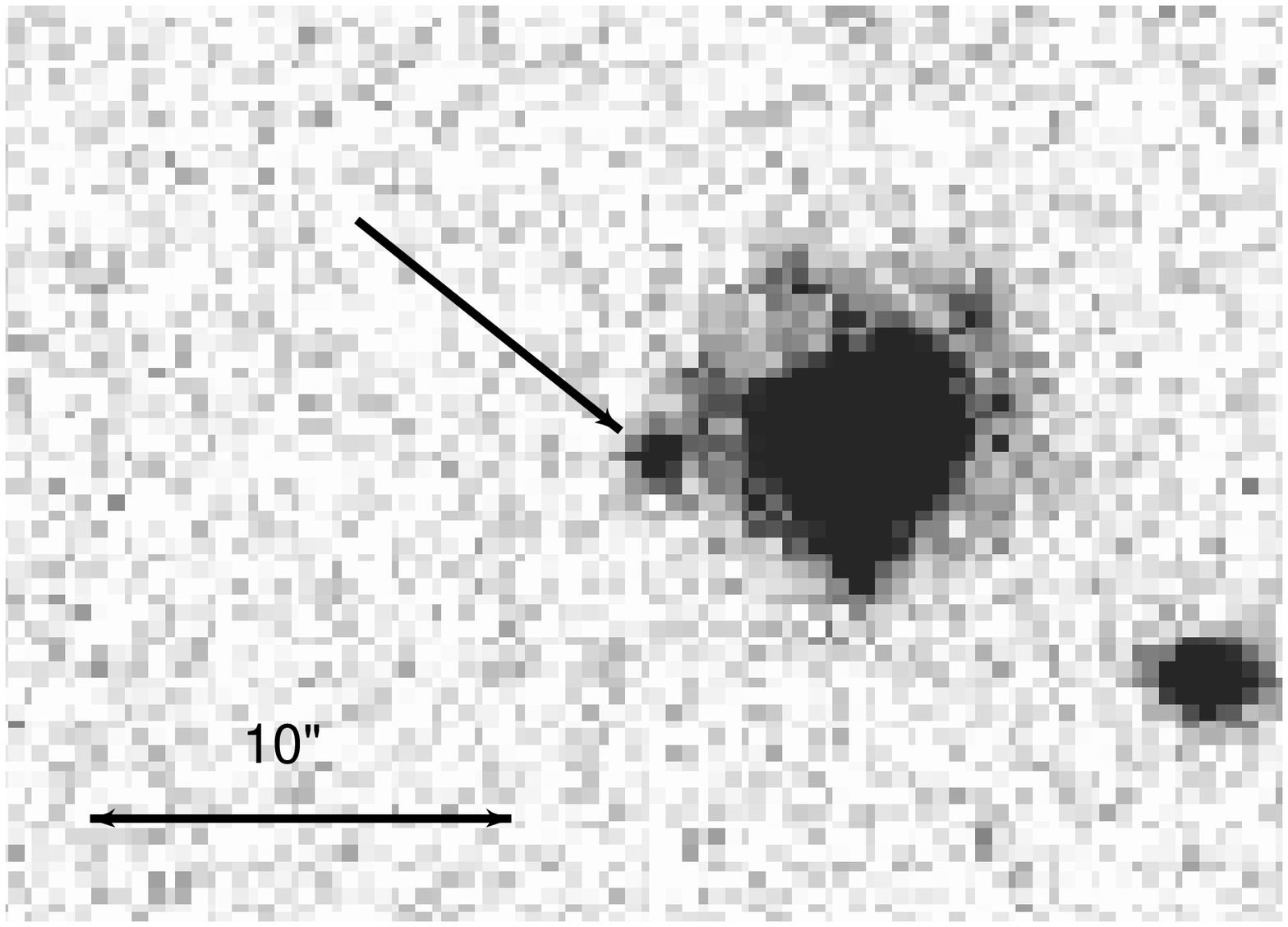}

\caption{ Top:
Finding chart of SN 2002lq. g' band image (exp. time : 240 sec) obtained on  
June 8, 2002 with the INT telescope + WFC).
Bottom: Same image zoomed to point at the supernova.}

         \label{Figure 7}
   \end{figure}


\clearpage

   \begin{figure}
   \centering
   \includegraphics[angle=-90,width=0.45\columnwidth]{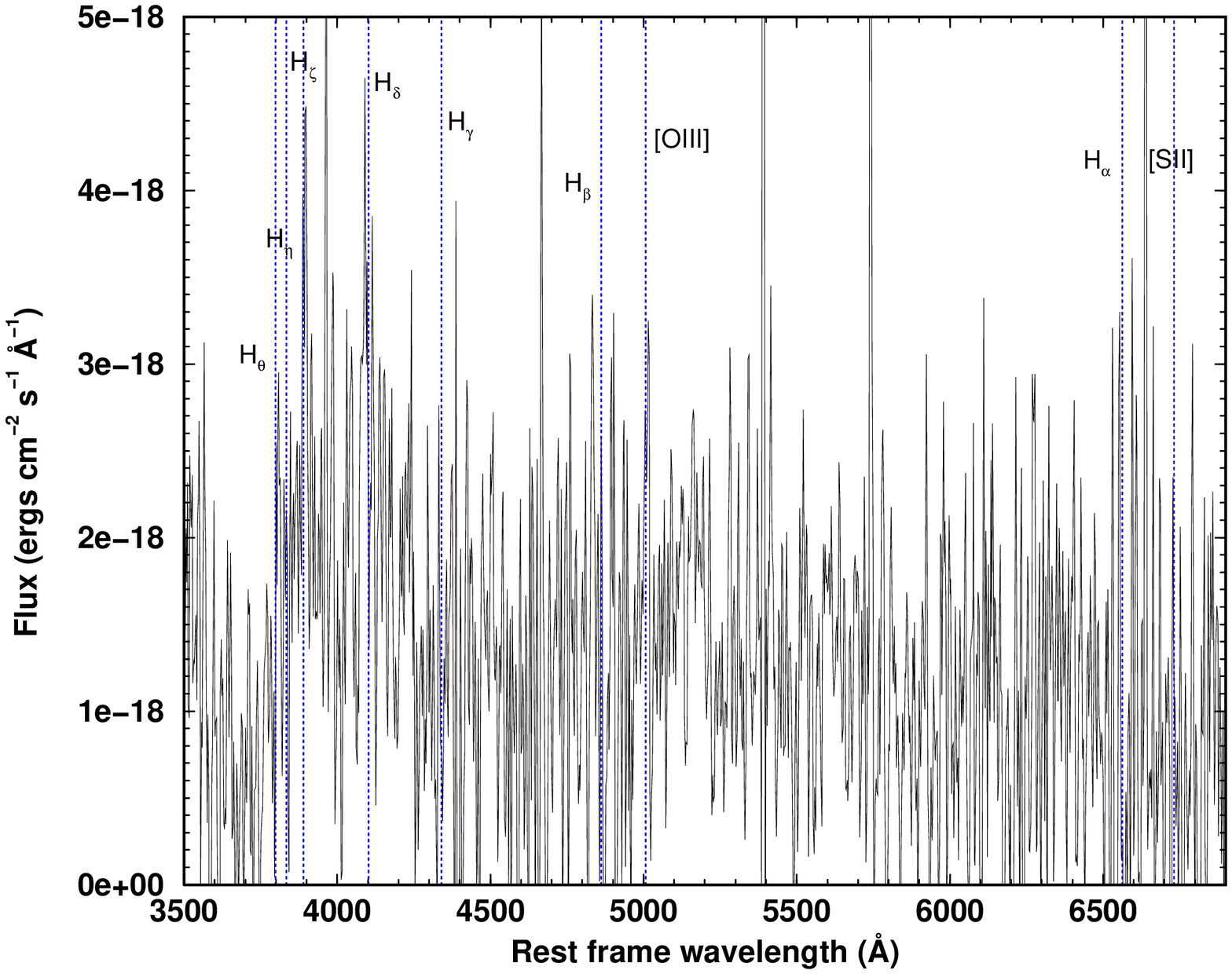}
        
   \centering
   \includegraphics[angle=-90,width=0.45\columnwidth]{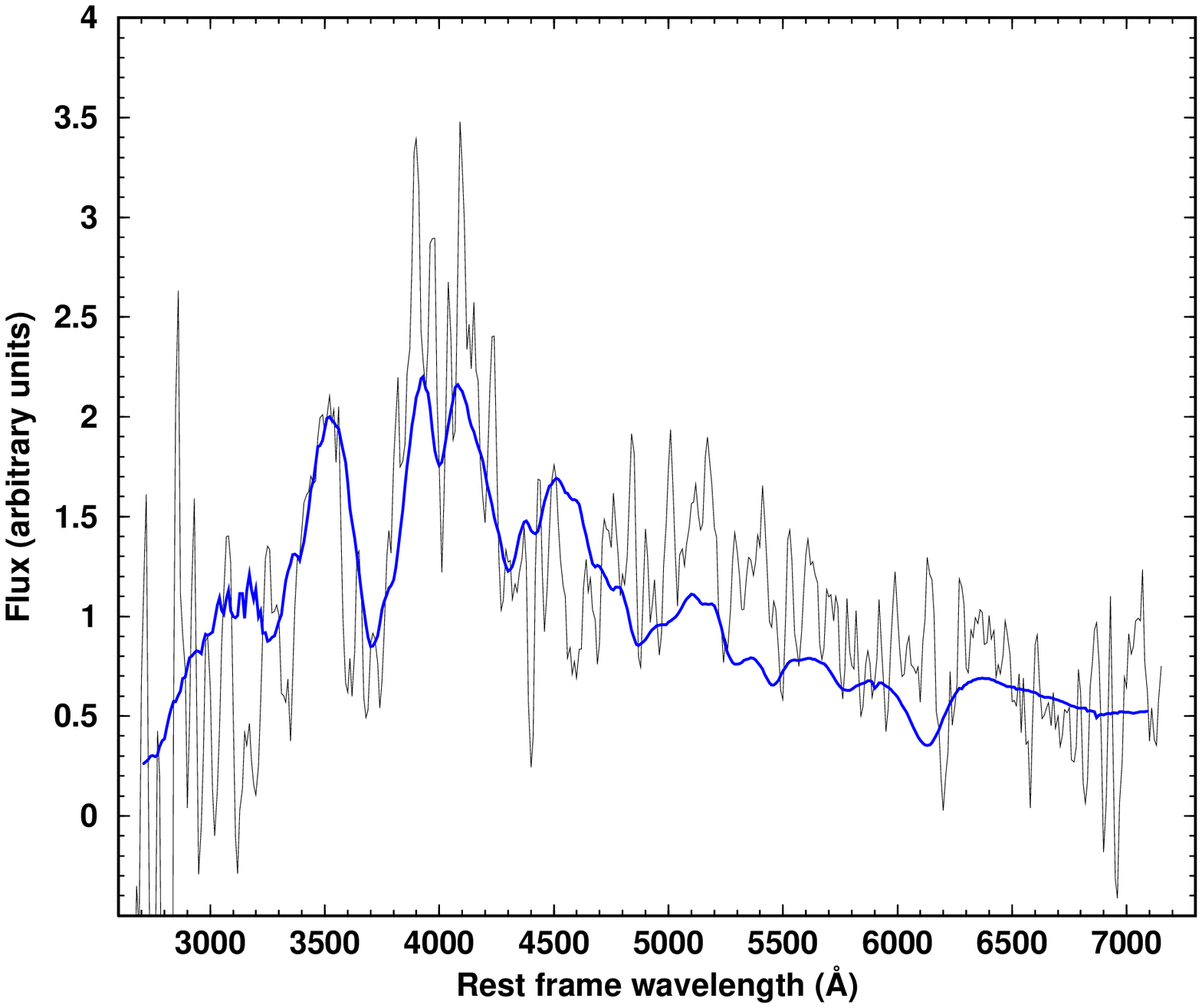}
        
   \centering
   \includegraphics[angle=-90,width=0.44\columnwidth]{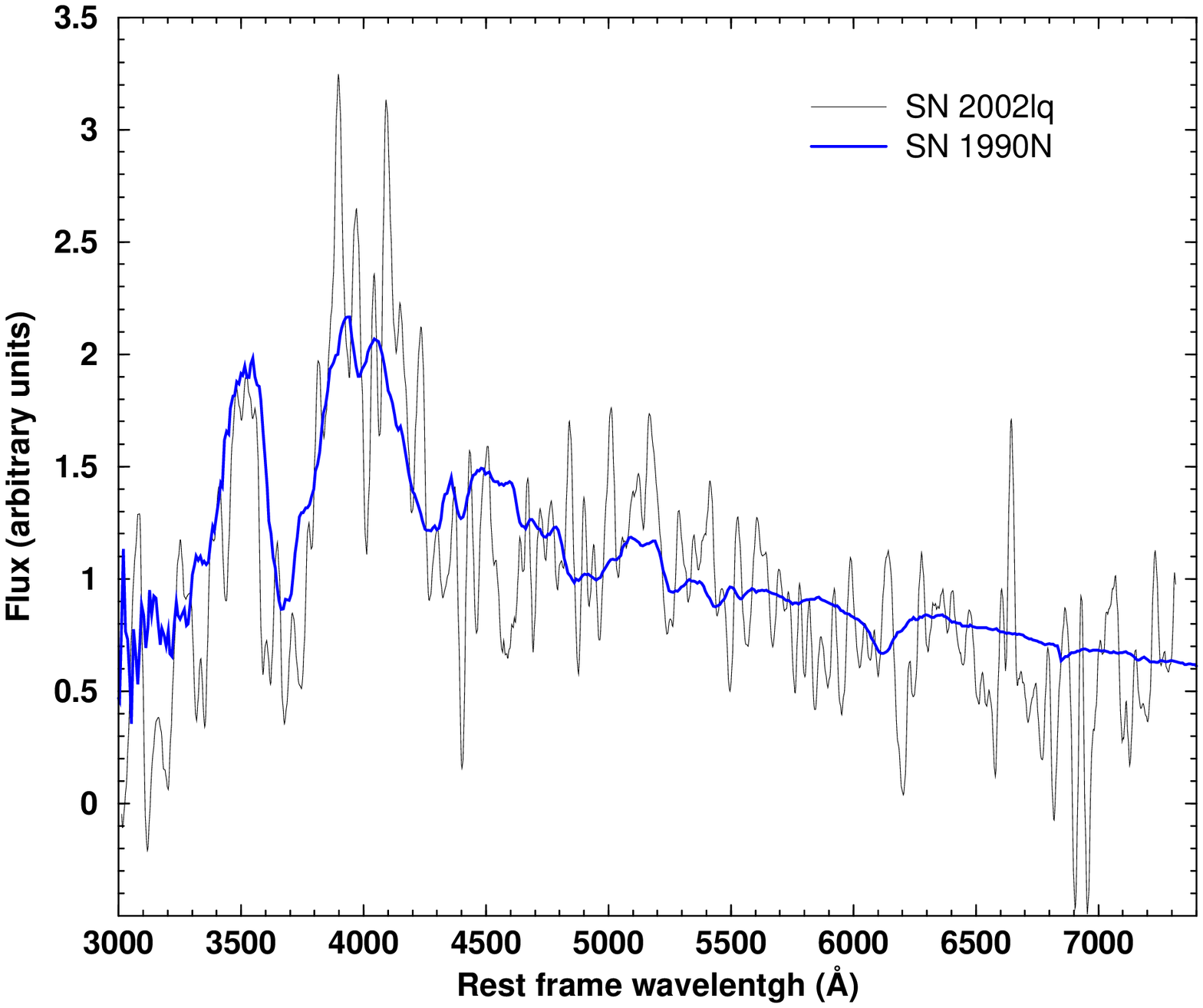}

\caption{
Top: Spectrum of SN 2002lq, including galaxy lines for 
the redshift determination.
Middle: Template fitting of SN 2002lq spectrum. Template epoch is 7
   days before maximum. SN 2002lq spectrum has been smoothed and dereddened.
Bottom: 
Comparison of SN 2002lq spectrum with that of SN 1990N 7 days before 
maximum. SN 2002lq spectrum has been smoothed and dereddened.}
         \label{Figure 8}
   \end{figure}

\clearpage

   \begin{figure}
   \centering

   \centering
   \includegraphics[width=0.8\columnwidth]{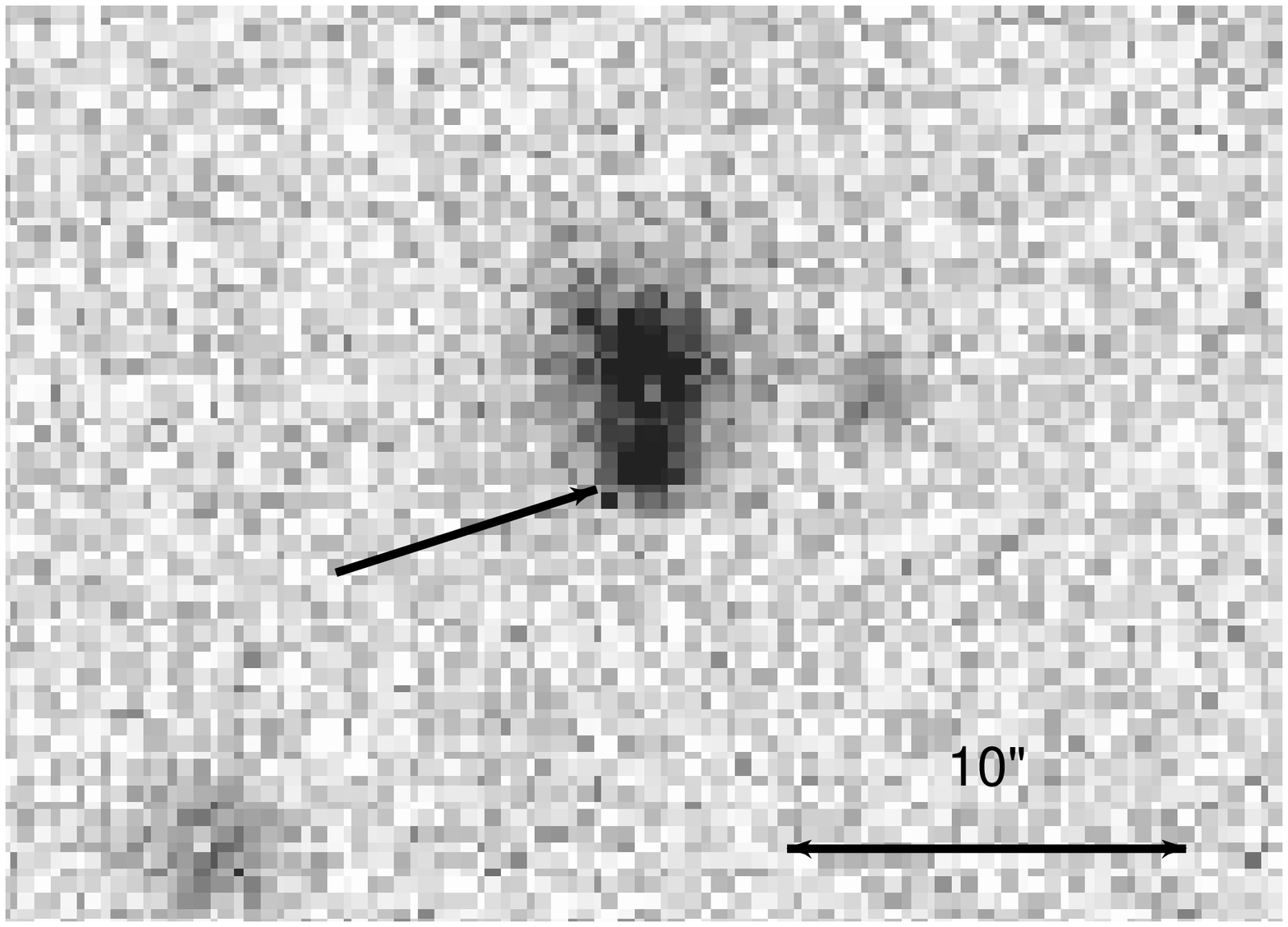}

      \caption{
Top: Finding chart of SN 2002lr. g' band image (exp. time : 240 sec)
   obtained on  June 6, 2002 with the INT telescope + WFC). 
Bottom: Same image zoomed to point out at the supernova.}

   \label{Figure 9}

   \end{figure}


\clearpage

   \begin{figure}
   \centering
   \includegraphics[angle=-90,width=0.45\columnwidth]{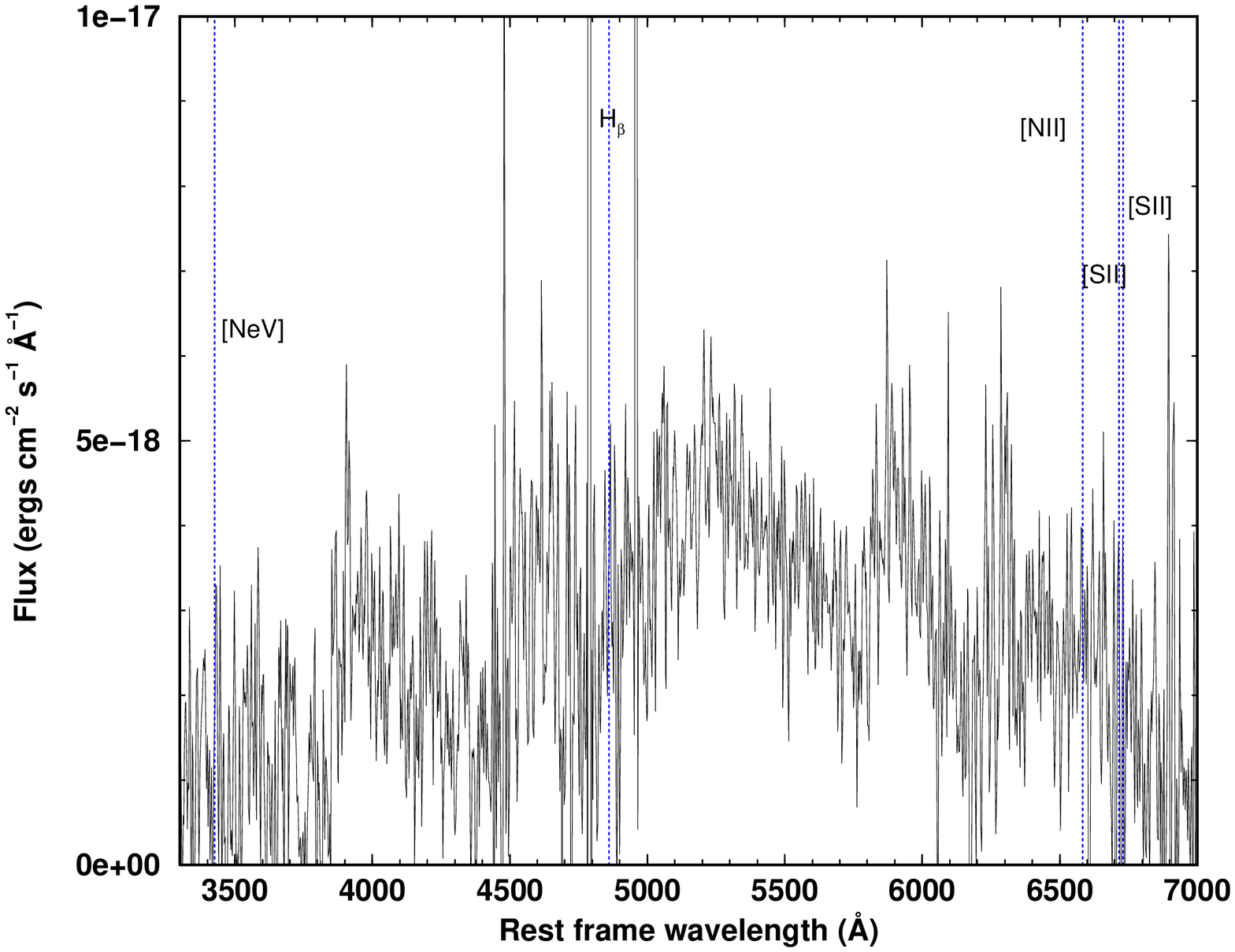}

   \centering
   \includegraphics[angle=-90,width=0.45\columnwidth]{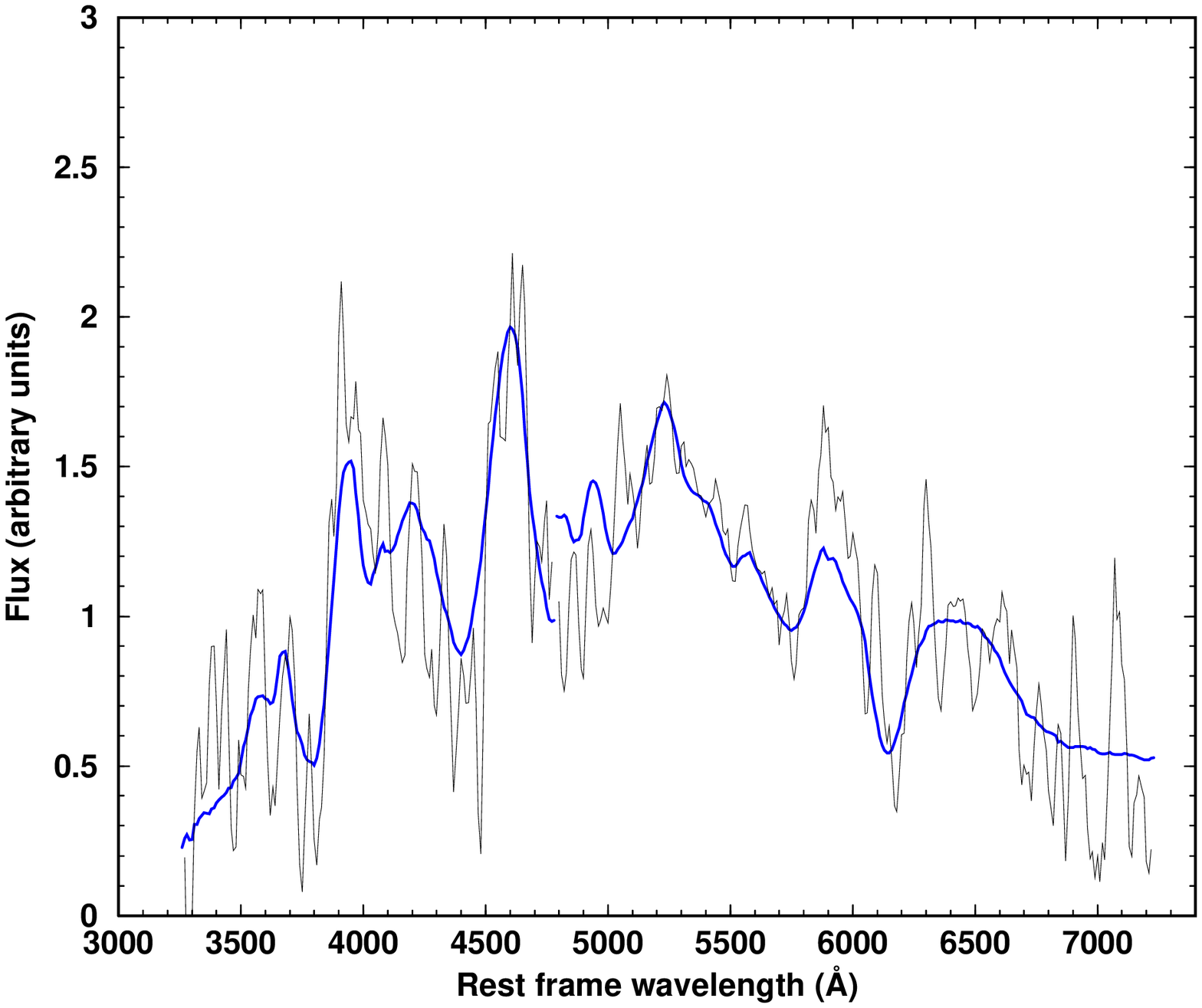}
        
   \centering
   \includegraphics[angle=-90,width=0.45\columnwidth]{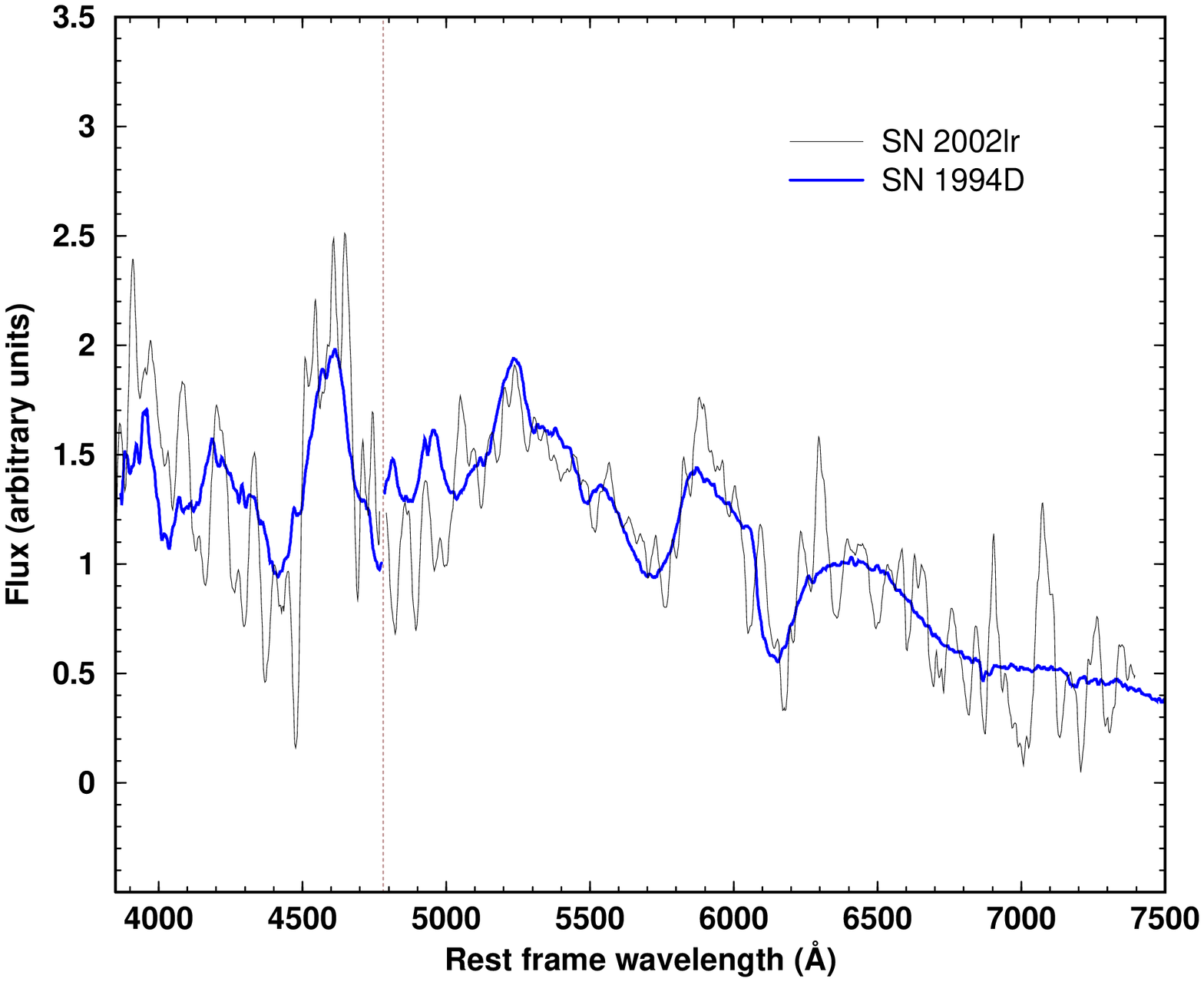}
      \caption{
Top: Spectrum of SN 2002lr, including galaxy lines for the redshift 
determination.
Middle: Template fitting of SN 2002lr smoothed spectrum. Template
 epoch is 10 days past maximum. 
Bottom: Comparison of SN 2002lr  spectrum with that of SN 1994D 10
days past maximum. Both spectra have been smoothed.}

         \label{Figure 10}
   \end{figure}

\clearpage

   \begin{figure}
   \centering
   \includegraphics[width=0.6\columnwidth]{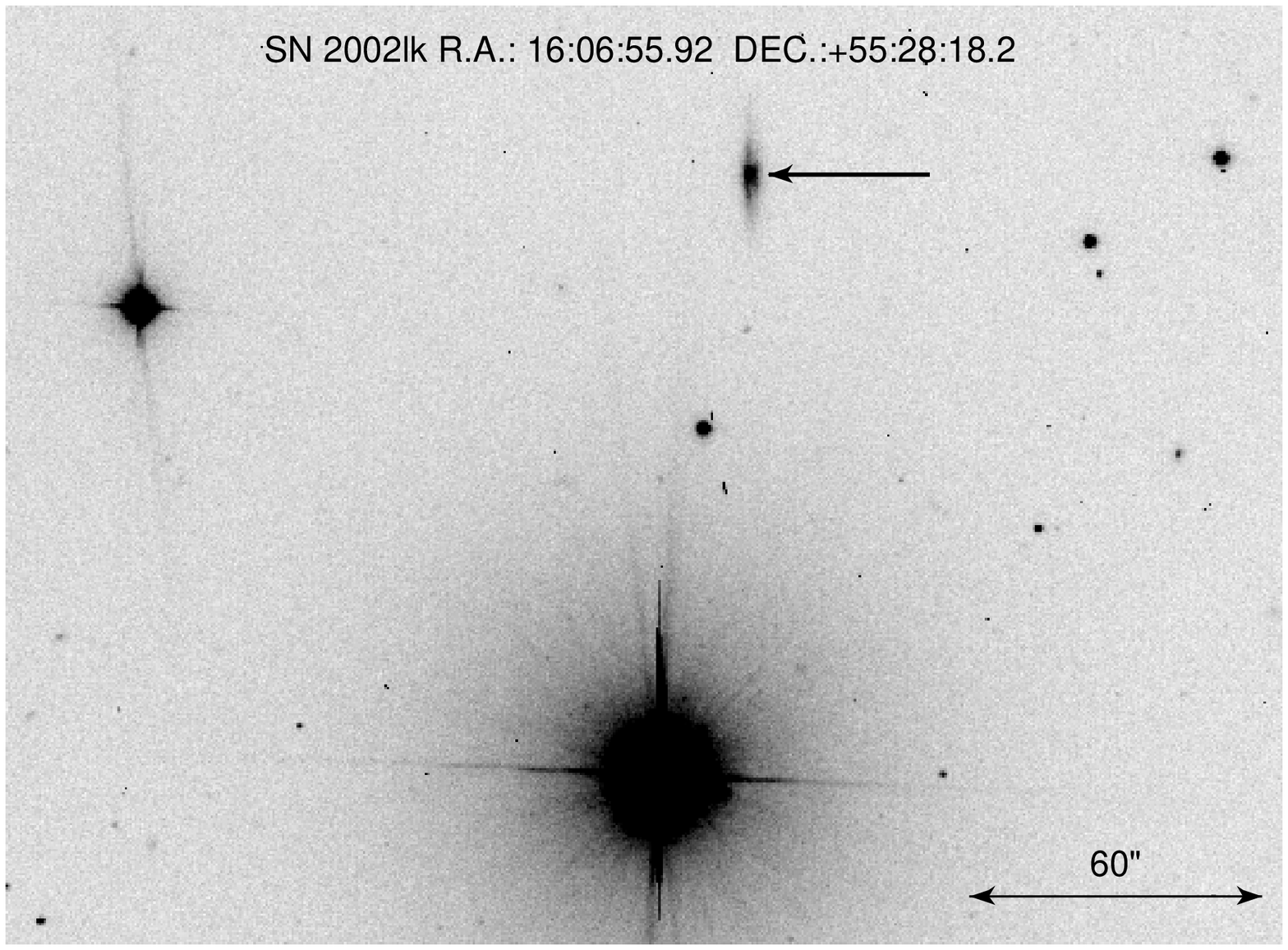}
  
   \centering

 \includegraphics[width=0.6\columnwidth]{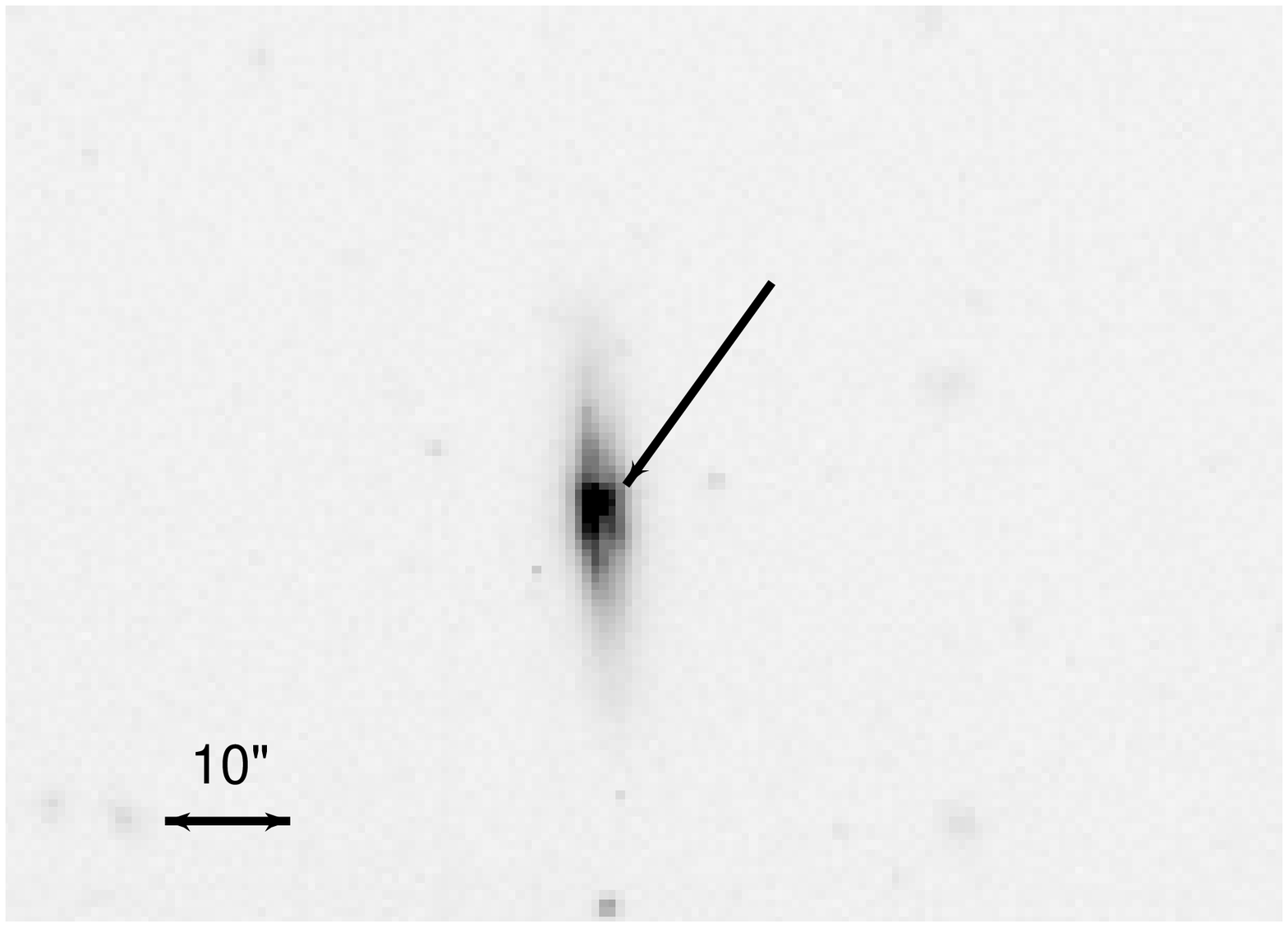}
\caption{ Top: Finding chart of SN 2002lk. g' band image 
(exp. time : 240 sec) obtained on  June 7, 2002 with 
the INT telescope + WFC). Bottom: 
Same image zoomed to point out the supernova.}
         \label{Figure 11}

   \end{figure}

\clearpage

   \begin{figure}
   \centering
   \includegraphics[angle=-90,width=0.45\columnwidth]{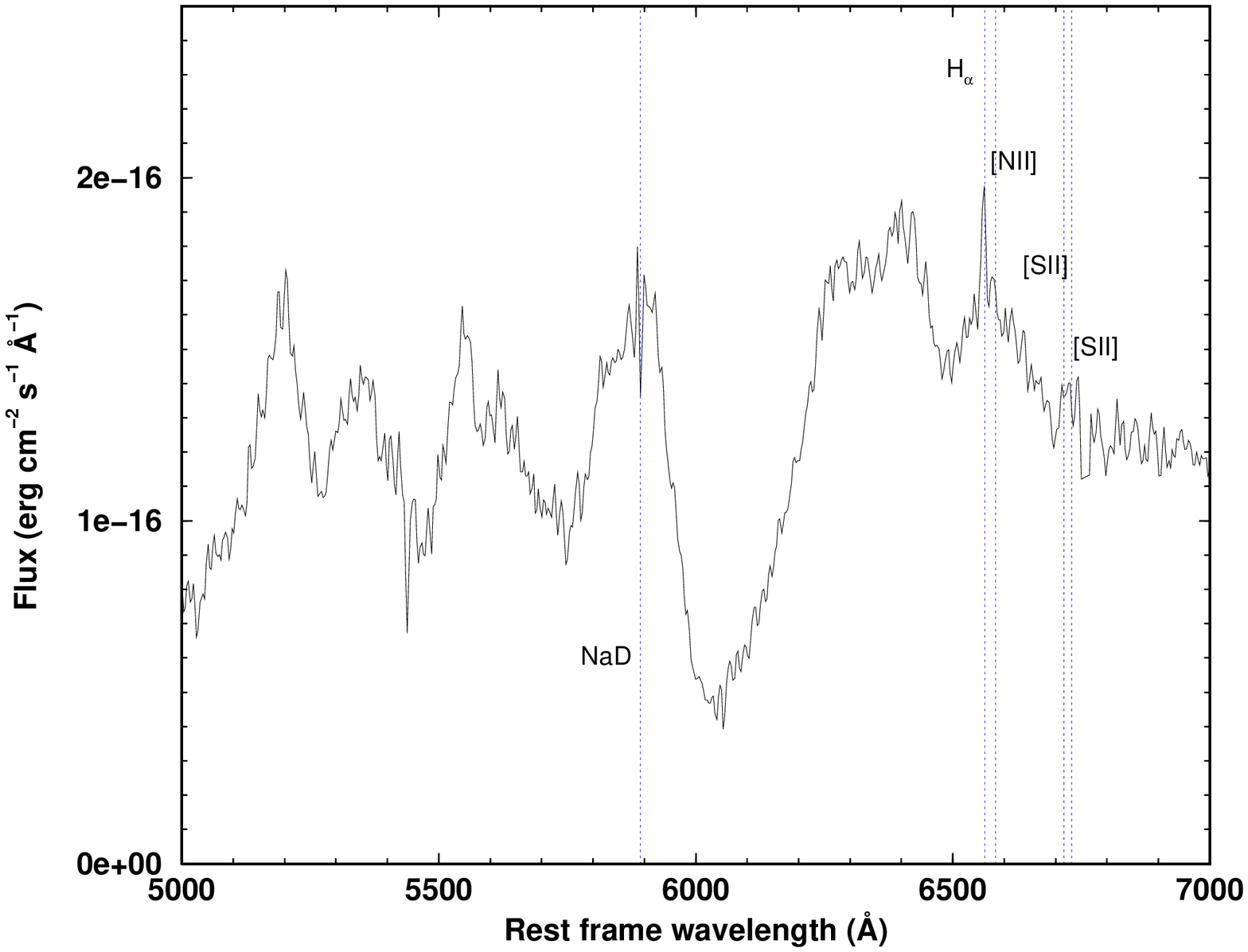}

   \centering
   \includegraphics[angle=-90,width=0.45\columnwidth]{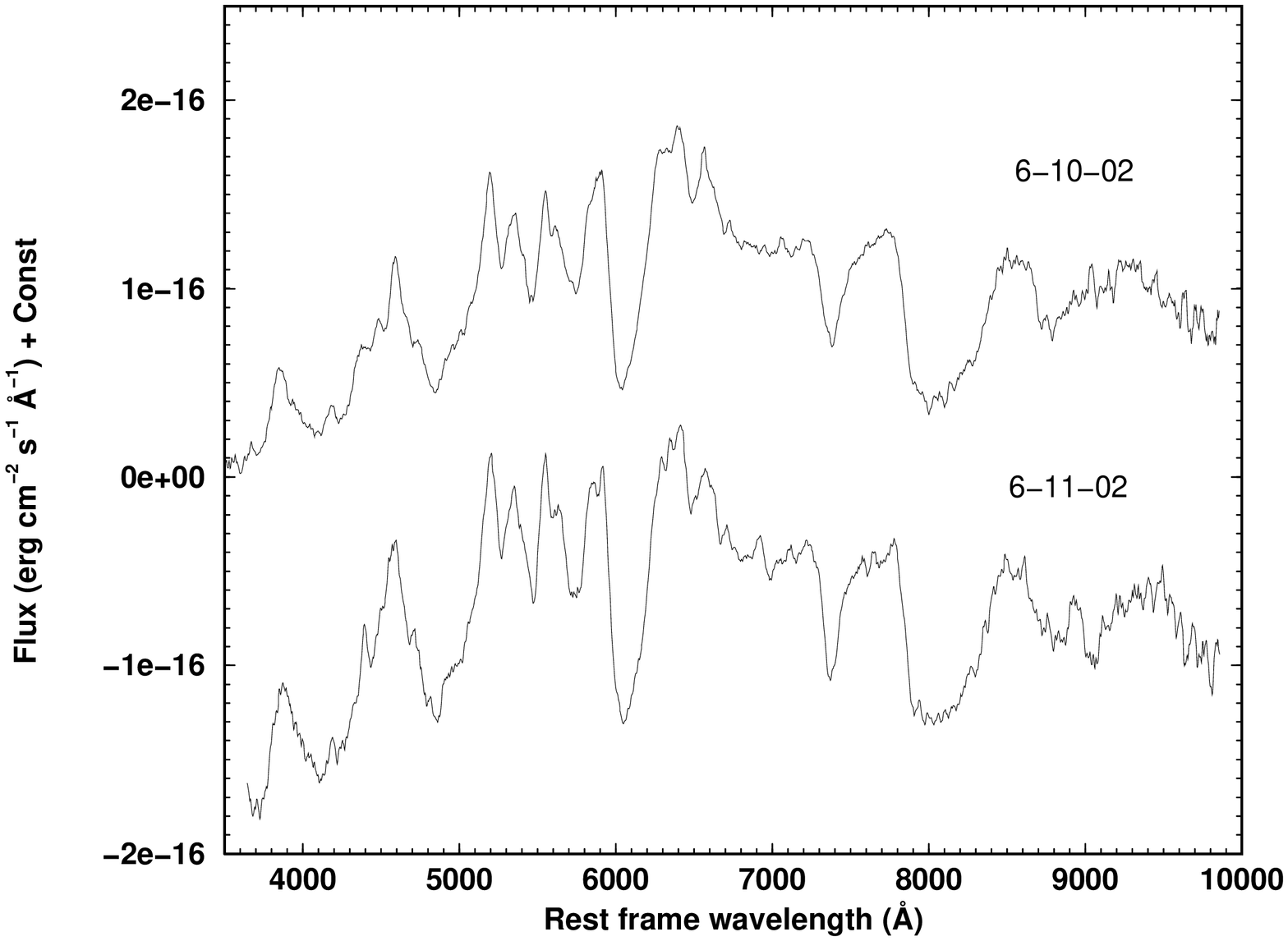}
    
   \centering
 \includegraphics[angle=-90,width=0.45\columnwidth]{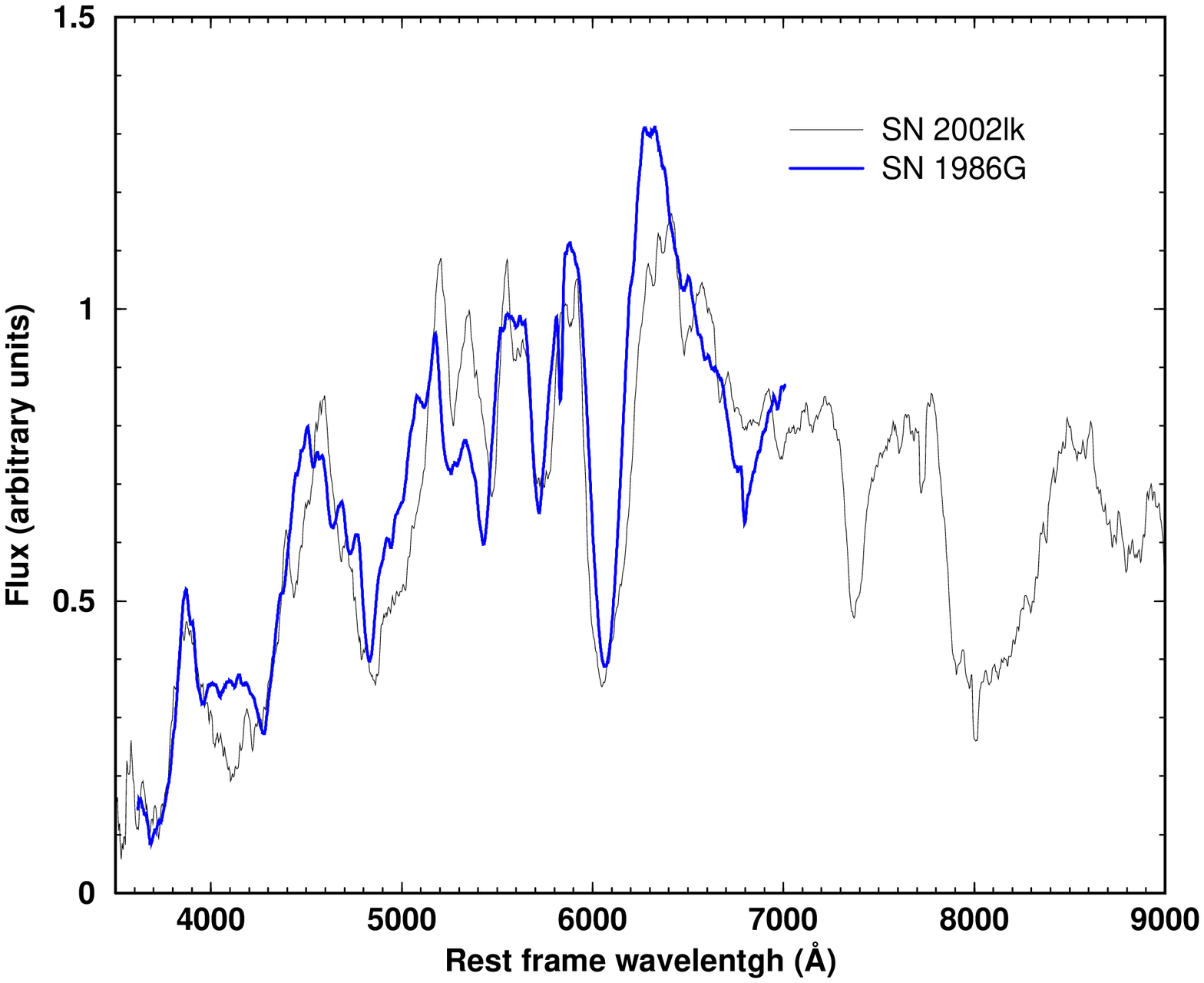}      

\caption{
Top: Spectrum of SN 2002lk on June 10, including galaxy lines 
for the redshift determination.
Spectra of SN 2002lk on June 10 and June 11 
($-2 \times 10^{-16} erg cm^{-2} s^{-1}$ \AA$^{-1}$).
Middle: 
Comparison of SN 2002lk spectrum on June 11 with that of SN 1991bg at
maximum.
Bottom: 
Comparison of SN 2002lk spectrum on June 11 with that of SN 1986G  
2 days before maximum. SN 1986G spectrum has been smoothed.}

\label{Figure 12}

\end{figure}

\clearpage

   \begin{figure}
   \centering
   \includegraphics[angle=0,width=\columnwidth]{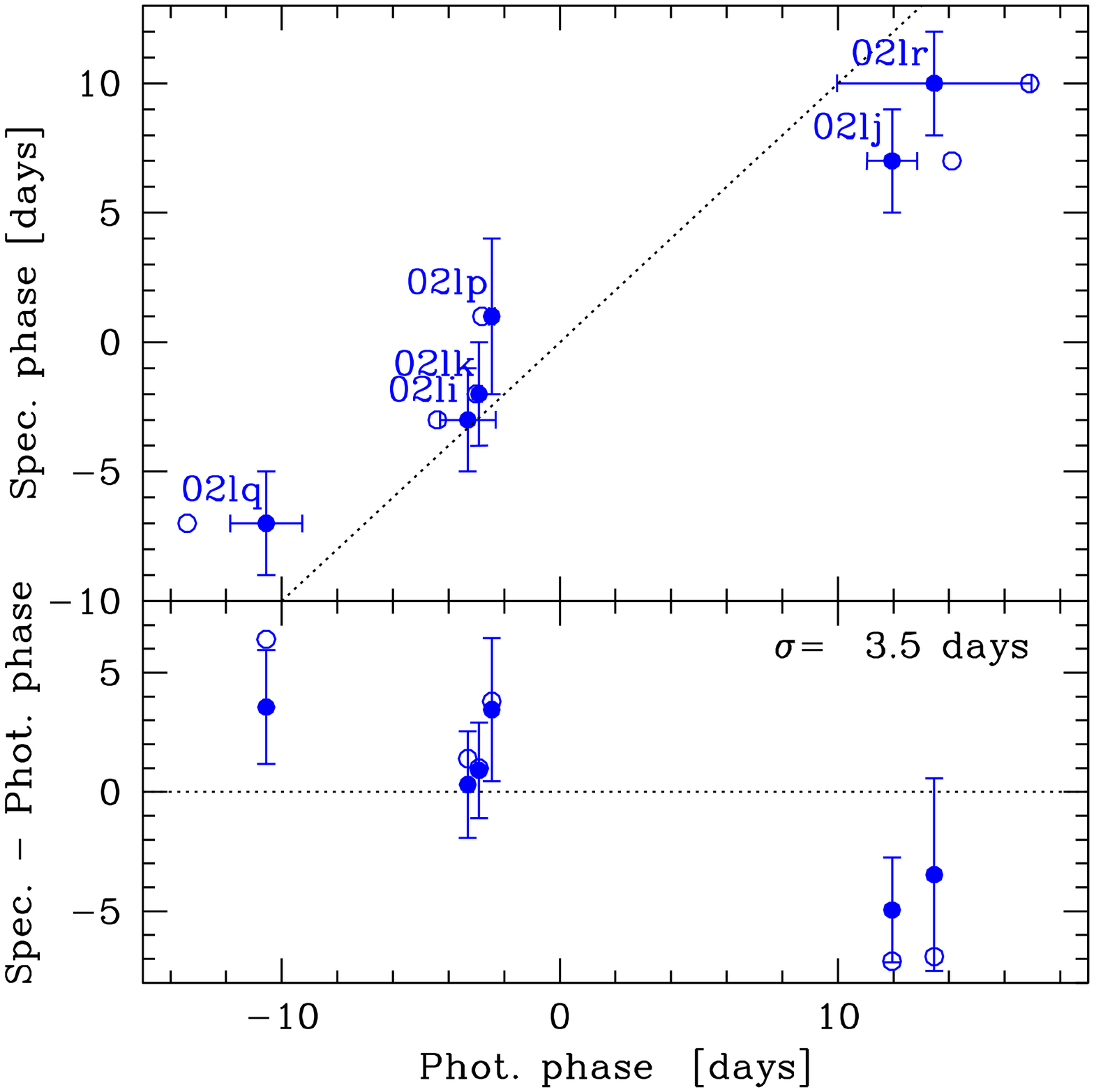}
      \caption{Comparison of the spectrum epochs as determined from the
 spectrum fits ($\tau_{spec}$) and the light curve fits ($\tau_{pho}$).
 Filled dots:  $\tau_{pho}$ corrected for time dilation;
 empty dots: $\tau_{pho}$ not corrected for time dilation.
 Epochs are relative to the $B$ band  maximum. The dispersion $\sigma$ is
relative to the corrected points.
  The dotted line corresponds to $\tau_{spec}=\tau_{pho}$.  }
         \label{Figure 13}
   \end{figure}

   \begin{figure}
   \centering
   \includegraphics[angle=-90,width=0.45\columnwidth]{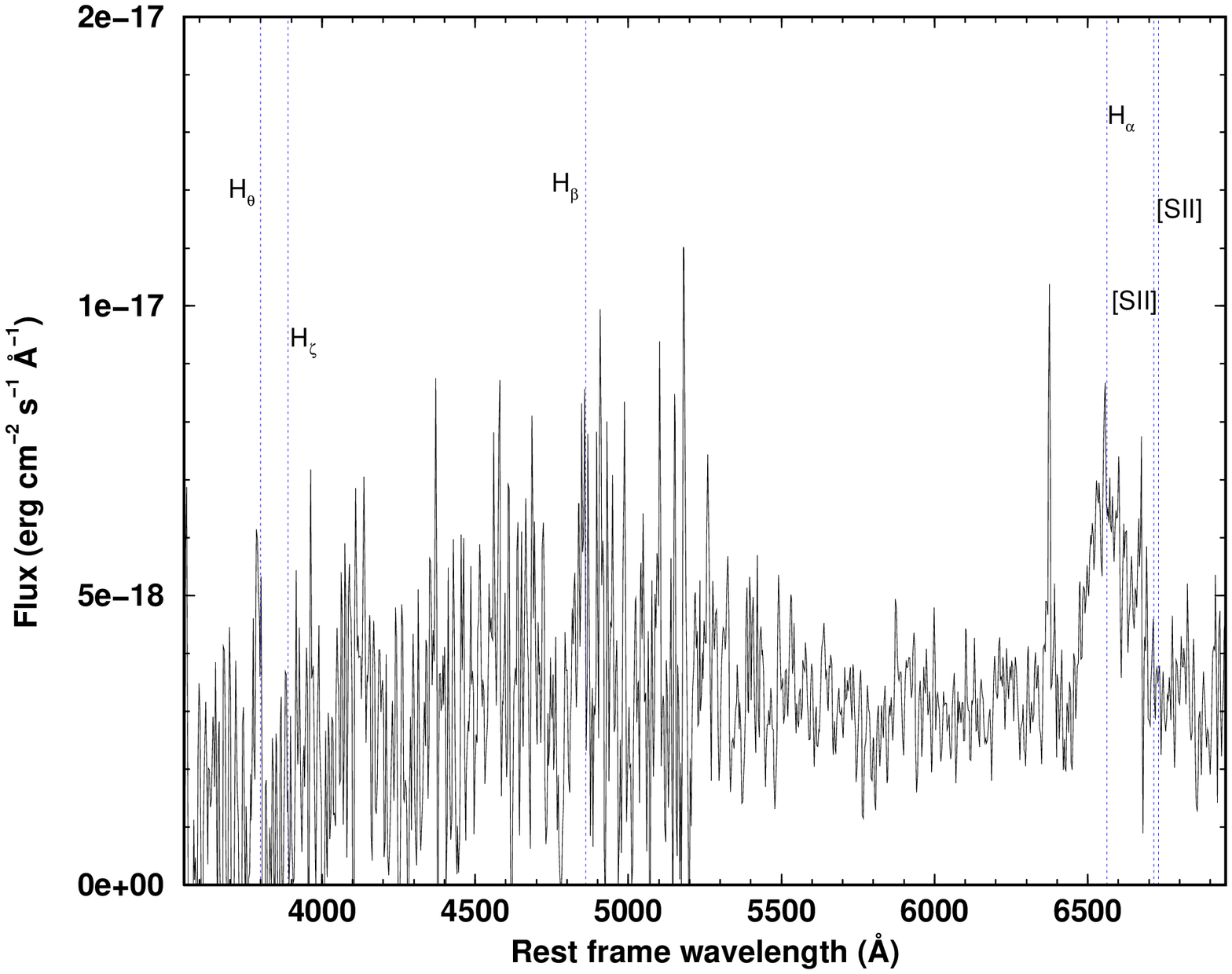}
   \includegraphics[angle=-90,width=0.45\columnwidth]{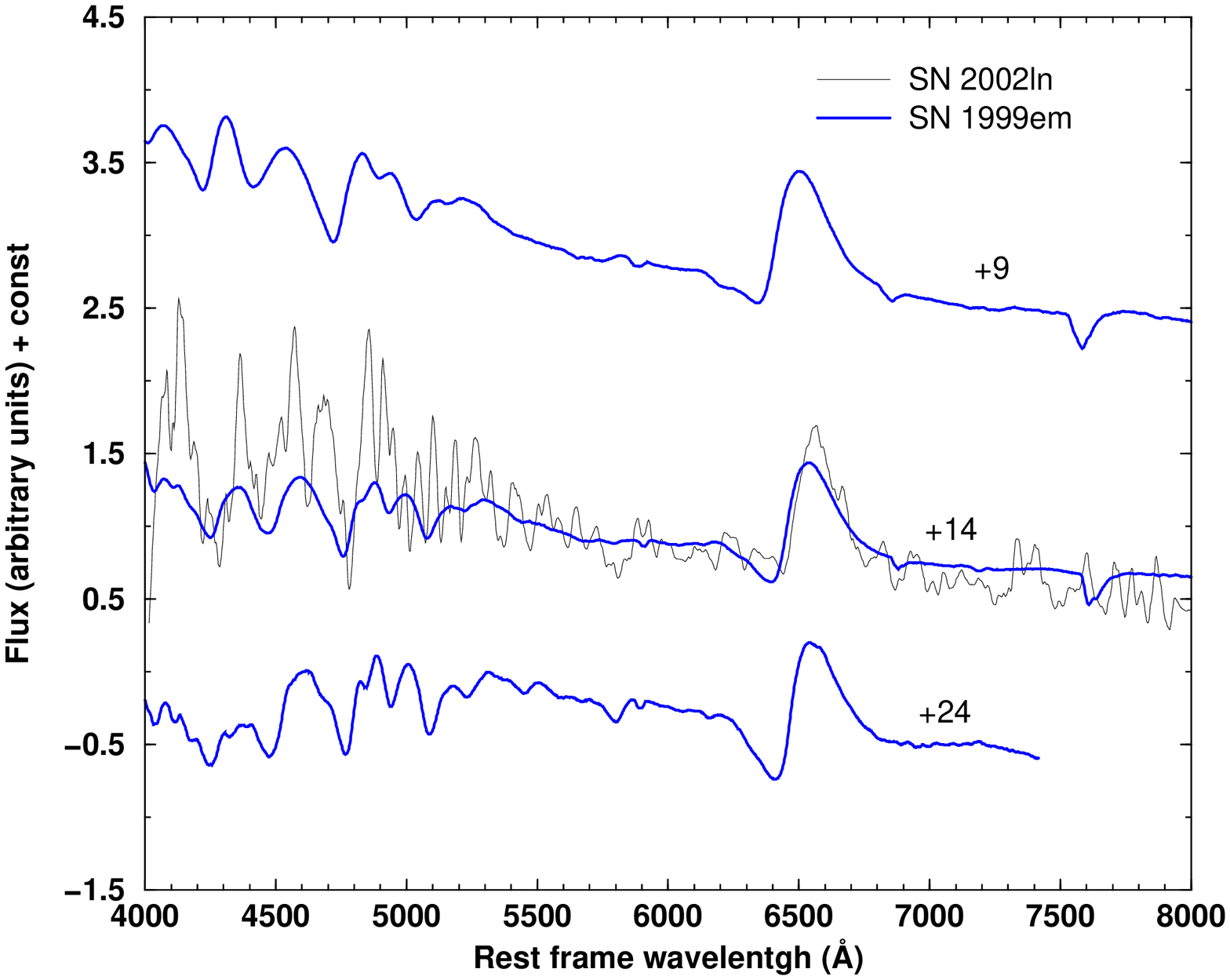}
      \caption{Left panel: Spectrum of the Type II supernova
 SN 2002ln, including galaxy lines for the redshift determination.
       Right panel: Comparison of SN 2002ln spectrum with that of SN
   1999em  about 9, 14, and 24 since $B$ maximum. 
Both SN 2002ln and SN 1999em spectra have been smoothed.}
   \label{Figure 14}
    
   \end{figure}

   \begin{figure}
   \centering
   \includegraphics[angle=-90,width=0.45\columnwidth]{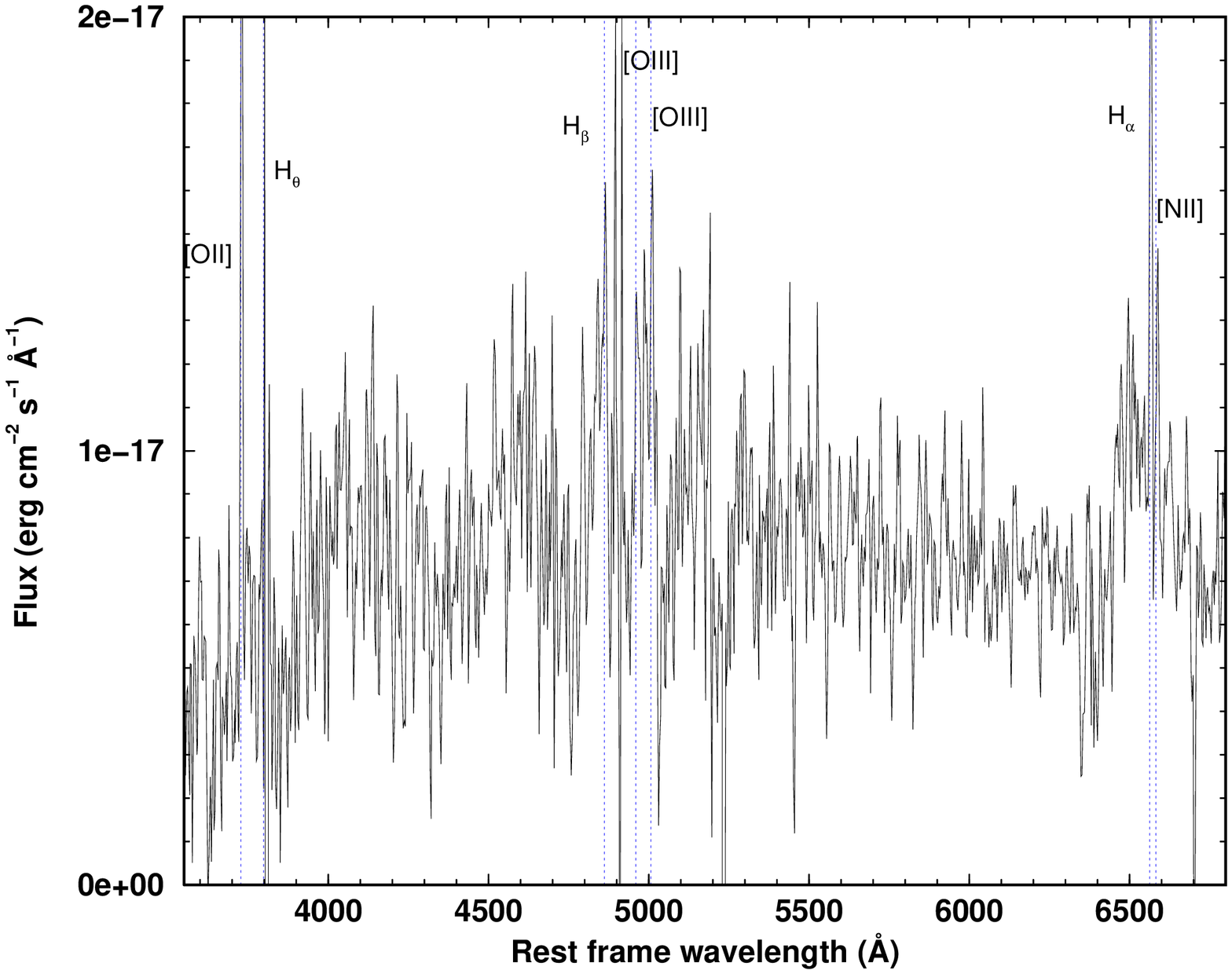}
   \includegraphics[angle=-90,width=0.45\columnwidth]{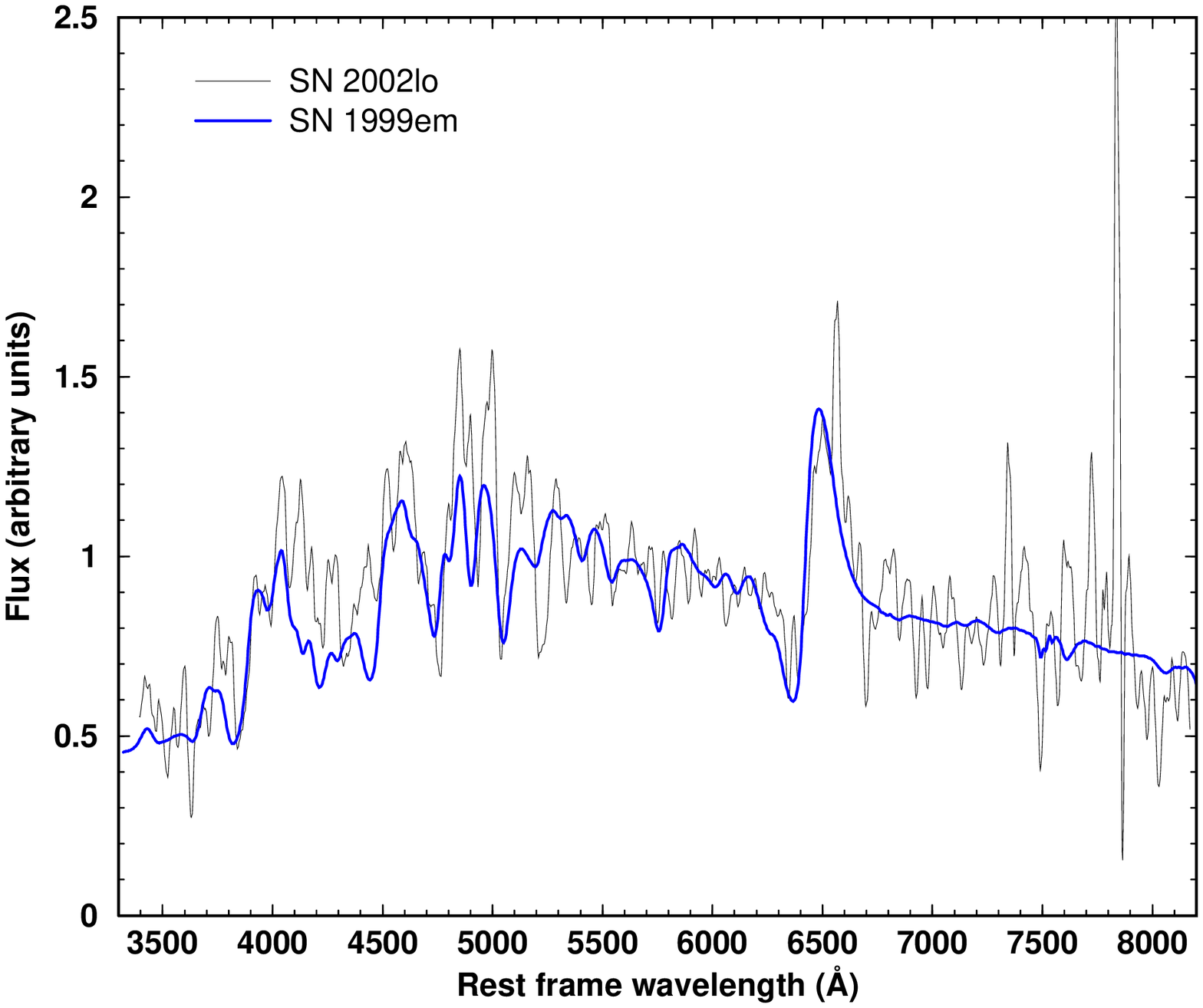}
      \caption{Left panel: Spectrum of the Type II supernova 
SN 2002lo, including galaxy lines for the redshift determination. Right panel: 
comparison of SN 2002lo spectrum with that of SN 1999em $\sim$35 days 
since $B$ maximum. Both SN 2002lo and SN 1999em spectra have been smoothed.}
        \label{Figure 15}
   \end{figure}

   \begin{figure}
   \centering
   \includegraphics[angle=0,width=\columnwidth]{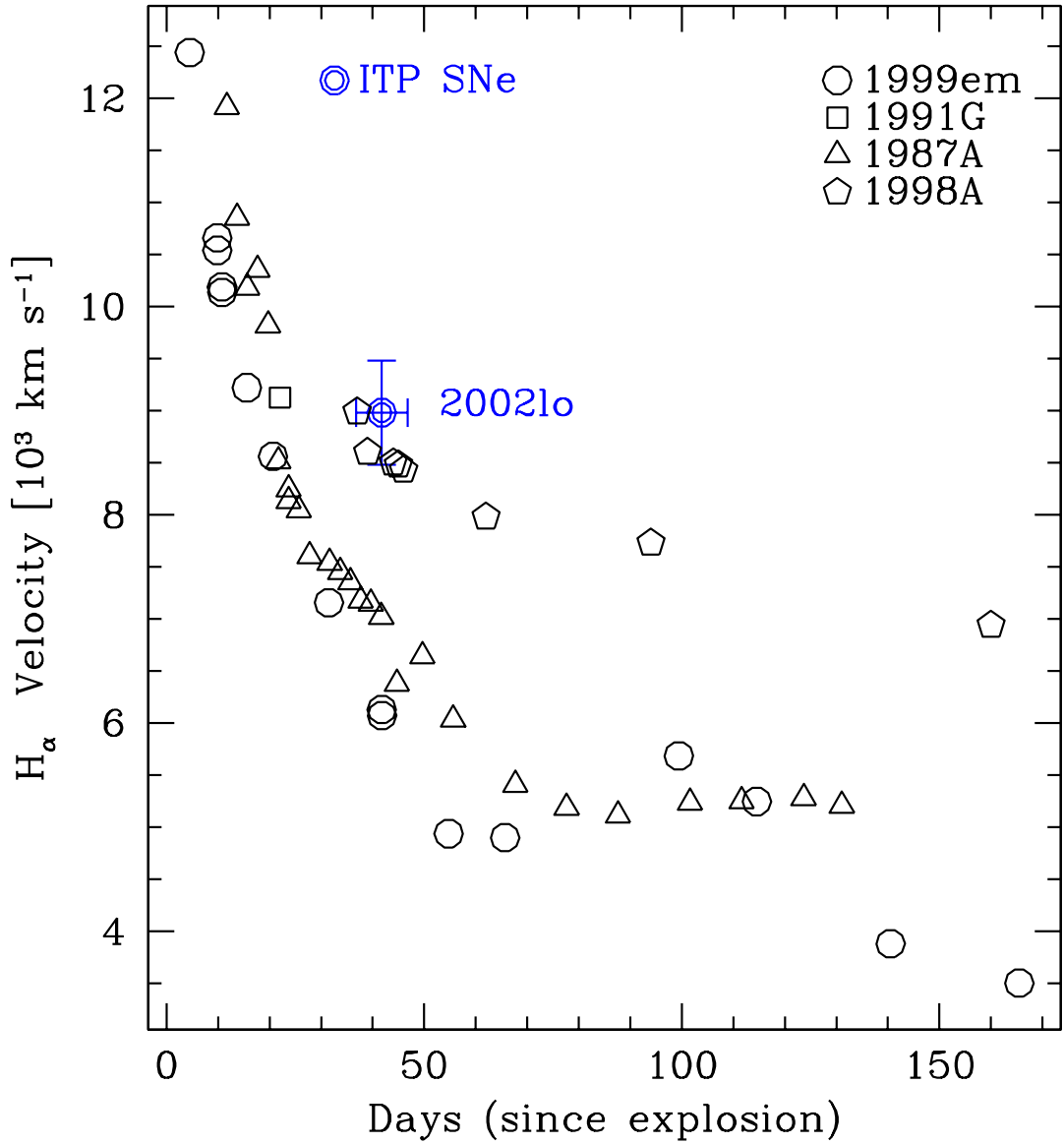}
        \caption{Expansion velocity for H$_{\alpha}$ (6562.8 \AA) as 
         deduced from its minimum, in SN 2002lo, compared with those
         of SN 1999em (Elmhamdi et al. 2003), 1991G (Blanton et al. 1995),
         1987A (Phillips et al. 1989), and 1998A (Pastorello et al. 2005).}
         \label{Figure 16}
   \end{figure}

   \begin{figure}
   \centering
   \includegraphics[angle=0,width=\columnwidth]{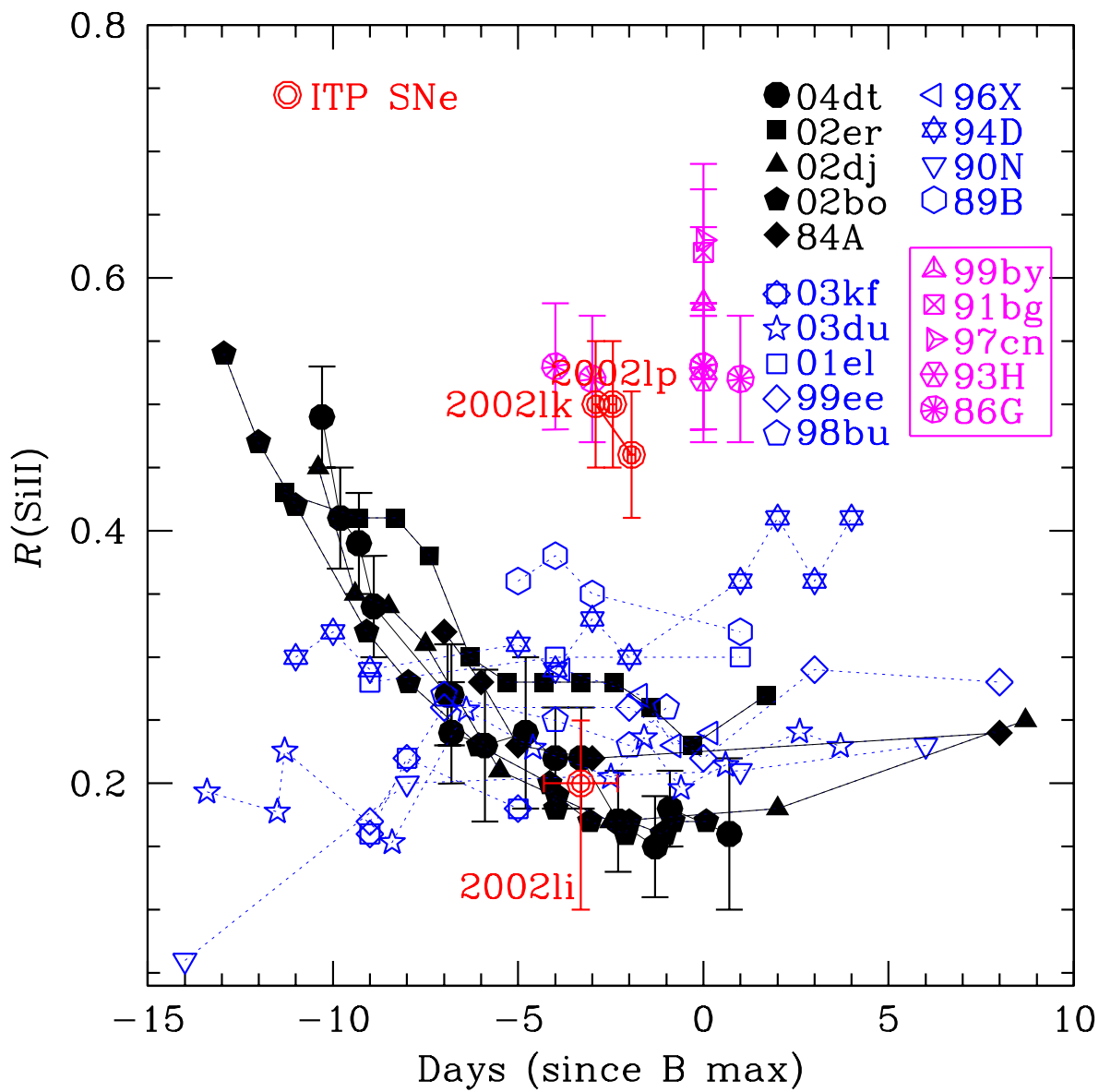}
  \caption{$\mathcal{R}$(\SiII) parameter
for SN 2002lk, SN 2002lp and SN 2002li, and for a  sample of 
nearby SNe shown as comparison.
Filled symbols refer to High Velocity Gradient SNe, open symbols to Low 
Velocity Gradient SNe as defined in  Benetti et al. (2005). FAINT SNe, 
listed in the box, are also shown.
}
\label{Figure 17}
   \end{figure}


   \begin{figure}
   \centering
 XC  \includegraphics[angle=0,width=\columnwidth]{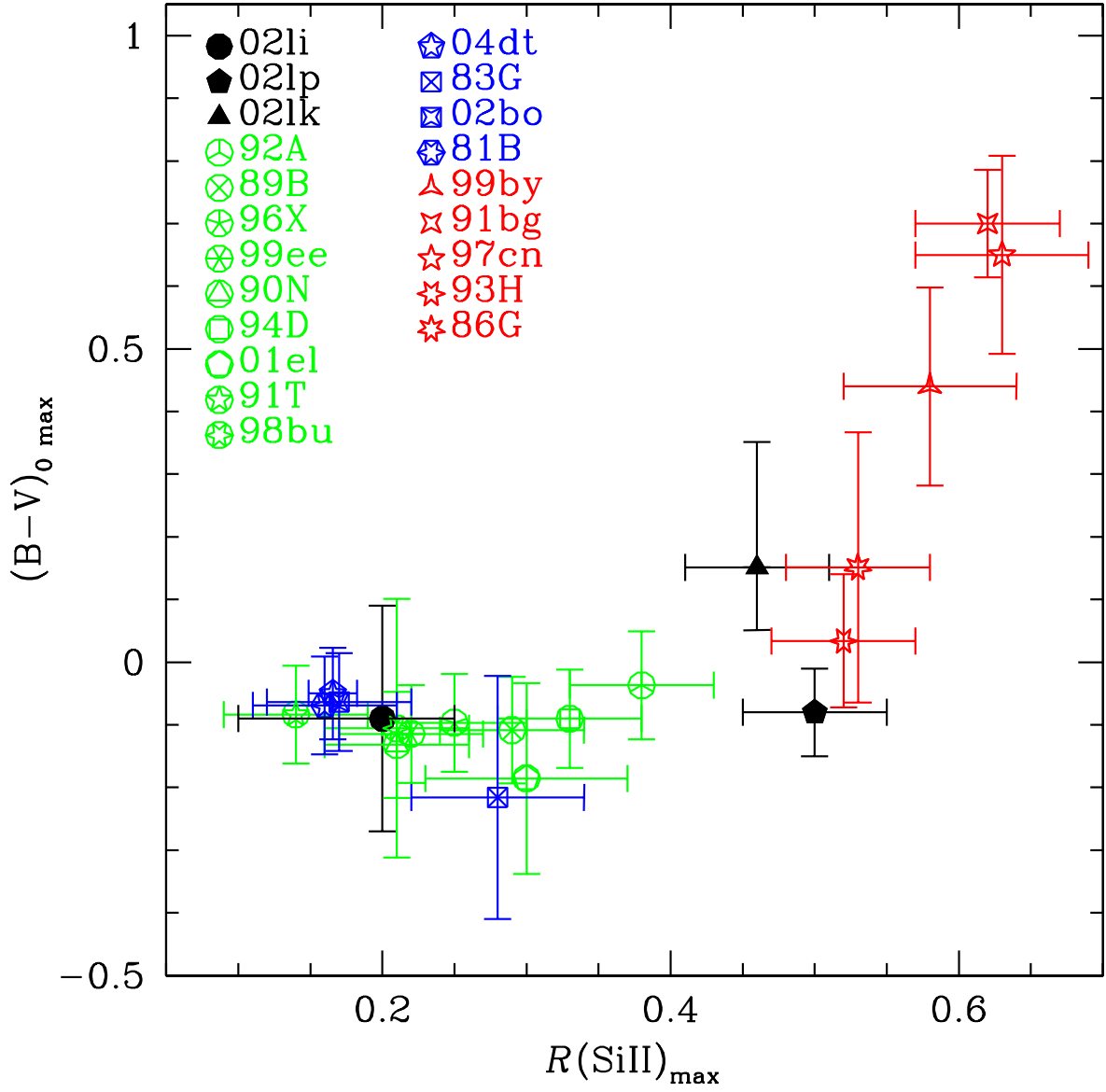}
  \caption{$\mathcal{R}$(\SiII) parameter and intrinsic (B-V).}

\label{Figure 18}
   \end{figure}

   \begin{figure}
   \centering
   \includegraphics[angle=0,width=\columnwidth]{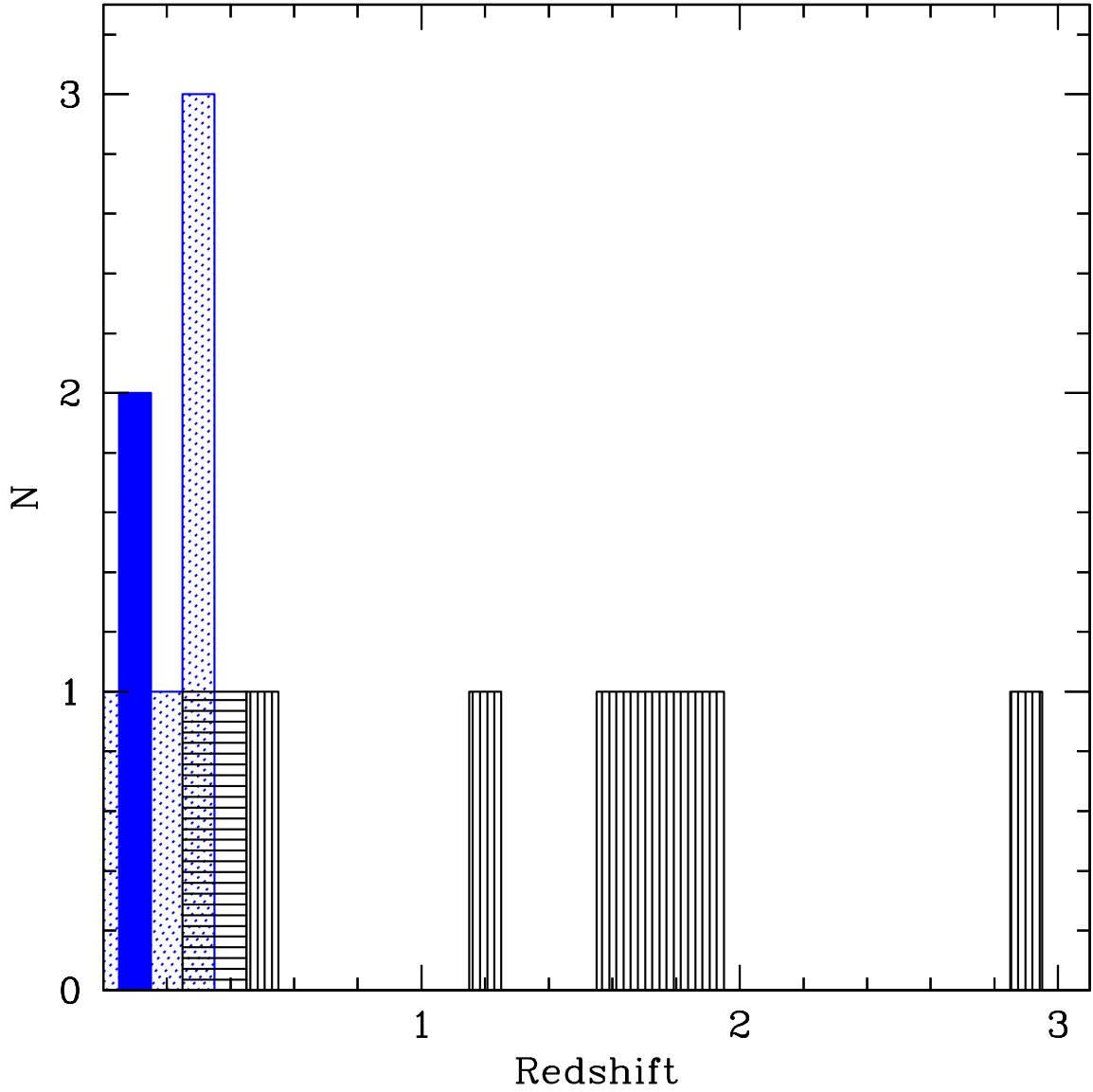}
  \caption{Redshift distribution of the observed objects. Filled histogram: 
Type II SNe; dotted histogram: Type Ia SNe; horizontal line histogram: 
Seyfert galaxies; vertical lines histogram: QSOs. redshift bin: 0.1.}
\label{Figure 19}
   \end{figure}



   \begin{figure*}
   \centering
   \includegraphics[angle=0,width=0.6\columnwidth]{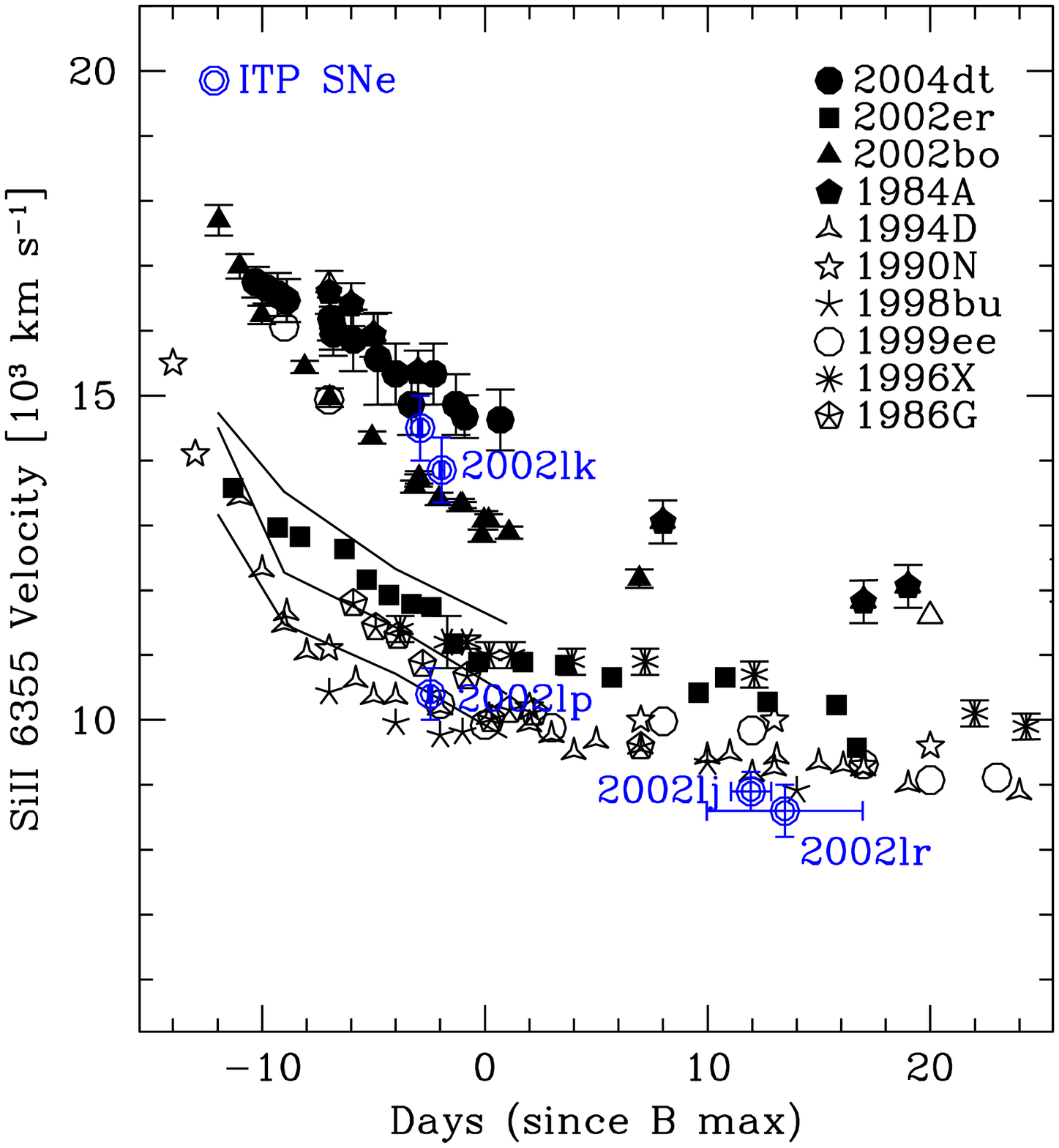}
   \includegraphics[angle=0,width=0.6\columnwidth]{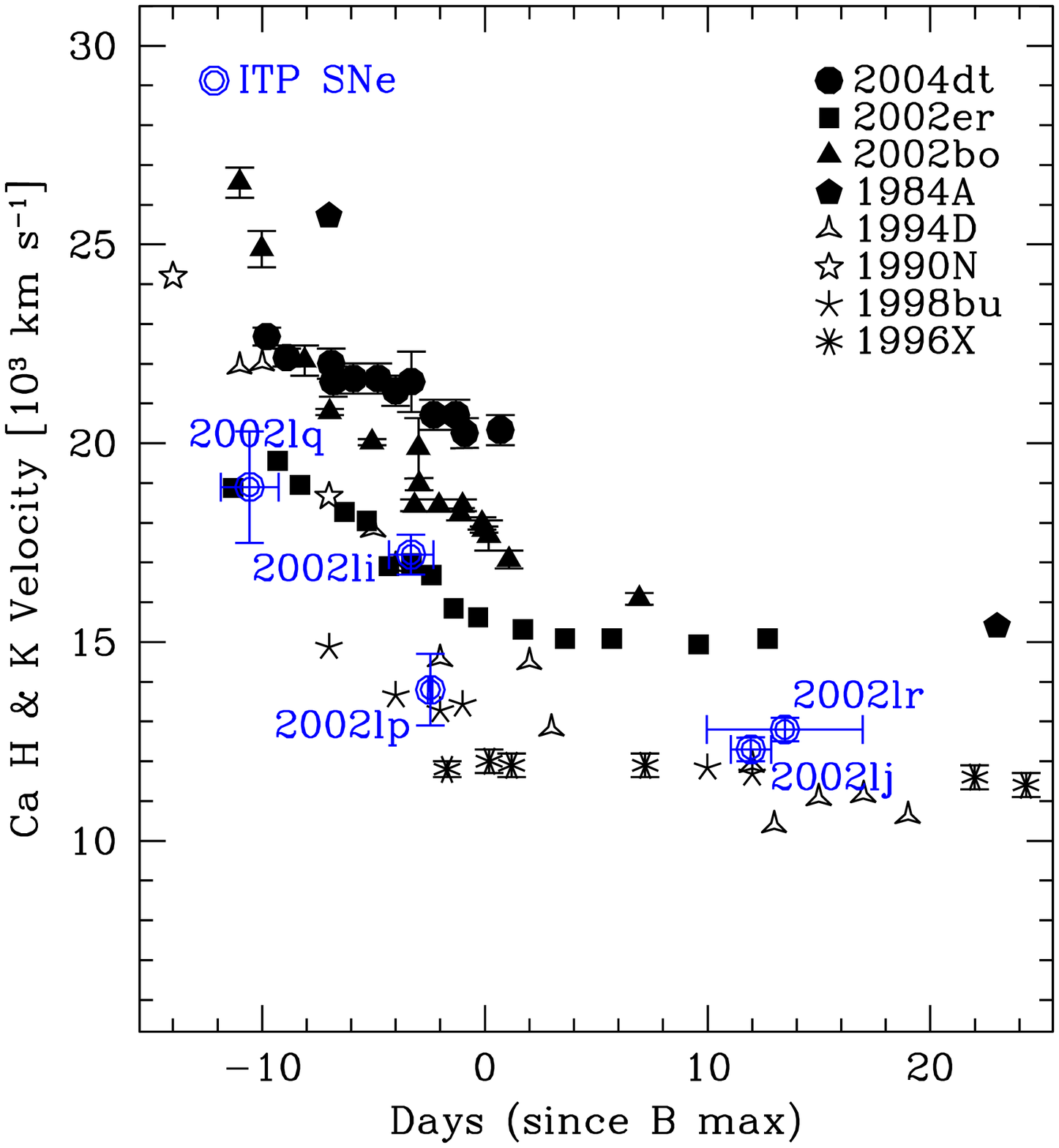}

      \caption{Expansion velocities for \SiII\ (6355 \AA) (upper panel) and  
           \CaII\ (3950 \AA) (lower panel) 
   as deduced from their minima, compared with those of other SNe.
   The \SiII\ evolution expected for different metallicities (solid line, top:
$\times 10$ solar metallicity; middle: $\times 1$;
 bottom: $\times 0.1$) is also shown (Lentz et al. 2000).
 The phase has been determined from the light curve fitting.}
         \label{Figure 20}
   \end{figure*}

   \begin{figure}
   \centering
   \includegraphics[angle=0,width=0.80\columnwidth]{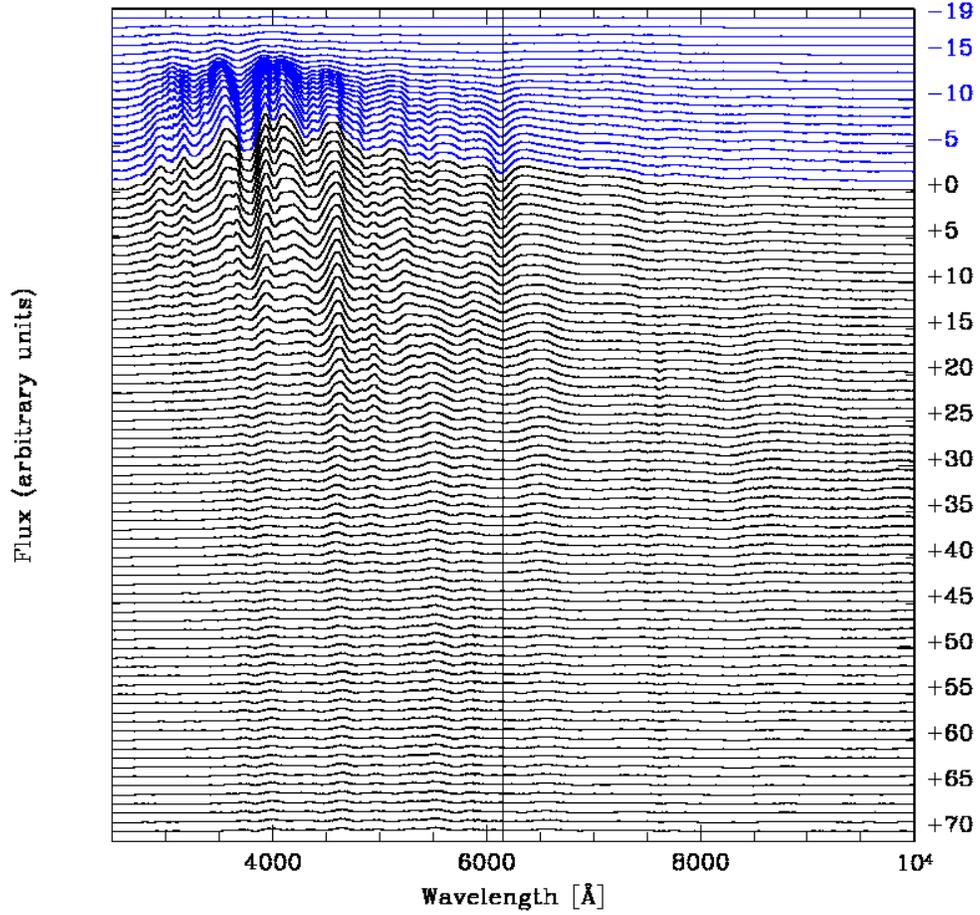}
      \caption{Synthetic spectra by Nobili et al. (2003). A vertical line 
marks the
 position of the \SiII\ $\lambda 6150$ absorption. Wavelength range limited 
to 2500--10000 \AA.  Phases are shown on the right.}
         \label{Figure 21}
   \end{figure}

  \begin{figure}
   \centering
   \includegraphics[angle=0,width=0.50\columnwidth]{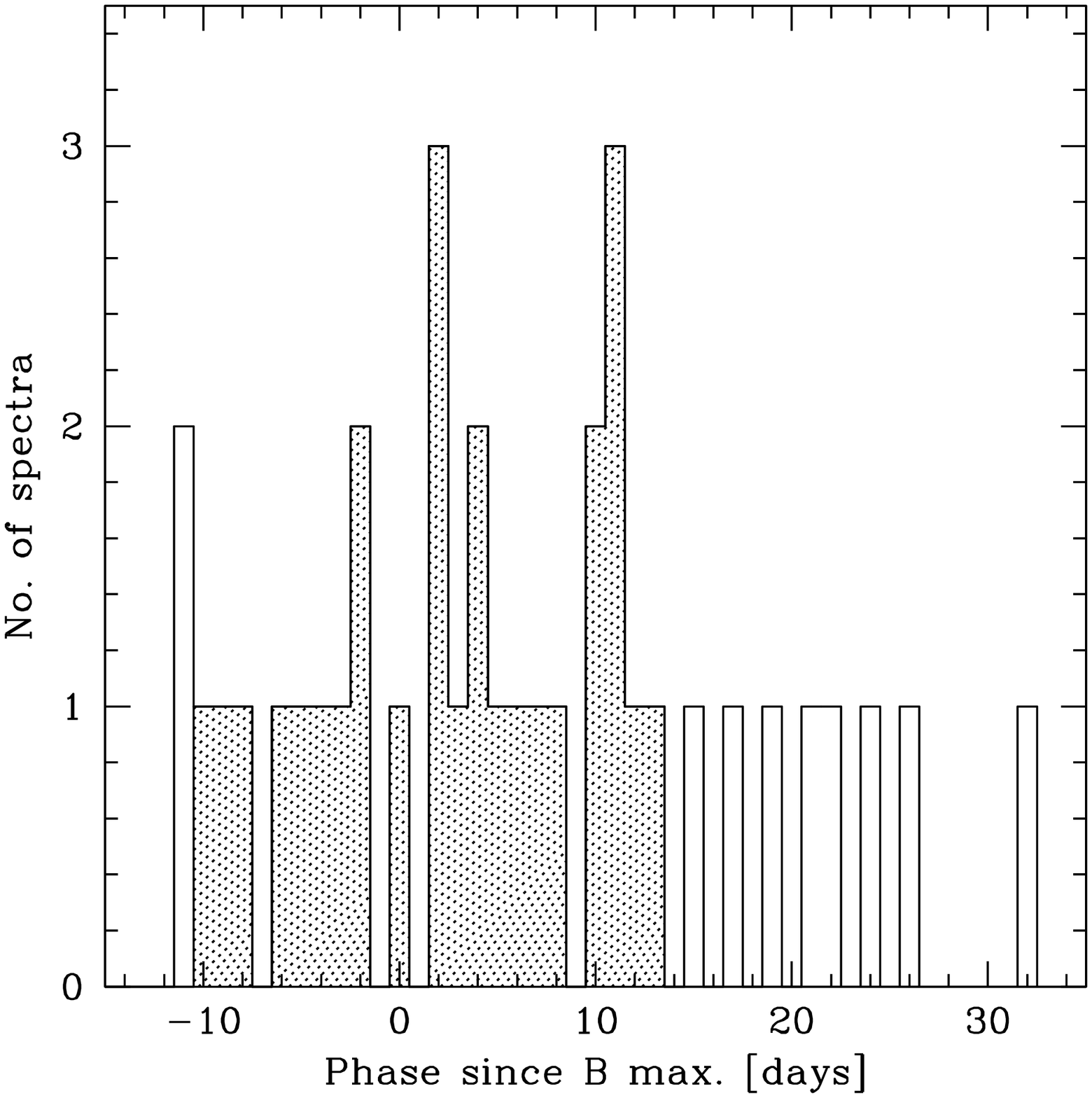}
      \caption{Phase distibution of the normal SN Ia templates listed
   in Table 5 (limited to 35 days since $B$ maximum). The
   dotted
 area shows the phase range of our SN Ia sample. Phase bin: 1 day.}
         \label{Figure 22}
   \end{figure}


\begin{thebibliography}{}
\bibitem{alta04}
Altavilla G. et al. 2005, in Turatto M,  Benetti S.,  Zampieri L.,  Shea
W., eds, ASP Conf. Ser. Vol. 342, 1604-2004: Supernovae as
Cosmological Lighthouses. Astron. Soc. Pac., San Francisco, p. 486

\bibitem{aldersnfac} 
Aldering, G. et al. 2004, Proceedings of SPIE, 4836, 61 

\bibitem{astier}
Astier, P. et al. (The SNLS Collaboration) 2005, A \& A (in press)

\bibitem{bala04} 
Balastegui A. et al. 2005, in Turatto M,  Benetti S.,  Zampieri L.,  
Shea W., eds, ASP Conf. Ser. Vol. 342, 1604-2004: Supernovae as Cosmological 
Lighthouses. Astron. Soc. Pac., San Francisco, p. 490

\bibitem{balland06} 
Balland C. et al., 2006, A\&A, 445, 387

\bibitem{barbon1990} 
Barbon R., Benetti S., Cappellaro E.,  Rosino L.,  Turatto M. 1990, 
A\&A, 237, 79

\bibitem{barris2004} 
Barris B.J. et al., 2004, ApJ, 602, 571 

\bibitem{benettibo} 
Benetti S. et al., 2004, MNRAS, 348, 261

\bibitem{Bene05} 
Benetti S. et al., 2005, ApJ, 623, 1011

\bibitem{blanton1995} 
Blanton E.L., Schmidt B.P., Kirshner R.P., Ford C.H., Chromey F.R., 
Herbst W. 1995, AJ, 110, 2868

\bibitem{branch1993b} 
Branch D., van den Bergh S. 1993, AJ, 105, 2231

\bibitem{branch1993a} 
Branch D., Fisher A., Nugent P. 1993, AJ, 106, 2383

\bibitem{branch06}
 Branch D., Dang, L. C., Hall, N., Ketchum, W., Melakayil, M., 
 Parrent, J., Troxel, M. A., Casebeer, D., Jeffery, D. J. \& Baron E.,
2006, astro-ph/0601048 

\bibitem{clocch} 
Clocchiatti A. et al. 2005, astro-ph/0510155, accepted for publication in ApJ 

\bibitem{deustua2000} 
Deustua S. et al. 2000, AAS, 196, 3212

\bibitem{dilday2005} 
Dilday, B. et al. 2005, AAS, 20718005D

\bibitem{elmhamdi2003} 
Elmhamdi A. et al. 2003 MNRAS, 338, 939

\bibitem{filippenko1989} 
Filippenko A.V. 1989, PASP, 101, 588

\bibitem{filippenko1997} 
Filippenko A.V., 1997, ARA\&A, 35, 309

\bibitem{garnavich1998} 
Garnavich P., Jha S.,  Kirshner R. 1998, IAU Circ. 6980

\bibitem{ham} 
Hamuy M., Phillips M.M., Suntzeff N.B., Nicholas B., Schommer R.A., 
Maza  J., Aviles R. 1996, AJ, 112, 2398

\bibitem{hamuy2001} 
Hamuy M. et al. 2001, ApJ, 558, 615

\bibitem{hamuy2002} 
Hamuy M. et al. 2002, AJ, 124, 417

\bibitem{hatano2000} 
Hatano K., Branch D., Lentz E.J.,  
Baron  E., Filippenko A.V., Garnavich P.M. 2000, ApJ, 543, 49

\bibitem{hille}
Hillebrandt, W., et al. 2005, 
 http://www.mpa-garching.mpg.de/~rtn 


\bibitem{hook05}  
Hook I.M. et al. 2005, AJ, 130, 2788

\bibitem{howell2001} 
Howell D.A. 2001, ApJ, 554, 193

\bibitem{jha2006}  
Jha S.,   Branch D.,  Chornock R., Foley R.J.,  Li W., Swift B.J., 
Casebeer D., Filippenko A.V. 2006, astro-ph/0602250, submitted to  AJ

\bibitem{kennicutt92} 
Kennicutt R.C. 1992, ApJSS, 79, 255

\bibitem{knop2003} 
Knop R.A. et al. 2003, ApJ, 598, 102

\bibitem{krisc}
Krisciunas K. et al. 2005, AJ, 130, 2453

\bibitem{leibundgut91} 
Leibundgut B., Kirshner R.P., Filippenko A.V., Shields, J.C.,  
Foltz  C.B.,  Phillips M.M., Sonneborn G. 1991, ApJ, 371, 23

\bibitem{li2003} 
Li W. et al. 2003, PASP, 115, 453

\bibitem{lira95} 
Lira, P. 1995, Master Thesis, Univ. Chile 

\bibitem{lentz2000} 
Lentz E.J., Baron E., Branch D.,  Hauschildt P.H.,  
Nugent P.E. 2000, ApJ, 530, 966 

\bibitem{leonard2002} 
Leonard D.C. et al., 2002, PASP, 114, 35

\bibitem{leonard2005} 
Leonard D.C., Li, W., Filippenko A.V., Foley R.J., Chornock R. 2005, 
ApJ, 632, 450

\bibitem{matheson2005}	
Matheson T. et al. 2005, AJ, 129, 2352

\bibitem{mazzali93}
Mazzali, P. A., Lucy, L.B, Danziger, I.J., Gouiffes, C., 
Cappellaro, E. \& Turatto, M., 1993, A \& A, 269, 423

\bibitem{mendez04} 
M\'endez J.  et al. 2005, in Turatto M,  Benetti S.,  Zampieri L.,  Shea W., 
eds, ASP Conf. Ser. Vol. 342, 1604-2004: Supernovae as Cosmological 
Lighthouses. Astron. Soc. Pac., San Francisco, p. 488

\bibitem{modjaz2001} 
Modjaz M., Li W., Filippenko A.V., King J.Y., 
Leonard D.C., Matheson  T., Treffers, R.R., Riess A.G. 2001, PASP, 113, 308 

\bibitem{serena}
Nobili S., Goobar A., Knop R., Nugent, P.  2003, A\&A, 404, 901

\bibitem{nugentemp}
Nugent, P., Kim. A. \& Perlmutter,S. 2002, PASP, 114, 803 

\bibitem{nugent1995} 
Nugent  P., Phillips M., Baron E.,  Branch. D., Hauschildt  P. 1995, 
ApJ, 455, 147

\bibitem{pastorello2005} 
Pastorello A. et al. 2005, MNRAS, 360, 950 

\bibitem{patat1994} Patat F., Barbon R., Cappellaro E., Turatto M.
1994, A\&A, 282, 731

\bibitem{patat} 
Patat F., Benetti S., Cappellaro E., Danziger, I.J., della Valle M., 
Mazzali P.A., Turatto M. 1996, MNRAS, 278, 111

\bibitem{perl98} 
Perlmutter S. et al. 1998, Nature, 391, 51

\bibitem{perl} 
Perlmutter S. et al. 1999, ApJ, 517, 565

\bibitem{phillips1987}  
Phillips M.M. et al. 1987, PASP, 99, 592

\bibitem{phillips1989} 
Phillips M.M., Heathcote S.R., Hamuy M., Navarrete M. 1989, 
AJ, 95, 1087 

\bibitem{phillips1999}
Phillips M.M., Lira, P., Suntzeff, N.R., et al. 1999, ApJ, 118, 1766 

\bibitem{pritchet2005} 
Pritchet C.J., For The SNLS Collaboration 2005, 
in  Wolff S.C., Laurer T.R., eds,  ASP Conf. Ser. Vol. 339, Observing 
Dark Energy. Astron. Soc. Pac., San Francisco, p.  60 

\bibitem{riessepoch} 
Riess A.G.  et al. 1997, ApJ, 114, 722

\bibitem{riess} 
Riess A.G. et al. 1998, AJ, 116, 1009

\bibitem{riess99} 
Riess A.G., Filippenko A.V., Li W., Schmidt B.P. 1999, AJ, 118, 2668

\bibitem{riess05} 
Riess A.G. et al. 2004, ApJ, 607, 665

\bibitem{richmond1994} 
Richmond M.W., Treffers R.R., Filippenko A.V., Paik Y., Leibundgut B., 
Schulman E., Cox C.V. 1994, AJ, 107, 1022 

\bibitem{ruiz-lapuente06}
 Ruiz--Lapuente P. 2006, in Bernard's Cosmic Stories. Conference held in
Valencia, June 2006.  

\bibitem{sakoa} 
Sako, M. et al. 2005a (astro--ph/0504455)

\bibitem{sakob} 
Sako, M. et al. 2005b, AAS, 207, 15002

\bibitem{schlegel1998}  
Schlegel, D.J., Finkbeiner, D.P., Davis, M. 1998, ApJ, 500, 525 


\bibitem{stehle2004}
Stehle M., Mazzali, P. A., Benetti, S. \& Hillebrandt, W. 2004, MNRAS,
360, 1231

\bibitem{sullivan03} 
Sullivan M. et al. 2003, MNRAS, 340, 1057

\bibitem{tonry} 
Tonry J.L. et al. 2003, ApJ, 594, 1

\bibitem{turatto2003}  
Turatto M., Benetti S., Cappellaro E. 2003, in  Leibundgut B, Hillebrandt W., 
eds, Proc. of the ESO/MPA/MPE Workshop: From Twilight to Highlight - The 
Physics of Supernovae. Springer-Verlag, Berlin, p. 200 

\bibitem{valenti2005} 
Valenti S.  et al. 2005, in Turatto M,  Benetti S.,  
Zampieri L.,  Shea W., eds, ASP Conf. Ser. Vol. 342, 1604-2004: Supernovae 
as Cosmological Lighthouses. Astron. Soc. Pac., San Francisco, p.  505

\bibitem{valentini2003}  
Valentini G.  et al. 2003, ApJ, 595, 779

\bibitem{wang2004}  
Wang L.,  Baade D., Hoeflich P., Wheeler J.C., Kawabata K., Khokhlov A., 
Nomoto K., Patat F. 2004,  astro-ph/0409593, submitted to ApJ

\bibitem{wells1994} 
Wells L.A. et al. 1994, AJ, 108, 2233
\end{thebibliography}
\end{document}